\documentclass[
	aps, prd, reprint,
	10pt, notitlepage, a4paper,
        floats, floatfix,
	amsmath, amssymb, amsfonts, eqsecnum,
	superscriptaddress,
	showpacs, showkeys,
	nofootinbib,longbibliography,
]{revtex4-2}

\usepackage[usenames,dvipsnames]{xcolor}
\usepackage{xspace} 
\usepackage{bm} 
\usepackage[utf8]{inputenc}
\usepackage{mathrsfs}
\usepackage{graphicx}
\usepackage{float}
\usepackage[normalem]{ulem}
\graphicspath{ {Figures/} }
\usepackage{aas_macros}
\usepackage{ragged2e}

\xdefinecolor{mylinkcolor}{rgb}{0,0,0.5}
\usepackage[
	bookmarksnumbered, bookmarksopen, bookmarksopenlevel=2,
	breaklinks=true,
	colorlinks=true, filecolor=mylinkcolor, citecolor=mylinkcolor,
	linkcolor=mylinkcolor, urlcolor=mylinkcolor, menucolor=mylinkcolor,
]{hyperref}

\def\ee{\end{equation}}
\def\eea{\end{eqnarray}}
\def\be{\begin{equation}}
\def\bea{\begin{eqnarray}}
\def\p{\partial}

\allowdisplaybreaks

\begin{document}

\newcommand{\AEI}{\affiliation{Max-Planck-Institute for Gravitational Physics (Albert-Einstein-Institute),\\ Am M{\"u}hlenberg 1, 14476 Potsdam-Golm, Germany, European Union}}
\newcommand{\UU}{\affiliation{Institute for Theoretical Physics, Utrecht University,\\ Princetonplein 5, 3584 CC Utrecht, Netherlands, European Union}}

\title{Tidal properties of neutron stars in scalar-tensor theories of gravity}

\date{\today}

\author{Gast\'on Creci}\email{g.f.crecikeinbaum@uu.nl}
\UU
\author{Tanja Hinderer}
\UU
\author{Jan Steinhoff}\AEI

\begin{abstract}
A major science goal of gravitational-wave (GW) observations is to probe the nature of gravity and constrain modifications to General Relativity. An established class of modified gravity theories are scalar-tensor models, which introduce an extra scalar degree of freedom. This affects the internal structure of neutron stars (NSs), as well as their dynamics and GWs in binary systems, where distinct novel features can arise from the appearance of scalar condensates in parts of the parameter space.
To improve the robustness of the analyses of such GW events requires advances in modeling internal-structure-dependent phenomena in scalar-tensor theories. We develop an effective description of potentially scalarized NSs on large scales, where information about the interior is encoded in characteristic
Love numbers or equivalently tidal deformabilities. We demonstrate that three independent tidal deformabilities are needed to characterize the configurations: a scalar, tensor, and a novel 'mixed' parameter, and develop the general methodology to compute these quantities. We also present case studies for different NS equations of state and scalar properties and  
provide the mapping between 
the deformabilities in different frames often used for calculations. Our results have direct applications 
for future GW tests of gravity and studies of potential degeneracies with other uncertain physics such as the equation of state or presence of dark matter in NS binary systems. 
\end{abstract}

\maketitle

\makeatletter
\def\l@subsubsection#1#2{}
\makeatother

\tableofcontents
\newpage
\section{Introduction}
Binary systems of compact objects such as neutron stars (NSs) or black holes are key sources of gravitational waves (GWs). The GW signals depend in a very specific way on the parameters of the system, for example the masses, spins, and eccentricity~\cite{Cutler:1994ys,LIGOScientific:2016wyt}. Furthermore, GWs also contain unique information on the fundamental physics of strong-field gravity and the interior composition of compact objects~\cite{LIGOScientific:2021sio,LIGOScientific:2018cki,LIGOScientific:2017vwq}. Among the GW signatures that are especially sensitive to the nature and internal structure of the objects are tidal effects. The dominant adiabatic effects are parametrized by a tidal deformability, or Love number, which characterizes the body's response to a tidal field~\cite{Love1908}. This is familiar from Newtonian gravity, where, for the same applied tidal field, a body will deform differently depending on its composition or Equation of State (EoS), which in turn impacts its exterior gravitational potential. A relativistic generalization of these concepts~\cite{Hinderer:2007mb,Flanagan:2007ix,Damour:2009vw,Binnington:2009bb,Steinhoff:2016rfi} underpins gravitational-wave tests of the nature of compact objects~\cite{Cardoso:2017cfl,Sennett:2017etc,Cardoso:2019rvt,Ciancarella:2020msu,Herdeiro:2020kba,DeLuca:2021ite} and ways to search for dark matter signatures~\cite{Nelson:2018xtr,Ellis:2018bkr,Quddus:2019ghy,Das:2020ecp,Ciancarella:2020msu,Collier:2022cpr,Leung:2022wcf,Lourenco:2022fmf,Karkevandi:2021ygv,Hippert:2022snq,Diedrichs:2023trk,Thakur:2023aqm}. The most prominent role of tidal effects is for GW probes of the EoS of nuclear matter at high densities in NSs~\cite{Hinderer:2009ca,Han:2018mtj,LIGOScientific:2018cki,Pereira:2020jgv}, which at present and despite recent progress from multimessenger observations~\cite{LIGOScientific:2017ync,LIGOScientific:2017zic,LIGOScientific:2017pwl,Raaijmakers:2019qny,Riley:2019yda,Miller:2019cac,Jiang:2019rcw,Dietrich:2020efo,Zimmerman:2020eho} remains among the major future science goals of nuclear astrophysics~\cite{AlvesBatista:2021eeu,DOE,NuPEEC}.

The tidal deformability associated with a matter configuration further depends on the theory of gravity~\cite{Saffer:2021gak,Shao:2022koz,Ajith:2022uaw}. This has, for instance, been used jointly with multimessenger data to set unprecedentedly stringent constraints on higher-curvature extensions of General Relativity~\cite{Silva:2020acr}. Recent work has also demonstrated the use of tidal deformability to constrain scalar-tensor theories~\cite{Pani:2014jra,Brown:2022kbw}, which introduce an extra scalar degree of freedom. The presence of the scalar field gives rise to a rich phenomenology, for instance, depending on the parameters of the theory and properties of the NS, scalar condensates may form in and around NSs either in isolation~\cite{Damour_1992,PhysRevLett.70.2220,Damour:1996ke,Damour:1998jk,Andreou:2019ikc,Ventagli:2022fgg} or dynamically during an inspiral~\cite{Palenzuela:2013hsa,Khalil:2022sii,Doneva:2022ewd}. At large separation, the most striking effect in such inspiraling binaries is that they radiate dipolar scalar waves, in addition to GWs, which accelerates the inspiral. However, the dipole emission may be degenerate with tidal effects~\cite{Ma:2023sok,Mehta:2022pcn}. Hence it is necessary to work out accurate predictions for the strain that would be observed in GW detectors for specific gravity theories, including effects such as tides, although they might be expected to be small effects compared to scalar dipole radiation. However, tidal phenomena in scalarized systems include not only gravitational tides but also scalar tidal phenomena, where the scalar condensates respond to the gradients of the companion's scalar field across the matter distribution, which leads to further distinctive GW imprints~\cite{Bernard:2019yfz,vanGemeren:2023rhh}. 
In this paper, we advance the description of tidal effects during the inspiral of compact binaries when the gravitational theory is modified by an additional scalar field.
This complements calculations for the binary dynamics and radiation to higher orders in post-Newtonian theory~\cite{Bernard:2022noq,Bernard:2018hta,Bernard:2018ivi,Julie:2019sab,Shiralilou:2020gah,Shiralilou:2021mfl,Julie:2022qux}.
However, while the post-Newtonian approximation is a weak-field expansion, tidal effects crucially depend on the strong-gravity regimes inside and around the compact object, among other properties such as the uncertain EoS.
In particular, we demonstrate the importance of accounting for the details of the coupling between scalar and tensor modes in the strong-field regime.

Specifically, we reexamine the identification and numerical calculation of the various tidal deformability parameters from the perspective of an effective or skeletonized action description whereby the object is reduced to a worldline augmented with multipole moments. Such an effective description underpins computations of the dynamics and GWs in the post-Newtonian approximation. The coefficients appearing in the effective action must be carefully matched to the tidal response of a relativistic compact object, which we consider to linear order. That is, the tidal coefficients are extracted from linear perturbations of fully nonlinear solutions for the isolated stationary compact object, depending on both the EoS and strong-field modification of gravity. Here, we establish the methodology for facilitating this connection for scalarized configurations, showing that it involves three different deformabilities. We demonstrate how these are extracted from the perturbative information for the case of NSs in scalar-tensor theories. We also provide the mapping of these quantities between two frames commonly used in the calculations: the Jordan frame, where matter couplings to the metric are as in the standard model but the equations of motion for perturbed compact objects are highly complex, and the Einstein frame obtained after a conformal transformation and more convenient for calculations. 

This paper is organized as follows. We introduce scalar-tensor theories in Sec.~\ref{sec:STTheories}, where we distinguish between Jordan and Einstein frames and provide the equations of motion. In Sec.~\ref{sec:EFT} we provide the effective action for compact objects in scalar-tensor theories at orbital scales. In particular, we introduce a novel tidal deformability parameter needed to characterize the object's multipole moments. This introduces subtleties for the numerical extraction of multipolar and tidal moments. In Sec.~\ref{sec: ComputingLoveNumbers} we address such subtleties and show how to extract the information needed to compute the tidal deformabilities. In Sec.~\ref{sec:CaseStudy} we apply the framework to scalarized neutron stars and obtain the tidal deformabilities for configurations with different masses, radii, and scalar charges. Section~\ref{sec:SummaryDiscussion} summarizes our methodology and main findings and Section~\ref{sec:Conclusions} contains the conclusion. Additional technical details are delegated to several Appendixes, with Appendix~\ref{app:JordantoEinsteinAction} providing all the derivations relevant for the effective action and Appendices~\ref{app:JustCoordApp} and \ref{app: ADMMass}
giving rederivations and reviews of relevant calculations in our notation and conventions. Finally,  Appendix~\ref{app:PlotsLambdak} includes plots of the dimensionless Love number coefficients and the adimensional tidal deformability parameter commonly used in data analysis for different multipolar orders $\ell=1,2,3$. 

The notation and conventions we use are the following. We denote spacetime quantities by greek letters $\alpha$, $\beta$, $\dots$, and spatial components by latin indices $i$, $j$, $\dots$. We use $\nabla_\mu$ to denote the covariant derivative and $\p_\mu$ for the partial derivative. Capital-letter super- and subscripts, with the exception of the labels $T$, $S$, and $ST$, denote a string of indices on a symmetric and trace-free (STF) tensor (see e.g.~\cite{RevModPhys.52.299} for more details). For instance, for a unit three-vector ${n}^i$, the STF tensor $n^{L=2}=n^in^j-\frac{1}{3}\delta^{ij}$, where $\delta^{ij}$ is the Kronecker delta. We adopt the Einstein summation convention on all
types of indices, i.e., any repeated indices are summed over. Throughout the paper we use units where $G=c=1$ unless stated otherwise.

\section{Scalar-tensor theories}\label{sec:STTheories}

In this section, we briefly review the basics of scalar-tensor theories of gravity. 
The action is given by
\begin{align}\label{eq: STaction}
S_{\rm ST}^{\rm (J)}=&\int_{\mathcal{M}}d^4x \sqrt{-g}\left[K_R F(\phi) R-K_{\phi}\frac{\omega(\phi)}{\phi}\p^\mu\phi\p_\mu\phi-V(\phi)\right]\nonumber\\&+S_{\rm matter}\left[\psi_m,g_{\mu\nu}\right]~,
\end{align}
where $\phi$ is a scalar field, $\omega(\phi)$ is its self-coupling and $V(\phi)$ is a potential. The scalar field is coupled to the Ricci curvature scalar $R$ via a field-dependent function $F(\phi)$ and $K_R$ and $K_{\phi}$ are normalization constants\footnote{We keep the normalization generic here to encompass different choices in the literature.}. Throughout the paper we set $V(\phi)=0$. The matter action $S_{\rm matter}$  is a functional of the matter fields $\psi_m$ and metric $g_{\mu\nu}$. The action~\eqref{eq: STaction} is formulated in the so-called Jordan frame. Performing a (local) conformal transformation
\begin{align}\label{eq:MetricConformalTransf}
g_{\mu\nu}=A(\varphi)^2 g^\ast_{\mu\nu}~,
\end{align}
transforms the action to the so-called Einstein frame (see Appendix~\ref{app:JordantoEinsteinAction} for a detailed derivation). Here, we denote the metric in the Jordan frame by $g_{\mu\nu}$ and that in the Einstein frame by $g^\ast_{\mu\nu}$.\footnote{For convenience, we will place the asterisk wherever the indices are not placed, for example $g^{\ast\mu\nu}=g_\ast^{\mu\nu}~.$} In the Einstein frame,~\eqref{eq: STaction} becomes
\begin{align}
\label{eq:SST}
S_{\rm ST}^{\rm (E)}=&\int_{\mathcal{M}}d^4x\sqrt{-g^\ast}\left[K_R R_\ast- K_{\varphi} g_\ast^{\mu\nu}\p_\mu\varphi\p_\nu\varphi\right]\nonumber\\&+S_{\rm matter}\left[\psi_m,A^2(\varphi)g^\ast_{\mu\nu}\right]~.
\end{align}
Here $K_\varphi$ is the normalization constant of a new scalar field $\varphi$, defined by
\begin{align}
	&\frac{d\varphi}{d\phi}=\sqrt{\Delta}~,\\
	&\p_\alpha\phi=\frac{1}{\sqrt{\Delta}}\p_\alpha\varphi~,	
\end{align}
with
\begin{align}
\Delta\equiv\frac{3}{2}\frac{K_R}{K_{\varphi}}\left(\frac{F'}{F}\right)^2+\frac{K_\phi}{K_{\varphi}}\frac{\omega(\phi)}{\phi F}~.
\end{align}
The action~\eqref{eq:SST} is the action of a free scalar field $\varphi$ that is decoupled from the Ricci scalar. 
The field-dependent matter action in the second line of~\eqref{eq:SST} indicates that this transformation has led to a coupling between the scalar field $\varphi$ and matter through the conformal factor $A(\varphi)$, which is related to the coupling function $F(\phi)$ in~\eqref{eq: STaction} by 
\begin{align}
A(\varphi)=\exp\left(-\int d\varphi \frac{F'}{2F\sqrt{\Delta}}\right)~,
\end{align}
where a prime denotes a derivative with respect to the argument, $F'=dF/d\phi$. Analogously, we define\footnote{Note the minus sign in front of $1/A$. Depending on the different conventions in the literature, this parameter may be defined with a plus sign instead of a minus sign. For our purposes, the minus sign is more convenient for the transformations between frames.}
\begin{align}\label{eq:STalpha}
\alpha(\varphi)=-\frac{1}{A}\frac{dA}{d\varphi}=\frac{F'}{2F\sqrt{\Delta}}~,
\end{align}
where all the quantities are understood as functions of the Einstein frame field $\varphi$. The equation of motion for the metric and the scalar field in the Einstein frame are then given by
\begin{subequations}
\label{eq:eomEinsteinframe}
\begin{align}
G^\ast_{\mu\nu}&=\frac{1}{2K_R}T^{\ast}_{\mu\nu}+\frac{K_\varphi}{K_R}T^{\ast \varphi}_{\mu\nu}~,\label{eq:EoMMetric}\\
\Box\varphi&=\frac{1}{2K_\varphi}\alpha(\varphi) T^\ast~,\label{eq:EoMScalar}
\end{align}
\end{subequations}
where 
\begin{align}
G^\ast_{\mu\nu}=R^\ast_{\mu\nu}-\frac{1}{2}g^\ast_{\mu\nu}R^\ast
\end{align}
is the Einstein tensor in the Einstein frame, i.e. corresponding to the Einstein frame metric $g^\ast_{\mu\nu}$,
\begin{align}
    T^{\ast \varphi}_{\mu\nu}=\p_\mu\varphi\p_\nu\varphi-\frac{1}{2}g^\ast_{\mu\nu}\p^\ast_\gamma\varphi\p_\ast^\gamma\varphi
\end{align}
is the scalar field energy-momentum tensor,  \be
{T^{\ast}}^{\mu\nu}=\frac{2}{\sqrt{-g^\ast}}\frac{\delta{S_{\rm matter}}}{\delta{g^\ast_{\mu\nu}}}\label{eq:matterTmunu}
\ee
is the matter energy-momentum tensor, and $T^\ast=g^\ast_{\mu\nu}{T^{\ast}}^{\mu\nu}$ its trace. For practical calculations of compact object configurations and their perturbations, it is easiest to work in the Einstein frame and only transform to quantities in the Jordan frame at the end. 

For black holes ${T^\ast}^{\mu\nu}=0$ and the scalar equation of motion \eqref{eq:EoMScalar} becomes sourceless. This leads to the same solutions as in General Relativity (GR) that obey the no-hair theorem~\cite{Bekenstein:1995un,Sotiriou_2015}. For NSs, however, the matter configuration that entangles the metric and the scalar field that circumvents the no-hair theorem introduces interesting phenomena, e.g., depending on the parameters, a scalar condensate may appear. To describe NSs, we assume the matter energy-momentum tensor to have a perfect-fluid form. In the Einstein frame, we parametrize it as
\begin{align}
T^\ast_{\mu\nu}=\left(p^\ast+\rho^\ast\right)u^\ast_\mu u^\ast_\nu+p^\ast g^\ast_{\mu\nu}~,
\end{align}
with the four-velocity $u^\mu_*$ normalized as $u^*_\mu u_*^\mu=-1$. 
From ~\eqref{eq:matterTmunu}  with~\eqref{eq: STaction} -- ~\eqref{eq:SST}, we find that the energy-momentum tensor in the Einstein frame is related to its Jordan frame counterpart by
\begin{align}
T^\ast_{\mu\nu}=T_{\mu\nu}\frac{\delta{g_{\mu\nu}}}{\delta{g^\ast_{\mu\nu}}}=A^2(\varphi)T_{\mu\nu}~.
\end{align}
This relation, together with the additional details reviewed in the Appendix \eqref{eq:SqrtConformal} and \eqref{eq:DifferentialTransf}, yields the following relation between pressures and densities:
\begin{align}
p^\ast&=A(\varphi)^4p~,\label{eq:PressureToEinstein}\\
\rho^\ast&=A(\varphi)^4\rho~.\label{eq:DensityToEinstein}
\end{align}
We assume that the equation of state of the cold NS matter $p=p(\rho)$ is given in the original Jordan frame, where only the gravitational sector is modified while the description of subatomic matter according to the standard model of particle physics remains unaltered. Thus, to obtain the energy-momentum tensor in the Einstein frame, we use the relations above, \eqref{eq:PressureToEinstein} and \eqref{eq:DensityToEinstein}, to obtain
\begin{align}
T^\ast_{\mu\nu}=A(\varphi)^4\left[\left(p+\rho\right)u^\ast_\mu u^\ast_\nu+p g^\ast_{\mu\nu}\right]~.
\end{align}
In the Jordan frame, the energy-momentum tensor is conserved. In the Einstein frame, transforming the covariant derivative using~\eqref{eq: ChristoffelToEinstein} (see also Appendix D of \cite{Wald:1984rg}), the equation for energy-momentum conservation in the Einstein frame reads
\begin{align}
\nabla_\ast^\mu T^\ast_{\mu\nu}=-\alpha T^\ast \p_\nu\varphi~.\label{eq: EinsteinEMCons}
\end{align}
The system of equations \eqref{eq:EoMMetric} and  \eqref{eq:EoMScalar}, together with choices for the coupling and EoS, will be used in Sec.~\ref{sec: ComputingLoveNumbers} to compute NS configurations and their response to perturbations in scalar-tensor theories.

\section{Effective action }\label{sec:EFT}
We first analyze the above system of equations of motion to identify the connection between information on a perturbed NS and quantities impacting the orbital dynamics in a binary system. We consider nonspinning binaries at large separation, where there is a hierarchy of scales between the size of the objects, the orbital separation, and the wavelength of GWs. Here, in the case of scalarized NS configurations, the size of the objects includes the scalar condensate, which extends to much larger distances. Nevertheless, during the early inspiral at large separation, this setting is still amenable to an effective field theory (EFT) description, where the model for the binary at scales larger than the size of the bodies is obtained by integrating out the internal degrees of freedom. At the most coarse-grained order, each body reduces to a worldline, which is then augmented by information on its interior contained in effective (or Wilsonian) coefficients. This is often referred to as the skeletonization of the body~\cite{Dixon:1974xoz}. 

An example of such a connection in the context of adiabatic tidal effects in General Relativity is the following. When considering linearized perturbations to a compact object the time-time component of the metric $g_{00}$ can be written in terms of an effective potential $U_{N}$, whose asymptotic behavior at spatial infinity in coordinates whose origin is at the center of mass of the object reads
\begin{align}\label{eq:NewtonianPotential}
\lim_{r\to \infty}U_{N}=-\lim_{r\to \infty}\frac{g_{00}+1}{2}\sim \frac{M}{r}+\sum_{\ell=2}^\infty\frac{(2\ell-1)!!}{\ell!}\frac{Q_Ln^L}{r^{\ell+1}}~.
\end{align}
Here, each term of the series corresponds to a correction to a point particle, encoded in the multipole moment $Q_L$, contracted with STF multilinears of unit vectors $n^L$. Similarly, in the relativistic skeletonization approach from the EFT, we can describe a body as a worldline corresponding to a point particle,
plus corrections containing information about size effects, 
spin-orbit couplings, etc. Such considerations lead to an effective action of the form
\begin{align}
S_{\rm EFT}=S_{g}+S_{\rm pm}+S_{\rm tidal}+\dots~,
\end{align}
with $S_{g}$ the underlying gravitational action and $S_{\rm pm}$ the action of a point mass. Focusing only on size effects, analogous to \eqref{eq:NewtonianPotential} the EFT reads 
\begin{align}\label{eq:NewtonianEFT}
    S_{\rm EFT}=S_{\rm pm}+\sum_{\ell=2}^\infty\int d\sigma \sqrt{-u_\mu u^\mu} \frac{\lambda_\ell}{2\ell!}E_LE^L~,
\end{align}
with $\sigma$ a worldline evolution parameter, $u^\mu=dx^\mu/d\sigma$ the four-velocity, $E_L$ an external tidal field, and $\lambda_\ell$ the tidal deformability, defined as the ratio between the induced multipole moment and the tidal field,
\begin{align}
Q_L=-\lambda_\ell E_L~,
\end{align}
related to the Love number $k_\ell$ by
\begin{align}\label{eq:LoveNumvsTidalDef}
k_\ell=\frac{(2\ell-1)!!}{2}\frac{\lambda_\ell}{R^{2\ell+1}}~,
\end{align}
with $R$ the radius of the star. In this study we focus on static size effects and, in particular, on electric-type perturbations (also called even or polar) of the form of~\eqref{eq:NewtonianEFT}, although the framework can also be applied to magnetic (odd or axial) perturbations in the case of dynamic tides~\cite{Gupta:2020lnv}. We give the EFT in both Jordan and Einstein frames, and the transformation between them. As in the full theory, we take advantage of the conformal transformation in order to match the EFT coefficients in the mathematically simpler Einstein frame. By using the transformations between frames, we then relate the Jordan and Einstein frame EFT coefficients and obtain all our coefficients in terms of Einstein frame quantities. Although we focus here on scalar-tensor theories, the methodology and framework can be applied to other contexts such as generalized scalar-tensor theories \cite{GastInPrep}.

\subsection{Effective action in the Jordan frame}
In the class of scalar-tensor theories considered here, the effective action in the Jordan frame including the skeletonized size effects reads 
\begin{align}
\label{eq:SEFT}
S_{\rm EFT}^{\rm (J)}=S_{\rm ST}^{\rm (J)}+S_{\rm pm} +S_{\rm tidal}~,
\end{align}
with the point-mass action~\cite{Damour:1998jk}
\begin{align}
S_{\rm pm}=-\int d\sigma~z~m(\phi)~,\label{eq:Spp}
\end{align}
where $\sigma$ is a worldline evolution parameter and $z=\sqrt{-u_\mu u^\mu}$ is the redshift factor. Note that the point-particle action~\eqref{eq:Spp} contains a field-dependent mass term. This is because the binding energy in the Jordan frame depends on the scalar field and, since the mass is related to the energy, it is likewise a function of the field~\cite{1975ApJ...196L..59E}.

An effective action describing tidal effects in scalar-tensor theories has been considered previously for the case of dipolar tides~\cite{Damour:1998jk,Bernard:2019yfz}. This demonstrated that in general, two kinds of tidal fields arise in the system, namely the scalar (S) and tensor (T) fields defined by
\begin{subequations}
\label{eq:tidalfields}
\begin{align}
E_L^S&=\mathcal{P}_L^{\Pi}E_\Pi^S=-\underset{r\rightarrow0}{\text{FP}}\mathcal{P}_L^\Pi\nabla_\Pi\phi~,\\ 
E_L^T&=\mathcal{P}_L^\Pi E_\Pi^T=\underset{r\rightarrow0}{\text{FP}}\mathcal{P}_L^\Pi \nabla_{\Pi-2}E_{\mu\nu}^T~,
\end{align}
where
\begin{align}
E_{\mu\nu}^T=\frac{1}{z^2}C_{\mu\alpha \nu\beta}u^\alpha u^\beta~,
\end{align}
\end{subequations}
with $C_{\mu\alpha\nu\beta}$ the Weyl tensor.
Here, FP denotes the finite part of the field evaluated at the worldline $r=0$, i.e. the external, regular field obeying $\Box\phi^{\rm ext}=0$. The projection operator
\begin{align}\label{eq:Projector}
\mathcal{P}_L^\Pi=\mathcal{P}^{\alpha_1\alpha_2\dots\alpha_\pi}_{\beta_1\beta_2\dots\beta_\ell}=\prod_{\ell=1}^L\prod_{\pi=1}^\Pi\mathcal{P}^{\alpha_\pi}_{\beta_\ell}=\prod_{\ell=1}^L\prod_{\pi=1}^\Pi\left(\delta^{\alpha_\pi}_{\beta_\ell}+u^{\alpha_\pi} u_{\beta_\ell}\right)~,
\end{align}
with $P_{\Pi}^\Xi u^\Pi=0$ projects to the physical degrees of freedom encapsulated in the symmetric and trace-free (STF) spatial pieces of the covariant tensors in the rest frame of the worldline, see e.g.~\cite{Vines:2016qwa} for more details. 
Here, capital greek letters denote strings of four-dimensional covariant indices, while capital latin letters from the middle of the alphabet denote their three-dimensional STF counterparts in the rest frame. Recall that here, the letters $S$, $T$ and $ST$ are labels rather than STF indices. Thus, we expect the effective action to involve corresponding scalar and tensor tidal deformabilities, $\lambda_\ell^S$ and $\lambda_\ell^T$ characterizing a scalar- or tensor-induced multipole moment, respectively. However, since the equations of motion \eqref{eq:eomEinsteinframe} are coupled, a scalar tidal field may also induce a tensor response and vice versa. Thus, we expect the action to require additional parameters to distinguish between a scalar response to $E_L^S$ or to $E_L^T$, and similarly for the tensor case. Specifically, we find that these considerations lead to the following form of the tidal action up to quadratic order in the tidal fields
\begin{align}\label{eq:TidalactionJordan}
S_{\rm tidal}=&\sum_{\ell} \int d\sigma~z~g^{LP}\times\nonumber\\&\left(\frac{\lambda^T_\ell}{2\ell!}E_L^{T}E_P^{T}+\frac{\lambda^S_\ell}{2\ell!}E_L^{S}E_P^{S}+\frac{\lambda^{ST}_\ell}{\ell!}E_L^{T}E_P^{S}\right),
\end{align}
where 
\be
\label{eq:gLP}
g^{L P}=\prod_{n=1}^{\ell} g^{l_n p_n}.
\ee
The last term in the tidal action~\eqref{eq:TidalactionJordan} 
has not been considered before. It contains a new type of tidal deformability, the scalar-tensor tidal deformability $\lambda_\ell^{ST}$, and characterizes a scalar/tensor  multipole moment induced by a tensor/scalar tidal field. To better understand the properties of the scalar-tensor deformability and connect with the microphysics of tidally perturbed scalarized NS configurations, we work in the Einstein frame, where the scalar field and the metric are only coupled through matter and the equations of motion are simpler to solve.

\subsection{Effective action in the Einstein frame}
The effective action in the Einstein frame is formally analogous to that in the Jordan frame~\eqref{eq:SEFT}, except that it is a functional of the Einstein frame scalar field $\varphi$ and conformal metric $g^\ast_{\mu\nu}$ instead of their Jordan frame counterparts $\phi$ and $g_{\mu\nu}$. Specifically, the effective action is given by
\begin{align}
\label{eq:SEFTEinstein}
S_{\rm EFT}^{\rm (E)}=S_{\rm ST}^{\rm (E)}+S_{\rm pm}^{\rm (E)} +S_{\rm tidal}^{\rm (E)}~,
\end{align}
with the tidal action
\begin{align}\label{eq:TidalactionEinstein}
S_{\rm tidal}^{({\rm E})}=&\sum_{\ell} \int d\sigma^\ast~z^\ast~g_\ast^{LP}\times\nonumber\\&\left(\frac{\lambda^{\ast T}_\ell}{2\ell!}E_L^{\ast T}E_P^{\ast T}+\frac{\lambda^{\ast S}_\ell}{2\ell!}E_L^{\ast S}E_P^{\ast S}+\frac{\lambda^{\ast ST}_\ell}{\ell!}E_L^{\ast T}E_P^{\ast S}\right)~,
\end{align}
and all quantities such as the tidal fields defined similar to as in~\eqref{eq:tidalfields} with the above-mentioned replacements $(\phi, g_{\mu\nu})\to (\varphi, g{\ast}_{\mu\nu})$ and the connection and curvature quantities associated with the conformal metric. Although the action~\eqref{eq:TidalactionEinstein} is expressed in terms of spatial STF tensors, one can obtain the covariant version having a similar structure using the inverse projection operator~\eqref{eq:Projector}.

\subsubsection{Role of the scalar-tensor deformability}
\begin{figure*}[htpb!]
\begin{center}
{\includegraphics[width=\textwidth,clip]{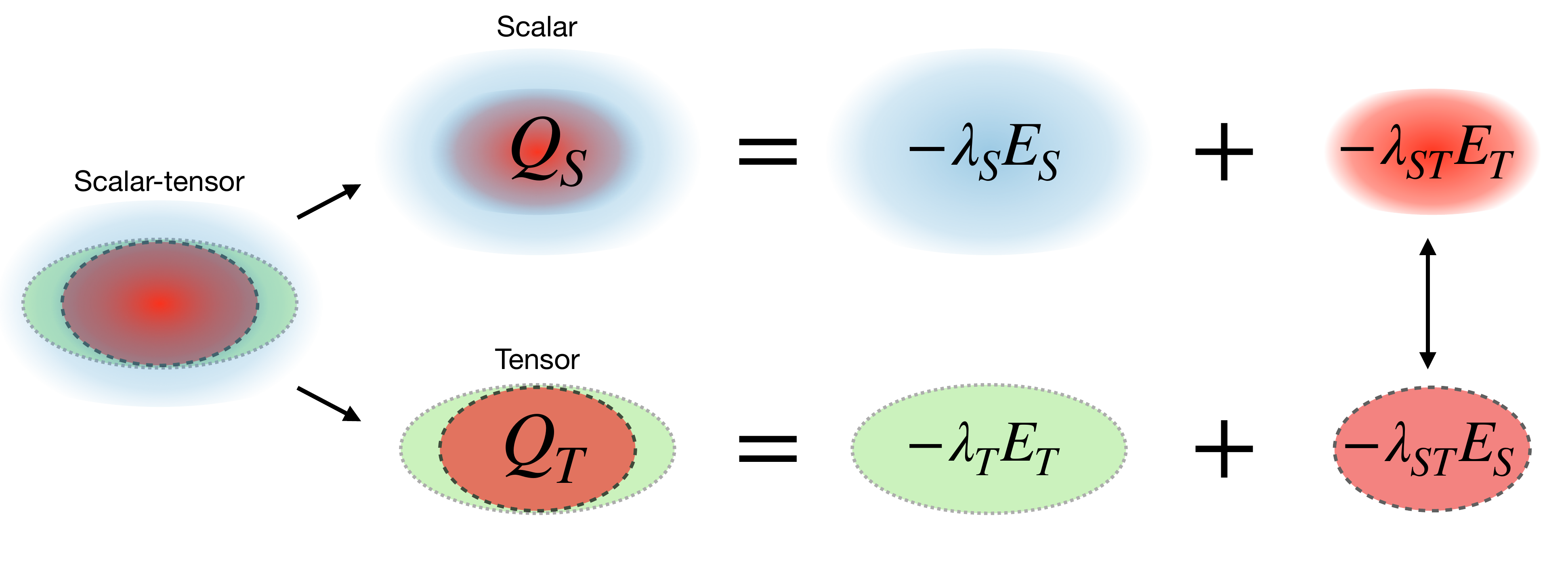}}
\caption{The scalar and tensor multipole moments consist of two contributions. The first contribution comes from the response to an external tidal field produced by the same kind of field, scalar or tensor. The second contribution accounts for a multipole moment induced by the other field, resulting from the coupling between matter and scalar field. }\label{fig:ScalarTensorLoveFig}
\end{center}
\end{figure*}

To study the effect of $\lambda_\ell^{\ast ST}$ we compute the equations of motion derived from the EFT action~\eqref{eq:SEFTEinstein} in vacuum and far away from the body, where spacetime is nearly Minkowski. To derive the equations of motion  for the scalar field, we use the definitions~\eqref{eq:tidalfields} and compute the variation
\begin{align}
\frac{\delta}{\delta\varphi}\left(E_L^{\ast S}E^L_{\ast S}\right)&=\underset{r\rightarrow0}{\text{FP}}\, \frac{\delta}{\delta\varphi}\left(\p_L\varphi\p^L\varphi\, \delta(x)\right)\nonumber\\&=2(-1)^\ell\underset{r\rightarrow0}{\text{FP}}\p_L\left(\p^L\varphi\, \delta(x)\right)\nonumber\\&=2(-1)^\ell\underset{r\rightarrow0}{\text{FP}}\left[\Box\varphi\delta(x)+\p^L\varphi\p_L\delta(x)\right]\nonumber\\&=2(-1)^{\ell+1}E_{\ast S}^L\p_L\delta(x)~,
\end{align}
where in the second equality we integrated by parts and in the last equality used that $\underset{r\rightarrow0}{\text{FP}}\Box\varphi=\underset{r\rightarrow0}{\text{FP}}\Box\varphi^{\rm ext}=0$ by definition. Performing similar calculations for the other pieces leads to equations of motion in the asymptotic limit given by
\begin{subequations}
\label{eq:Minkeom}
\begin{align}
\Box\varphi=&\frac{m_\ast'(\varphi_\infty)}{2K_\varphi}\delta(x)\nonumber\\&+\sum_{\ell=1}^\infty\frac{(-1)^{\ell+1}}{\ell!2K_\varphi}\left[\lambda^{\ast S}_{\ell}E^L_{\ast S}\p_L+\lambda^{\ast ST}_{\ell}E^L_{\ast T}\p_L\right]\delta(x)~,\\
\Box U^\ast_{N}=&\frac{m_\ast(\varphi_\infty)}{4K_R}\delta(x)\nonumber\\&+\sum_{\ell=2}^\infty\frac{(-1)^{\ell+1}}{\ell!4K_R}\left[\lambda^{\ast T}_{\ell}E^L_{\ast T}\p_L+\lambda^{\ast ST}_{\ell}E^L_{\ast S}\p_L\right]\delta(x)~,
\end{align}
\end{subequations}
where $\varphi_\infty$ is the value of the scalar field at infinity. For the metric functions, we focus on the $00$ component as it conveniently contains the tidal information. This is because $g^\ast_{00}$ is asymptotically related to the potential $U^\ast_N$ similar to  \eqref{eq:NewtonianPotential} in the Einstein frame. In the flat-space limit relevant here, $E^{\ast T}_L=-\p_LU^\ast_N$ \cite{Levi:2018nxp,Charalambous:2021mea}. The dominant terms in the solutions to the equations of motion~\eqref{eq:Minkeom} at infinity read 
\begin{subequations}\label{eq:EFTEoMEinstein}
\begin{align}
\varphi\sim &\frac{m_\ast'(\varphi_\infty)}{8\pi K_\varphi r}+\sum_{\ell=1}^\infty\frac{(2\ell-1)!!}{\ell!{8\pi K_\varphi}}\frac{\left(-\lambda^{\ast S}_{\ell}E^L_{\ast S}-\lambda^{\ast ST}_{\ell}E^L_{\ast T}\right)n_L}{r^{\ell+1}}\nonumber\\&+\varphi_{\rm tidal}~,\\
U^\ast_N\sim &\frac{m_\ast(\varphi_\infty)}{16\pi K_R r}+\sum_{\ell=1}^\infty\frac{(2\ell-1)!!}{\ell!{16\pi K_R}}\frac{\left(-\lambda^{\ast T}_{\ell}E^L_{\ast T}-\lambda^{\ast ST}_{\ell}E^L_{\ast S}\right)n_L}{r^{\ell+1}}\nonumber\\&+{U^\ast_N}_{\rm tidal}
\end{align}
\end{subequations}
with
\begin{align}
    \varphi_{\rm tidal}&=-\sum_{\ell=1}^\infty\frac{1}{\ell!8\pi K_\varphi}E^L_{\ast S}n_L r^\ell~,\\
    {U^\ast_N}_{\rm tidal}&=-\sum_{\ell=1}^\infty\frac{1}{\ell!16\pi K_R}E^L_{\ast T}n_L r^\ell
\end{align}
as the homogeneous solutions, corresponding to the external tidal fields $E^L_{\ast S}$ and $E^L_{\ast T}$, respectively. Comparing the solutions with the definition of the multipole moments from the asymptotic limit~\eqref{eq:NewtonianPotential}, we identify the induced $\ell$-th order multipole moment from the coefficient associated with the $r^{-\ell-1}$ falloff, which leads to 
\begin{subequations}
    \label{eq:multipolesST}
\begin{align}
    Q^{\ast S}_L=&-\lambda^{\ast S}_{\ell}E_L^{\ast S}-\lambda^{\ast ST}_{\ell}E_L^{\ast T}~,\label{eq:ScalarMultipole}\\
    Q^{\ast T}_L=&-\lambda^{\ast T}_{\ell}E_L^{\ast T}-\lambda^{\ast ST}_{\ell}E_L^{\ast S}~.\label{eq:TensorMultipole}
\end{align}
\end{subequations}
These relations formalize the effect of the scalar-tensor tidal deformability. Specifically, as seen in \eqref{eq:ScalarMultipole} and \eqref{eq:TensorMultipole}, and illustrated in Fig.~\ref{fig:ScalarTensorLoveFig}, $\lambda_{\ell}^{\ast ST}$ characterizes the scalar/tensor-induced multipole moment in the presence of an external tensor/scalar field.
Using~\eqref{eq:multipolesST}, we can compute the tidal deformabilities as
\begin{subequations}
\label{eq:lambdaSTST}
\begin{align}
\lambda^{\ast T}_\ell&=-\left.\frac{Q^{\ast T}_L}{E^{\ast T}_L}\right|_{E^{\ast S}_L=0}~,\\
\lambda^{\ast S}_\ell&=-\left.\frac{Q^{\ast S}_L}{E^{\ast S}_L}\right|_{E^{\ast T}_L=0}~,\\
\lambda^{\ast ST}_\ell&=-\left.\frac{Q^{\ast T}_L}{E^{\ast S}_L}\right|_{E^{\ast T}_L=0}=-\left.\frac{Q^{\ast S}_L}{E^{\ast T}_L}\right|_{E^{\ast S}_L=0}~.
\end{align}
\end{subequations}
This will be useful for identifying the information contained in these parameters from perturbation theory in Sec.~\ref{sec: ComputingLoveNumbers}.

\subsubsection{Parity symmetry  and $\lambda_\ell^{ST}$}\label{ssec:ParityST}
\label{subsect:Paritygeneral}
For generic couplings $A(\varphi)$ the EFT presented in~\eqref{eq:TidalactionEinstein} is the most generic one. However, depending on the choice of $A(\varphi)$, additional symmetries may emerge in the action. An example is parity symmetry, which refers here to the transformation 
\be
\varphi\rightarrow-\varphi
\ee 
and arises for couplings that yield scalarized configurations. However, the term 
\begin{align}
    \frac{\lambda_\ell^{\ast ST}}{\ell!}E^{\ast S}_L E_{\ast T}^L
\end{align}
in the action is not parity symmetric, since from~\eqref{eq:tidalfields}
\be
 E_{\ast T}^L\rightarrow{E_{\ast T}^L}~,\qquad
E_{\ast S}^L\rightarrow-{E_{\ast S}^L}~.
\ee
 To preserve the overall parity symmetry of the action assumed here requires the scalar-tensor tidal deformability $\lambda_\ell^{\ast ST}$ to develop a dependence on the scalar field. This is because the only term that respects the parity symmetry at quadratic order in the tidal fields in the EFT is 
\begin{align}
    \frac{\tilde{\lambda}_\ell^{\ast ST}}{\ell!}w(\varphi_{\infty},Q)E^{\ast S}_L E_{\ast T}^L~.
\end{align}
where $\tilde \lambda^{ST}$ is the 'bare', i.e. $\varphi_\infty$- and $Q$-independent, deformability parameter, and $w(\varphi_{\infty},Q)$ captures its dependence on the asymptotic scalar characteristics $(\varphi_{\infty},Q)$ in a polynomial containing only odd powers of the scalar charge $Q=m_\ast'(\varphi_\infty)$ and the background scalar field at infinity $\varphi_\infty$. It has the form
\begin{align}\label{eq:wPolynomial}
w(\varphi_{\infty},Q)=\sum_{p=0}^{\infty}c_{p}\varphi_{\infty}^{2p+1}+\sum_{n=0}^{\infty}c_{n}Q^{2n+1}~,
\end{align}
where $c_{p,n}$ are constants. This implies that, for parity symmetric theories, $\lambda_\ell^{\ast ST}$ will scale with the polynomial $w(\varphi_{\infty},Q)$,
\begin{align}
\lambda_\ell^{\ast ST}=w(\varphi_{\infty},Q)\tilde{\lambda}_\ell^{\ast ST}~.
\end{align}
Note that in GR $\varphi_{\infty}=0=Q$ and hence $\lambda_\ell^{\ast ST}=0$. In Sec.~\ref{sec:CaseStudy} we will explicitly compute $w(\varphi_\infty,Q)$ and  $\tilde{\lambda}_\ell^{\ast ST}$ for a class of scalar-tensor theories.

\subsection{Relating tidal deformabilities between frames}

Making use of the conformal transformation \eqref{eq:MetricConformalTransf}, we can relate the EFT coefficients between frames at the level of the action. This enables connecting between the tidal deformability computed from perturbation theory and the EFT in the Einstein frame, and then relating the Einstein frame tidal deformability to its Jordan frame counterpart. The details of the calculations can be found in Appendix~\ref{app:TidalDefTransf} and yield
\begin{subequations}
\label{eq:lambdatransform}
\begin{align}
\lambda_\ell^T&=\lambda_\ell^{\ast T}~,\label{eq:ScalarTidalDefTransf}\\
\lambda_\ell^S&=\left(\frac{A_\infty^{2}{F'}_\infty}{2\alpha_\infty}\right)^2\lambda_\ell^{\ast S}~,\\
\lambda_\ell^{ST}&=\frac{A_\infty^{2}F'_\infty}{2\alpha_\infty}\lambda_\ell^{\ast ST}~,
\end{align}
\end{subequations}
where the subscript ``$\infty$" denotes evaluation at infinity. This result is in agreement with the special case with $A_\infty=1$ considered in~\cite{Yazadjiev:2018xxk}, where the vanishing of the scalar field at infinity implied that the tensor tidal deformabilities in both frames are the same. 

The transformation of the dimensionless Love numbers $k_\ell$ defined in~\eqref{eq:LoveNumvsTidalDef} is obtained by combining~\eqref{eq:lambdatransform} with the transformation of the radius. The latter follows from using that, for a spherically symmetric spacetime,
\begin{align}
g_{\theta\theta}&=r^2=A^2 g^\ast_{\theta\theta}=A^2r_\ast^2~,
\end{align}
which implies that the radius in the two frames is related by \cite{Yazadjiev:2018xxk}
\begin{align}
R&=A(\varphi_R)R_\ast~,\label{eq: RadiusEinteinToJordan}
\end{align}
with $\varphi_R\equiv\varphi(R)$. We can also obtain the transformation for the dimensionless tidal deformability,
\begin{align}\label{eq:LambdaDef}
\Lambda_\ell=\frac{\lambda_\ell}{M^{2\ell+1}}=\frac{2}{(2\ell-1)!!}k_\ell C^{-2\ell-1}~,
\end{align}
where
\be
C=M/R
\ee
is the compactness of the star, with $M$ its Arnowitt-Deser-Misner (ADM) mass \cite{Arnowitt:1962hi}. In order to transform $\Lambda_\ell$ we can use \eqref{eq: RadiusEinteinToJordan} for the radius and \eqref{eq: ADMMassJordanGeneric} for the ADM mass, together with \eqref{eq:lambdatransform}.

\section{Computing Love numbers}\label{sec: ComputingLoveNumbers}
Having established the relevant deformability coefficients in the EFT, the next step is to connect them with the information from detailed calculations of the perturbed configuration. The baseline for computing the Love numbers is to first compute the background equilibrium configuration. This is given by the modified Tolman-Oppenheimer-Volkoff (TOV) equations for both the spacetime $g_{\mu\nu}^{(0)}$, and the equation of motion for the background scalar field $\varphi_0(r)$. This configuration also yields the corresponding mass-radius relations for a chosen equation of state. Subsequently, we consider linearized perturbations to the spacetime and scalar fields,
\begin{align}
    g_{\mu\nu}^\ast&={g_{\mu\nu}^{\ast}}^{(0)}+\epsilon~\sum_{\ell,m}{h_{\mu\nu}^{\ast\ell m}}(r)~Y_{\ell}^m(\theta,\phi)~,\label{eq:MetricPert}\\
    \varphi&=\varphi_0(r)+\epsilon~\sum_{\ell,m}\delta\varphi^{\ell m}(r)~Y_{\ell}^m(\theta,\phi),\label{eq:ScalarPert}
\end{align}
where $\epsilon$ is a counting parameter for the perturbations and $Y_{\ell}^m(\theta,\phi)$ are the spherical harmonics. Similarly, the fluid pressure, density and four-velocity for the perturbed configuration are
\begin{align}
    p&=p_0+\epsilon~\sum_{\ell,m}\delta{p}^{\ell m}(r)~Y_{\ell}^m(\theta,\phi),\label{eq:PressurePert}\\
    \rho&=\rho_0+\epsilon~\sum_{\ell,m}\frac{d\rho_0}{dp_0}\delta{p}^{\ell m}(r)~Y_{\ell}^m(\theta,\phi),\label{eq:DensityPert}\\
    {u^\mu_\ast}&={u^\mu_{0\ast}}+ \epsilon~\delta u_\ast^\mu.
\end{align}
Substituting these ansaetze into the equations of motion~\eqref{eq:eomEinsteinframe} and keeping only terms up to linear order in $\epsilon$ leads to the system of differential equations that ultimately determine the Love numbers. Solving these in the interior of the star as well as in the exterior, where no NS matter is present, and using the definitions of the multipole moments~\eqref{eq:NewtonianPotential} and ~\eqref{eq:lambdaSTST} determines the Love numbers. To make this methodology concrete, we start by reviewing the computation of the background configuration in Sec.~\ref{ssec:BackgroundConf}. We then calculate the perturbed equations of motion in Sec.~\ref{ssec:STPerturbations} and describe a framework to extract the multipole and tidal moments. In Sec.~\ref{sec:CaseStudy} we compute the mass-radius curves for specific choices of the couplings that trigger scalarized stars, and use our framework to compute the Love numbers. Finally, we also provide the results transformed to the Jordan frame. For the numerical calculations we choose the normalization constants to be $K_{\phi}^{-1}=K_R^{-1}=16\pi{G}$ and $K_\varphi^{–1}=8\pi G$ with $G=c=1$ \footnote{Here, we have kept $G$ for reference, since as explained in Appendix~\ref{app: ADMMass} some of the scalar effects can be interpreted as an effective scalar field-dependent gravitational coupling $\tilde{G}(\varphi_\infty)$.}. The derivations presented here can also be found in \cite{Pani:2014jra,Brown:2022kbw}, though with different conventions for the normalizations \footnote{Note that Reference \cite{Brown:2022kbw} reported differences in results with Reference \cite{Pani:2014jra}, and considers scalar and tensor tidal deformabilities computed in a different way (see Sec.~\ref{sec:SummaryDiscussion} for a comparison). However, the modified TOV equations coincide.}.

\subsection{Background configuration}\label{ssec:BackgroundConf}
\subsubsection{Modified TOV equations}
We start by writing the background, unperturbed metric describing a static, spherically symmetric configuration as
\begin{align}
\label{eq:sphermet}
    ds_0^2={g_{\mu\nu}^{\ast}}^{(0)}dx^\mu dx^\nu=-e^\nu dt^2+e^\gamma dr^2 + r^2 d\Omega^2~,
\end{align}
with $d\Omega^2=d\theta^2+\sin^2(\theta)d\phi^2$. For better readability we drop the asterisks in this section, but note that the only quantities in the Jordan frame, i.e. with no asterisks, are the pressure and density. Substituting~\eqref{eq:sphermet} into~\eqref{eq:eomEinsteinframe} leads to
\begin{align}
G_{tt}^0&=8\pi e^{\gamma}\rho_0 A(\varphi_0)^4+{\varphi_0'}^2-\frac{\gamma'}{r}-\frac{e^{\gamma}}{r^2}+\frac{1}{r^2}=0~,\label{eq: Gtt0}\\
\frac{G_{rr}^0}{e^{\gamma-\nu}}&=8\pi e^{\gamma}p_0 A(\varphi_0)^4+{\varphi_0'}^2-\frac{\nu'}{r}+\frac{e^{\gamma}}{r^2}-\frac{1}{r^2}=0~,\label{eq: Grr0}\\
\varphi_0''+&\frac{4-r\left(\gamma'-\nu'\right)}{2r}\varphi_0'\nonumber\\&-4\pi A(\varphi_0)^4 e^\gamma\alpha(\varphi_0)(3p_0-\rho_0)=0~.
\end{align}
Assuming a spherically symmetric configuration implies
\begin{align}
\gamma=-\log\left(1-\frac{2m(r)}{r}\right)~,\label{eq: BackgroundGamma}
\end{align}
which upon using \eqref{eq: Gtt0} yields
\begin{align}
m'=4\pi r^2\rho_0 A(\varphi_0)^4 + \frac{r^2}{2}\left(1-\frac{2m}{r}\right){\varphi_0'}^2~.\label{eq: TOVMass}
\end{align}
Substituting into \eqref{eq: Grr0} we obtain
\begin{align}
\nu'=\frac{r^3\left(8\pi p_0 A(\varphi_0)^4+{\varphi_0'}^2\right)+2m\left(1-r^2{\varphi_0'}^2\right)}{r\left(r-2m\right)}~.\label{eq: TOVNu}
\end{align}
Next, using the conservation of the energy-momentum tensor \eqref{eq: EinsteinEMCons} and the normalization of the four-velocity $u_\mu u^\mu=-1$ leads to
\begin{align}
u_0^\mu&=\left(e^{-\nu/2},0,0,0\right)~,\\
p_0'&=\frac{p_0+\rho_0}{2}\left[2\alpha(\varphi_0)\varphi_0'-\nu'\right]~.\label{eq: TOVPressure}
\end{align}
The modified TOV equations \eqref{eq: TOVMass}, \eqref{eq: TOVNu} and \eqref{eq: TOVPressure}, together with an equation of state relating $p_0$ and $\rho_0$, fully describe the background configuration, given a chosen conformal factor. As a check, one can set $\varphi_0=0$, $A(\varphi_0)=1$, and $\alpha(\varphi_0)=0$ and see that, indeed, one recovers the general-relativistic TOV equations.

\subsubsection{Boundary conditions near the origin}
The solutions to \eqref{eq: TOVMass}, \eqref{eq: TOVNu} and \eqref{eq: TOVPressure} in the interior of the star can, in general, only be obtained numerically. They must satisfy the following boundary conditions
near the center of the star at $r_{\rm min}\rightarrow0$:
\begin{subequations}
    \label{eq:bcscenter}
\begin{align}
\rho_0(r_{\rm min})&=\rho_c~,\quad
m(r_{\rm min})=\frac{4}{3}\pi r_{\rm min}^3\rho_c~,\\
\varphi_0(r_{\rm min})&=\varphi_{0c}~.
\end{align}
\end{subequations}
In most applications, it is desirable to control the asymptotic value of the scalar field at infinity $\varphi_{0\infty}$ rather than $\varphi_{0c}$. This can be implemented by using a shooting method for obtaining the appropriate $\varphi_{0c}$ corresponding to a given $\varphi_{0\infty}$. 

\subsubsection{Scalar field outside the star}
This task can be simplified by using an exact solution for the field in the exterior that exists in Just coordinates\footnote{Note that this only applies for a vanishing scalar-field potential $V(\phi)$ in \eqref{eq: STaction}. For nonvanishing potentials the shooting method should be extended to infinity.} (see Appendix \ref{app:JustCoordApp} for details), which determines the field at the surface to be 
\begin{align}\label{eq: JustInfSurface}
\varphi_0(R)=\varphi_{0\infty}+q\frac{\nu_S}{2}~,
\end{align}
with
\be
\nu_S=-\frac{2}{\sqrt{1+q^2}}{\text{arctanh}}\left(\frac{\sqrt{1+q^2}}{1+\frac{2}{R\nu'_S}}\right)~,
\ee
and $q$ a parameter related to the scalar charge given by 
\begin{align}
q&=\frac{2 \varphi_0'(R)}{\nu'_S}~,\\
\nu'_S&=\frac{2m_S}{R\left(R-2m_S\right)}+R\varphi_0'(R)^2~,
\end{align}
and $m_S$ the mass at the surface of the star. 
Specifically, the quantity 
\be
q=-Q/M \label{eq:smallq}
\ee characterizes the scalar charge $Q$ per unit mass $M$ of the configuration. The scalar charge is defined as the coefficient of the $1/r$ falloff of the solution in an asymptotic expansion near spatial infinity, similarly to the ADM mass $M$ in the gravitational potential, with
\begin{align}
\lim_{r\to\infty}\varphi_0(r)=\varphi_{0\infty}+\frac{Q}{r}+\mathcal{O}\left(\frac{1}{r^2}\right)~.
\end{align}
Hence, $q$ is a measure of the strength of the scalar field compared to the gravitational field. 
Using the exact exterior solution for the scalar field from~\eqref{eq: JustInfSurface} has the advantage that the numerical integration only has to be performed up to the star's surface, rather than infinity.

\subsection{Scalar and tensor perturbations}\label{ssec:STPerturbations}
We now focus on the tensor and scalar perturbations, introduced in \eqref{eq:MetricPert}
 and \eqref{eq:ScalarPert}, and drop the labels $\ell,m$ in the radial functions \eqref{eq:MetricPert}-\eqref{eq:DensityPert}. In the class of scalar-tensor theories we consider, we can write the static, even parity (also known as polar or electric) metric perturbations in the Regge-Wheeler gauge \cite{PhysRevD.71.124038},
\begin{align}
h_{\alpha\beta}=\text{Diag}\left[- e^{\nu} H_0(r),e^{\gamma}  H_2(r),r^2   K(r),r^2  \sin^2(\theta) K(r) \right]~,
\end{align}
where $H_0$, $H_2$ and $K$ are functions characterizing the metric perturbation. Using this gauge and perturbing the Einstein Field Equations \eqref{eq:EoMMetric} and the scalar field equation of motion \eqref{eq:EoMScalar}, together with the fluid quantities at first order in $\epsilon$, we obtain\footnote{Note that there is an apparent minus sign difference in the source term with \cite{Pani:2014jra}, however, this is consistent with the different conventions where in \cite{Pani:2014jra} $H_0$ is defined such that $H_0=H_2$ whereas we have $H_0=-H_2$.}
\begin{align}
K'&+H_0'+\nu' H_0 + 4\varphi_0'\delta\varphi=0~,\\
H_0&=-H_2~,\\
H_0''&+ f_1 H_0' + f_0 H_0 = f_s \delta\varphi~,\label{eq:H0PertEq}\\
\delta\varphi''&+ g_1 \delta\varphi' + g_0 \delta\varphi = g_s H_0~.\label{eq:PhiPertEq}
\end{align}
The terms in the metric perturbation equation of motion are
\begin{align}
f_1=&\frac{4 \pi r^3 A(\varphi_0)^4 (p_0-\rho_0)+2(r-m)}{r (r-2 m)}~,\\
f_0=&\frac{1}{r^2 (r-2 m)^2}\{4 \pi  r^3 p_0 A(\varphi_0)^4 [r (\p\rho_0/\p p_0+9)\nonumber\\&-2m (\p\rho_0/\p p_0+13)]+4 \pi  r^3 \rho_0 A(\varphi_0)^4\nonumber\times\\& (\p\rho_0/\p p_0+5) (r-2 m) -4 r^2 (r-2 m) \varphi_0'^2 \nonumber\times\\&\left(4 \pi  r^3 p_0 A(\varphi_0)^4+m\right)-64 \pi ^2 r^6 p_0^2 A(\varphi_0)^8\nonumber\\&-\ell(\ell+1)r(r-2m)-r^4 (r-2 m)^2 \varphi_0'^4-4 m^2\}~,\\
f_s=&\frac{1}{r (r-2 m)}\{4 r^2 [2 \pi  A(\varphi_0)^4 (\alpha (\varphi_0)\times\nonumber\\&((\p\rho_0/\p p_0-9) p_0+(\p\rho_0/\p p_0-1)\rho_0)+4 r p_0\varphi_ 0')\nonumber\\&+(r-2 m) \varphi_0'^3]+8 m \varphi_0'\}~,
\end{align}
and the terms in the scalar perturbation equation of motion are
\begin{align}
    g_1&=f_1~,\\
    g_0&=\frac{1}{r-2 m}\{4\pi r A(\varphi_0)^4 [\alpha (\varphi_0)^2 ((\p\rho_0/\p p_0+9) p_0\nonumber\\&+(\p\rho_0/\p p_0-7) \rho_0)+(\rho_0-3 p_0) \alpha'(\varphi_0)]\}\label{eq:ScalarPertEqg0}\nonumber\\&-\frac{\ell(\ell+1)}{r(r-2 m)}-4 \varphi_0'^2~,\\
    g_s&=\frac{f_s}{4}~.
\end{align}
 In GR the source terms vanish, $f_s=g_s=0$, and therefore the perturbations decouple. In scalar-tensor theories, however, the perturbations are coupled as a result of the coupling between matter and the scalar field. This means that a tensor perturbation $H_0$ will induce a scalar perturbation $\delta\varphi$, and vice versa. This is also in agreement with the scalar-tensor term in the effective action, $\lambda_{ST}$, which quantifies the induced Love number on top of the pure tensor and scalar perturbations. 
 \subsubsection{Dipolar perturbations}
 \label{subsubsec:dipole}
 When specializing to tensor dipolar perturbations $\ell=1$, there are two equivalent ways to proceed. The first way is to fix $\ell=1$ in Einstein's equations, as in \cite{1970ApJ...159..847C}. This changes the combinations of components needed in order to decouple the different functions $H_0$, $H_2$, and $K$ and results in a first-order differential equation for $H_0$. The second way is by fixing $\ell=1$ at the level of the perturbation equation of motion \eqref{eq:H0PertEq}, which yields a second-order differential equation. However, for the particular case $\ell=1$, we can integrate the second-order differential equation and obtain the same first-order differential equation as with the first way,
 \begin{align}
     H_0' +d_0  H_0 =  s_1 \delta\varphi' + s_0 \delta\varphi~,\label{eq:H0DipolePertEq}
 \end{align}
 with
 \begin{align}
    d_0&=\frac{2(r-m)+8\pi A(\varphi_0)^4 p_0 r^3}{r(r-2m)}+r\varphi_0'^2\nonumber\\&-\frac{8\pi A(\varphi_0)^4r^3(3p_0+\rho_0)}{2m+8\pi A(\varphi_0)^4p_0r^3+r^2(r-2m){\varphi_0'}^2}~,\\
    s_1&=-\frac{4r(r-2m)\varphi_0'}{2m+8\pi A(\varphi_0)^4p_0r^3+r^2(r-2m){\varphi_0'}^2}~,\\
    s_0&=-\frac{1}{2m+8\pi A(\varphi_0)^4p_0r^3+r^2(r-2m){\varphi_0'}^2}\nonumber\\&\times\Big\{8(r-m)\varphi_0'+4r^2(r-2m){\varphi_0'}^3\nonumber\\&+16\pi A(\varphi_0)^4r^2\left[\alpha(\varphi_0)\left(\rho_0-3p_0\right)+2p_0r\varphi_0'\right]\Big\}~.
 \end{align}
In the exterior of the star, where $p_0=0=\rho_0$, the tensor and scalar perturbations decouple when defining a new function $\zeta(r)$ by
\begin{align}
\zeta=d_0^{ext}~H_0-s_0^{ext}~\delta\varphi~.
\end{align}
With this,~\eqref{eq:H0DipolePertEq} becomes
\begin{align}\label{eq:ZetaDipolarEq}
\zeta'+\zeta\left(d_0^{ext}-\frac{{d_0'}^{ext}}{d_0^{ext}}\right)=0~.
\end{align}
The asymptotic solution to~\eqref{eq:ZetaDipolarEq} for large $r$ has the form
\begin{align}
    \zeta=\zeta^{(-3)}/r^3+\mathcal{O}\left(\frac{1}{r^4}\right)~,
\end{align}
with $\zeta^{(-3)}$ a constant of integration. This yields for the metric function
\begin{align}\label{eq:H0ScalarSolDipole}
H_0=\frac{\zeta^{(-3)}}{2r^2}+2q\delta\varphi+\mathcal{O}\left(\frac{1}{r^3}\right)~.
\end{align}
Substituting~\eqref{eq:H0ScalarSolDipole} into~\eqref{eq:PhiPertEq} gives, asymptotically, 
\begin{align}
\delta\varphi=\delta\varphi^{(1)}r+\frac{\delta\varphi^{(-2)}}{r^2}+\mathcal{O}\left(\frac{1}{r^3}\right)~,
\end{align}
with $\delta\varphi^{(\ell)}$ the coefficients associated with the $r^{\ell}$ dependence. Hence, 
 \begin{align}
     H_0=\frac{H_0^{(-2)}}{r^2}+2q\delta\varphi^{(1)}r+\mathcal{O}\left(\frac{1}{r^3}\right)~,
 \end{align}
 where we have redefined $H_0^{(-2)}=\zeta^{(-3)}/2+2q\delta\varphi^{(-2)}$. This manifestly shows how a scalar tidal field ${E^\ast_1}^S\propto{\delta\varphi^{(1)}}$, can induce a tensor tidal field ${E^\ast_1}^T=2q{E^\ast_1}^S$. In GR, the scalar charge vanishes $q=0$, and we recover the result in \cite{1970ApJ...159..847C}. The constant in front of the $r^{-2}$ falloff, proportional to the mass dipole moment ${Q^{\ast}_1}^T$, can be set to zero by a gauge transformation. Specifically, the change in a perturbed scalar field $\Psi=\Psi_0+\epsilon~\delta\Psi$ due to an infinitesimal translation $\tilde{x}^\mu={x^\mu+\epsilon~\xi^\mu}$, is given by the Lie derivative $\pounds_\xi$ along the vector field $\xi^\mu$,
 \begin{align}
\tilde{\delta\Psi}=\delta\Psi+\pounds_{\xi}\Psi_0=\delta\Psi+\xi^\mu\p_\mu\Psi_0~,
 \end{align}
 where, for the static case \cite{1970ApJ...159..847C},
 \begin{align}
\xi_\mu=\delta_{\mu}^r\xi_r=a\frac{e^{\gamma}}{r}fY^1_m\delta_{\mu}^r~,
 \end{align}
 with $a$ an arbitrary constant,
 \begin{align}
     f=r \exp\left[-\int_r^\infty{\frac{1-e^\gamma}{r}}dr\right]~,
 \end{align}
 and $\gamma$ the background metric function \eqref{eq:sphermet}. Therefore, the metric and scalar perturbations, $H_0$ and $\delta\varphi$, will transform as
 \begin{align}
     \tilde{H_0}&=H_0+a\frac{\nu'}{r}f~,\\
     \tilde{\delta\varphi}&=\delta\varphi+a\frac{\varphi_0'}{r}f~,
 \end{align}
 which asymptotically reads
 \begin{align}
     \lim_{r\rightarrow\infty}\tilde{H_0}&=2q\delta\varphi^{(1)}r+\dots+
     \frac{H_0^{(-2)}}{r^2}+a\frac{2M}{r^2}+\mathcal{O}\left(\frac{1}{r^3}\right)~,\\
     \lim_{r\rightarrow\infty}\tilde{\delta\varphi}&=\delta\varphi^{(1)}r+\dots+\frac{\delta\varphi^{(-2)}}{r^2}+a\frac{qM}{r^2}+\mathcal{O}\left(\frac{1}{r^3}\right)~.
 \end{align}
 In order to match to the EFT, which is formulated around the center-of-mass worldline, we choose a gauge where the mass dipole vanishes. This corresponds to setting $a=-H_0^{(-2)}/2M$. In this gauge the tensor and scalar dipole moments now read
 \begin{align}
 \tilde{H_0}^{(-2)}&=0~,\\
 \tilde{\delta\varphi}^{(-2)}&=\delta\varphi^{(-2)}-\frac{q}{2}H_0^{(-2)}~.\label{eq:DipoleShift}
 \end{align}
 Hence, the mass dipole moment in scalar-tensor theories can still be made to vanish, which however shifts the scalar dipole moment.
 
\begin{table*}[t]
\hspace{2cm}
\centerline{\begin{tabular}{c|p{1.1cm}|c|c|c|c|c|} 
 \cline{2-7}
 &\multicolumn{2}{|c|}{Interior $r=r_{min}$}&\multicolumn{4}{|c|}{Exterior $r=r_{\infty}$ \& \eqref{eq:SolsInf}}\\
 \hline
 \multicolumn{1}{|c|}{Solution} & \hspace{0.45cm}1\hspace{0.4cm} & 2 & A  & B & C & D \\
 \hline
 \multicolumn{1}{|c|}{$\delta\varphi$} &  \multicolumn{1}{|c|}{$r_{\rm min}^\ell$} & 0 & $\begin{array}{lr}
         \delta\varphi^{(-\ell-1)}=1&\\
         \text{Rest~}=~0&
         \end{array}$ & $\begin{array}{lr}
          \delta\varphi^{(\ell)}=1&\\
         \text{Rest~}=~0&
         \end{array}$ & $\begin{array}{lr}
          H_0^{(-\ell-1)}=1&\\
         \text{Rest~}=~0&
         \end{array}$ & $\begin{array}{lr}
          H_0^{(\ell)}=1&\\
         \text{Rest~}=~0&
         \end{array}$ \\
 \hline
 \multicolumn{1}{|c|}{$H_0$}  & \multicolumn{1}{|c|}{0} & $r_{\rm min}^\ell$ & $\begin{array}{lr}
         \delta\varphi^{(-\ell-1)}=1&\\
         \text{Rest~}=~0&
         \end{array}$ & $\begin{array}{lr}
         \delta\varphi^{(\ell)}=1&\\
         \text{Rest~}=~0&
         \end{array}$ & $\begin{array}{lr}
          H_0^{(-\ell-1)}=1&\\
         \text{Rest~}=~0&
         \end{array}$ & $\begin{array}{lr}
          H_0^{(\ell)}=1&\\
         \text{Rest~}=~0&
         \end{array}$ \\
 \hline
\end{tabular}}
 \caption{Boundary conditions for the interior and exterior solutions.}
  \label{tab: BCs}
\end{table*}

\subsubsection{Extracting the multipole and tidal moments}\label{ssec:ExtractingMult}
To numerically extract the multipole and tidal moments, we construct series solutions around spatial infinity that enable imposing the appropriate boundary conditions, see also~\cite{Pani:2014jra}. For the background quantities, we obtain the series expansions 
\begin{subequations}
    \label{eq:asym22}
\begin{align}
\varphi_0(r)=&\varphi_{0\infty}-\frac{q M}{r}-\frac{q M^2}{r^2}+\mathcal{O}\left(\frac{1}{r^3}\right)~,\\
m(r)=&M-\frac{M^2q^2}{2r}-\frac{M^3q^2}{2r^2}+\mathcal{O}\left(\frac{1}{r^3}\right)~,
\end{align}
\end{subequations}
with $M$ the ADM mass and $q$ defined in~\eqref{eq:smallq} being minus the scalar charge per unit mass \footnote{For generic normalizations and $G\neq1$ we have
\begin{align}
    \varphi_0(r)=\varphi_{0\infty}-\sqrt{\frac{2G^2K_R}{K_\varphi}}\frac{q M}{r}~,\nonumber
\end{align} where $q=-\sqrt{\frac{K_\varphi}{2G^2K_R}}Q/M$. Recall that an adimensional scalar field is defined as $\varphi^{\rm adim}=\sqrt{\frac{K_\varphi}{2K_R}}\varphi$, such that
\begin{align}
\varphi^{\rm adim}=\varphi_{0\infty}^{\rm adim}-q^{\rm adim}\frac{G M}{r}~,\nonumber
\end{align} and $q^{\rm adim}=-Q/M$.}
(see Appendix \ref{app:JustCoordApp} for details). For the perturbed quantities, $\delta\varphi$ and $H_0$, the expansions near spatial infinity are of the form
\begin{subequations}\label{eq:SolsInf}
\begin{align}
\delta\varphi(r)=&\delta\varphi^{(\ell)}r^\ell\left(1-\frac{\ell M}{r}\right)+\dots+\frac{\delta\varphi^{(-\ell-1)}}{r^{\ell+1}}\nonumber\\
&+\mathcal{O}\left(\frac{1}{r^{\ell+2}}\right)~,\label{eq: ScalarSolInf}\\
H_0(r)=&H_0^{(\ell)}r^\ell\left(1-\frac{\ell M}{r}\right)+\dots+\frac{H_0^{(-\ell-1)}}{r^{\ell+1}}\nonumber\\
&+\mathcal{O}\left(\frac{1}{r^{\ell+2}}\right)~,\label{eq: TensorSolInf}
\end{align} 
\end{subequations}
where the omissions $\ldots$ denote $q$-dependent terms, some of which also contain a combination of $\delta\varphi^{(\ell)}, \delta\varphi^{(-\ell-1)}, H_0^{(\ell)}$, and $H_0^{(-\ell-1)}$. As explained above in Sec.~\ref{subsubsec:dipole}, for the dipolar case $\ell=1$, the tensor perturbation equation of motion is of first order and we have $H_0^{(\ell=1)}=2q\delta\varphi^{(1)}$, thus one less degree of freedom than for higher multipoles. Note that in GR, $q=0$ and we recover the same asymptotic expansion as~\eqref{eq: TensorSolInf} which results from an exact solution in terms of a combination of Legendre polynomials. Comparing with the EFT result~\eqref{eq:EFTEoMEinstein}, with $H_0=2U^\ast_N$, we can identify the multipole and tidal moments as
\begin{align}
Q^{\ast S}_Ln^L&\leftrightarrow\delta\varphi^{(-\ell-1)}\frac{\ell!8\pi K_\varphi}{(2\ell-1)!!}~,\\
Q^{\ast T}_Ln^L&\leftrightarrow{H_0^{(-\ell-1)}}\frac{\ell!8\pi K_R}{(2\ell-1)!!}~,\\
E^{\ast S}_Ln^L&\leftrightarrow-\delta\varphi^{(\ell)}{\ell!8\pi K_\varphi}~,\\
E^{\ast T}_Ln^L&\leftrightarrow-{H_0}^{(\ell)}{\ell!8\pi K_R}~,
\end{align}
such that, with our chosen normalizations and using~\eqref{eq:lambdaSTST} the tidal deformabilities read
\begin{align}
\lambda^{\ast T}_\ell&=\frac{1}{(2\ell-1)!!}\left.\frac{H_0^{(-\ell-1)}}{H_0^{(\ell)}}\right|_{\delta\varphi^{(\ell)}=0}~,\label{eq: TidalDefT}\\
\lambda^{\ast S}_\ell&=\frac{1}{(2\ell-1)!!}\left.\frac{\delta\varphi^{(-\ell-1)}}{\delta\varphi^{(\ell)}}\right|_{H_0^{(\ell)}=0}~,\label{eq: TidalDefS}\\
\lambda^{\ast ST}_\ell&=\frac{1}{(2\ell-1)!!}\left.\frac{H_0^{(-\ell-1)}}{2\delta\varphi^{(\ell)}}\right|_{H_0^{(\ell)}=0}\nonumber\\&=\frac{1}{(2\ell-1)!!}\left.\frac{2\delta\varphi^{(-\ell-1)}}{H_0^{(\ell)}}\right|_{\delta\varphi^{(\ell)}=0}~.\label{eq: TidalDefST}
\end{align}
To obtain explicit results, it is convenient to integrate the coupled system of differential equations, \eqref{eq:H0PertEq} and \eqref{eq:PhiPertEq}, from the singular points, i.e. the origin and infinity, and subsequently match them at the surface of the star. Extracting from this the multipole and tidal moments requires carefully disentangling the different moments corresponding to the different fields. This can be accomplished by constructing a generic solution as a linear combination of independent particular solutions. Particular solutions are computed by imposing certain boundary conditions, labeled $1$ and $2$ for the interior and A-D for the exterior of the star, and are listed in Table~\ref{tab: BCs}. Solutions A and C correspond to a nonzero scalar and tensor multipole moments respectively, i.e. $\delta\varphi^{(-\ell-1)}=1$ and ${H_0^{(-\ell-1)}}=1$, and the rest of the coefficients are set to zero, whereas solutions B and D correspond to nonzero tidal moments, $\delta\varphi^{(\ell)}=1$ and  $H_0^{(\ell)}=1$, and the rest of the coefficients are set to zero. This leads to six particular solutions with six associated constants of integration. Demanding continuity at the surface of the star fixes four of the constants. One of the remaining two constants can be fixed by choosing a normalization. Hence, one free constant remains that can be used to demand a zero scalar or tensor tidal field, $\delta\varphi^{(\ell)}=0$ or $H_0^{(\ell)}=0$. This disentangles the different contributions to the induced multipole moments and enables extracting the tidal deformabilities using~\eqref{eq: TidalDefT}-\eqref{eq: TidalDefST}.
 
 \subsubsection{Marginally stable solution}\label{ssec:StealthSol}
The scalar equation of motion for linearized perturbation \eqref{eq:ScalarPertEqg0} contains a term proportional to $\alpha'(\varphi)$. In cases where $\alpha'(\varphi)$ is constant and the background scalar field vanishes $\varphi_0=0$, the static scalar perturbations are solutions to
\begin{align}\label{eq:EoMStealth}
\delta\varphi''+&\frac{2\left[r-m+2 \pi  r^3 (p_0-\rho_0)\right]}{r (r-2 m)}\delta\varphi'\nonumber\\&-\left[\frac{\ell(\ell+1)}{r(r-2 m)}-\frac{4 \pi r \alpha'(0)  (\rho_0-3p_0)}{r-2m}\right]\delta\varphi=0~.
\end{align}
For this case most of the properties of the configuration are identical to those in GR, for instance, the equilibrium solutions and the fact that scalar and tensor perturbations decouple. However, an important difference is the presence of the  coupling term involving 
$\alpha'(0)$ and the matter variables in~\eqref{eq:EoMStealth}, which is absent in GR. Consequently, for $\alpha'(0)\neq 0$ the scalar tidal deformability may have a significantly different value than in GR. This is the case of a marginally stable solution. When perturbed, the system does not return to the GR state, but it is not unstable either, since the scalar charge does not grow unboundedly. Instead, due to nonlinearities, the system settles into a stable scalarized state. Specifically, the tidal deformabilities in this marginally stable case are given by the GR expressions
\begin{align}
\Lambda^T_\ell\mid_{\rm GR}=&\frac{e^{-i \ell \pi}(3+e^{2i\ell \pi})\Gamma(-\ell-\frac{1}{2})\Gamma(\ell+3)\Gamma(\ell-1)}{2^{2\ell+3}(2\ell-1)!!\Gamma(\ell+\frac{1}{2})}\times\nonumber\\&\frac{z^T_\ell P_{\ell}^2(1/C-1)+(\ell-1)CP_{\ell+1}^2(1/C-1)}{z^T_\ell Q_{\ell}^2(1/C-1)+(\ell-1)CQ_{\ell+1}^2(1/C-1)}~,\\
\Lambda^S_\ell\mid_{\rm GR}=&\frac{\pi\Gamma(\ell+1)^2}{2^{2\ell+1}(2\ell-1)!!\Gamma(\ell+\frac{1}{2})\Gamma(\ell+\frac{3}{2})}\times\nonumber\\&\frac{z^S_\ell P_{\ell}^0(1/C-1)+(\ell-1)CP_{\ell+1}^0(1/C-1)}{z^S_\ell Q_{\ell}^0(1/C-1)+(\ell-1)CQ_{\ell+1}^0(1/C-1)}~,
\end{align}
with
\be
y^T\mid_{\rm GR}=\frac{H_0'(R)}{H_0(R)}R~,\quad 
y^S\mid_{\rm GR}=\frac{\delta\varphi'(R)}{\delta\varphi(R)}R~,
\ee
the (adimensional) logarithmic derivative, $z^{T/S}_\ell=(\ell+1)(C-1)+(2C-1)y^{T/S}$ and $P^m_\ell(x)$ and $Q^m_\ell(x)$ the associated Legendre polynomials.

The condition $\alpha'(0)\neq0$ implies that only special cases of the coupling such as exponential couplings lead to the additional term in~\eqref{eq:EoMStealth}. The physical interpretation of this term is that, even though the equilibrium configuration is GR-like with a vanishing background scalar field, the scalar perturbation inherits the coupling between matter and the scalar field and differs from GR. This can be seen by considering the equation of motion for linearized perturbations of the scalar field \eqref{eq:EoMScalar}, 
\begin{align}
    \Box\delta\varphi&=\frac{1}{2K_\varphi}\alpha'(\varphi_0)T^\ast\delta\varphi\nonumber\\&=4\pi\alpha'(\varphi_0)A(\varphi_0)^4(3p_0-\rho_0)\delta\varphi~,
\end{align}
which for $\varphi_0=0$ and $A(0)=1$ yields the $\alpha'(0)$-dependent term in \eqref{eq:EoMStealth}. Therefore, this case represents a situation in which test particles follow geodesics independent of the scalar field, hence satisfying the weak equivalence principle as in GR.
This class of solutions are the marginally stable ones with zero charge, $q=0$, in the regime where spontaneous scalarization can occur (i.e., scalarized solutions are the stable ones). These solutions, albeit marginally stable, will have a scalar Love number that could be induced by a scalarized companion during the inspiral. Thus, they are relevant in the context of dynamical scalarization, in particular for the transition between nonscalarized and scalarized objects (see, e.g., Ref.~\cite{Khalil:2022sii} for a discussion in the context of the worldline EFT). Since in this paper we focus on isolated objects, we will leave this study for further work.

 \section{Numerical results for exemplary case studies}\label{sec:CaseStudy}
 \subsection{Setup}
To compute NS configurations we consider piecewise polytropic approximations to tabulated equations of state~\cite{Read:2008iy}. In particular, we choose WFF1, SLy and H4 since they cover a significant range of NS masses and radii and they have also been considered in the literature \cite{Damour:2009vw,Pani:2014jra,Yazadjiev:2018xxk}, which allow us to check and compare results. 

For the scalar coupling function, we choose here the concrete case of 
\begin{align}
A(\varphi)=e^{\frac{1}{2}\beta\varphi_0(r)^2},
\end{align}
 known as the Damour-Esposito-Farèse model \cite{PhysRevLett.70.2220}. This choice yields scalarized configurations, depending on the cosmological value $\varphi_{0\infty}$, which is related to $\beta$, $\omega(\phi)$, and $F(\phi)$ through \eqref{eq:STalpha} or explicitly 
\begin{align}
\varphi_{0\infty}=-\frac{1}{\beta}\sqrt{\frac{K_\varphi}{2K_R}}\frac{F'_\infty}{\sqrt{3 F_\infty'^2+2\omega_\infty F_\infty/\phi_\infty}}~.
\end{align}
For the numerical studies, we choose the values $\varphi_{0\infty}=(10^{-6},10^{-3})$. The latter is a common choice in the literature as it lies within the experimental bounds from binary-pulsar observations \cite{Damour:1998jk} and yields scalarized NSs throughout the entire mass range. The smaller value of $\varphi_{0\infty}$ leads to a more sharp transition into scalarized states, which is useful for comparing against the GR configurations. Scalarized configurations exist only for $\beta\leq-3.5$ \cite{PhysRevLett.70.2220,Palenzuela:2013hsa}. Furthermore, pulsar timing observations have discarded the parameter regimes $\beta\leq-4.5$ \cite{Freire:2012mg,Shao:2017gwu,Kramer:2021jcw,Zhao:2022vig}. However, we will study the cases $\beta=-4.5$ and $\beta=-6$ as in the literature to gain qualitative insights into parameter dependencies. We impose the boundary conditions near the center of the star~\eqref{eq:bcscenter} for
 $r_{\rm min}=10^{-10}$ in the geometric units we are using. For the asymptotic expansions of the solutions near spatial infinity, we consider 22 orders in the series~\eqref{eq:asym22}.
\begin{figure}
\begin{center}
{\includegraphics[width=0.49\textwidth,clip]{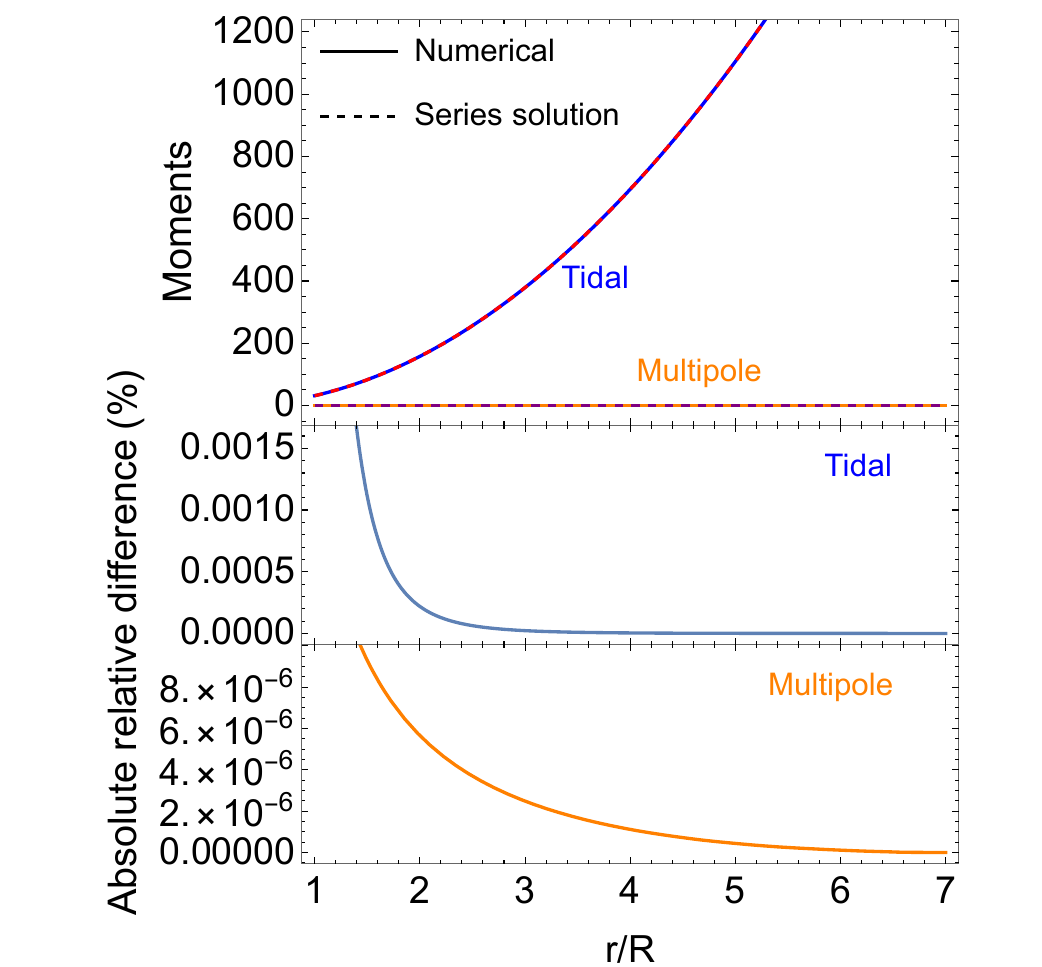}}
\caption{Agreement between series expansions near infinity and numerical solutions for the quadrupolar tidal field and induced quadrupole moment. The top panel shows the overlap between the numerical (solid lines) and series solutions (dashed lines) for the tidal (blue and red) and multipole moment (orange and purple). The middle and bottom panels show the percent absolute relative difference for the quadrupolar tidal field (blue) and the induced quadrupole moment (orange) respectively. This example corresponds to a configuration with the WFF1 EoS, $M=1.16 M_{\odot}$, $R=10.16$km, $\beta=-4.5$, and $\varphi_\infty=10^{-3}$; results for other configurations are similar. }\label{fig:OverlapPlot}
\end{center}
\end{figure}

\begin{figure}
\begin{center}
{\includegraphics[width=0.49\textwidth,clip]{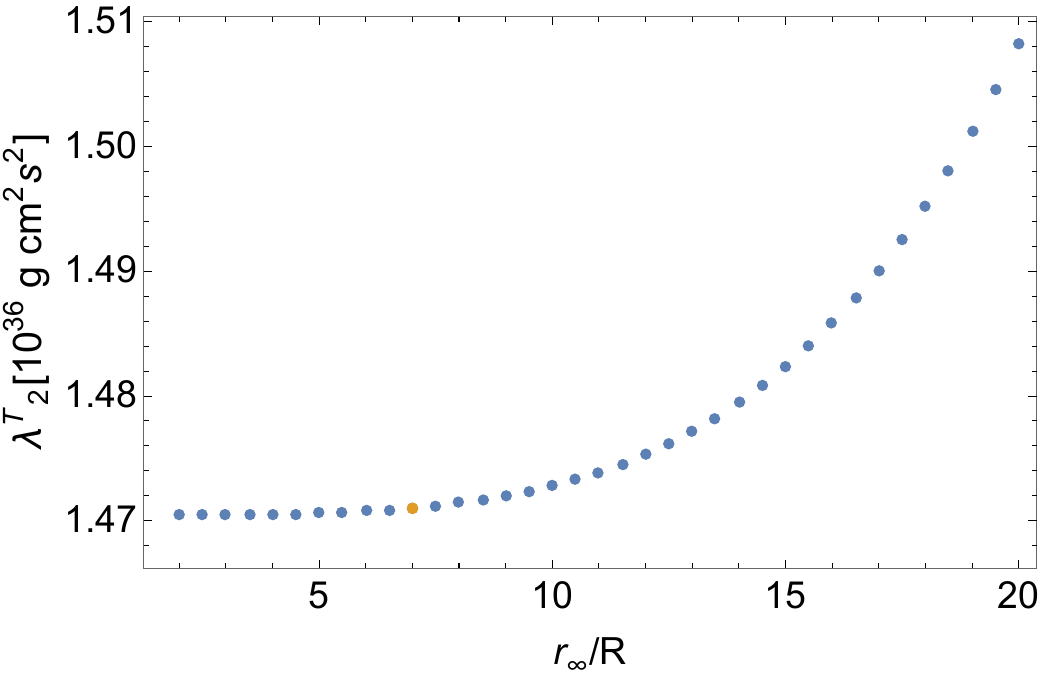}}
\caption{Quadrupolar tensor tidal deformability $\lambda^{T}_2$ in the Einstein frame as a function of $r_\infty$. The orange point is the chosen one for the integration. For illustrative purposes we choose a configuration with WFF1 EoS, $M=1.16 M_{\odot}$, $R=10.16$km, $\beta=-4.5$, and $\varphi_\infty=10^{-3}$; other choices yield similar results. }\label{fig:TidalDefVsrinf}
\end{center}
\end{figure}
\begin{figure*}[htpb!]
\begin{center}
{\includegraphics[width=0.49\textwidth,clip]{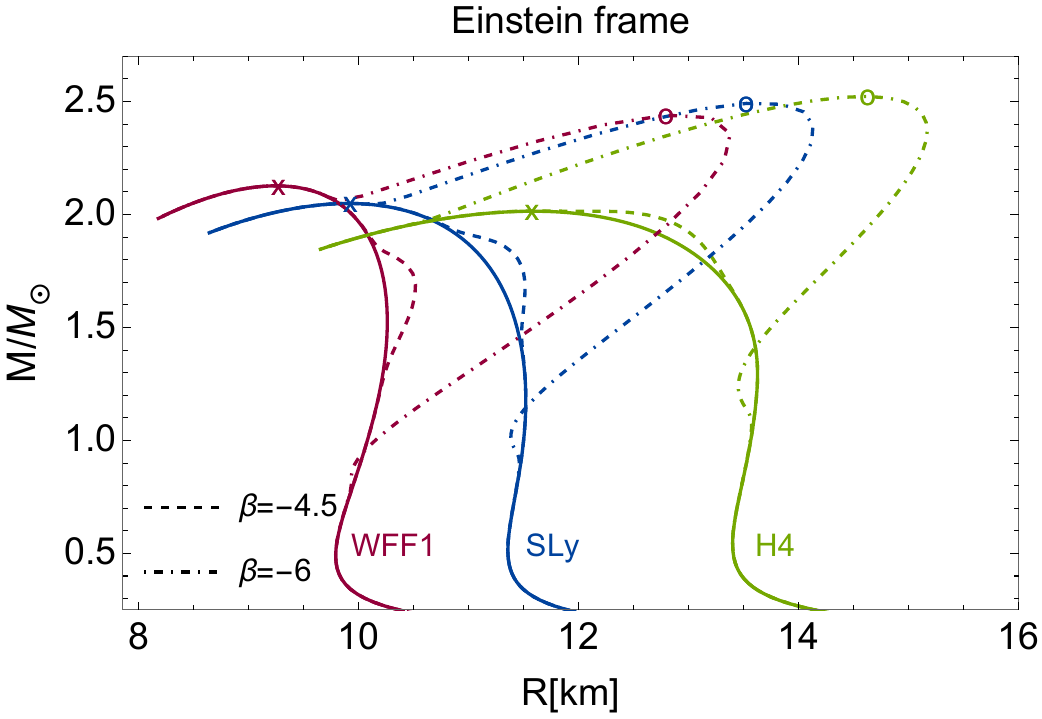}}
{\includegraphics[width=0.49\textwidth,clip]{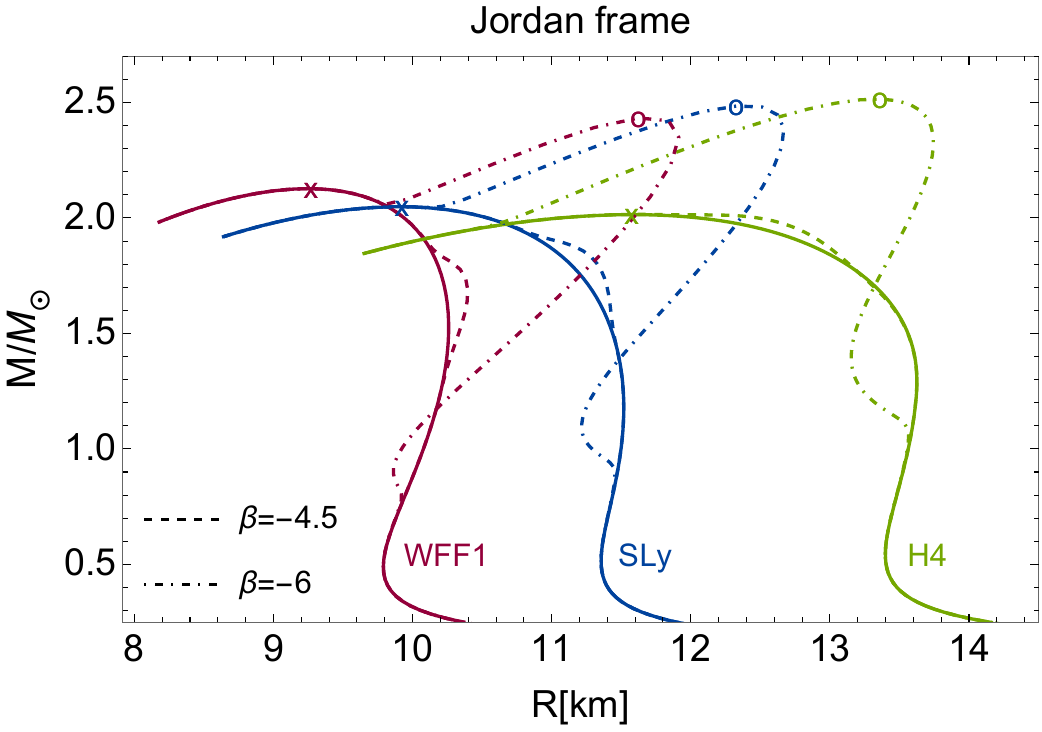}}
\caption{\textbf{\emph{Mass-radius curves}} in the Einstein (left) and Jordan frames (right) for three equations of state (WFF1, SLy, and H4). The solid lines represent the GR configurations $\beta=0$ and the dashed and dot-dashed lines are the scalarized configurations with $\beta=-4.5$ and $\beta=-6$, respectively. The cross represents the maximum mass configuration for $\beta=0,-4.5$, and the circle for $\beta=-6$. Both plots correspond to a scalar field at infinity $\varphi_{0\infty}=10^{-3}$.}\label{fig:MassRadius}
\end{center}
\end{figure*}
\begin{figure*}[htpb!]
\begin{center}
{\includegraphics[width=0.49\textwidth,clip]{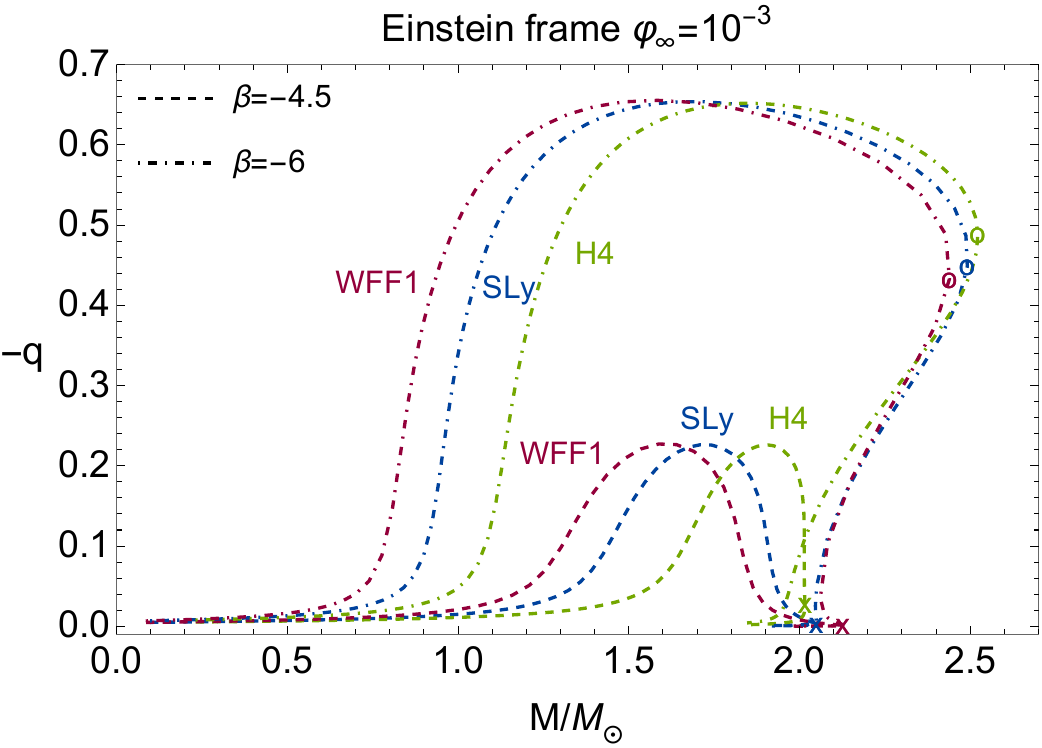}}
{\includegraphics[width=0.49\textwidth,clip]{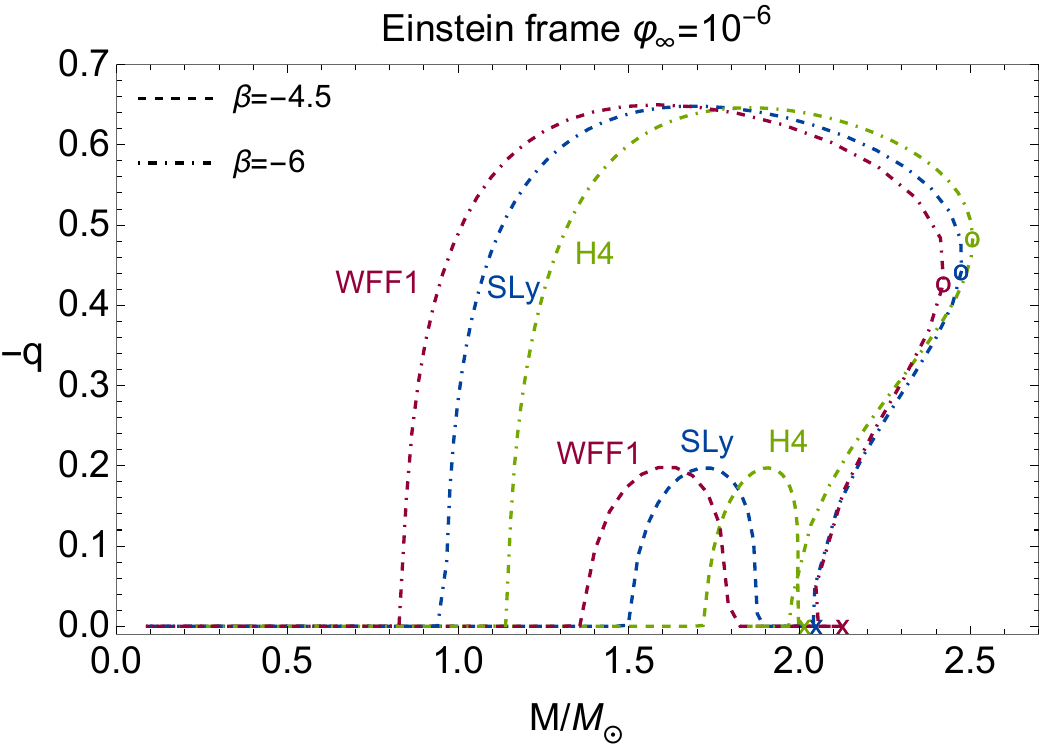}}
\caption{\textbf{\emph{Charge-mass curves}} in the Einstein frame for three equations of state (WFF1, SLy, and H4) and a scalar field at infinity $\varphi_{0\infty}=(10^{-3},10^{-6})$. The dashed and dot-dashed lines are the scalarized configurations with $\beta=-4.5$ and $\beta=-6$, respectively. The cross represents the maximum mass configuration for $\beta=0,-4.5$, and the circle for $\beta=-6$. }\label{fig:ChargeRadius}
\end{center}
\end{figure*}
To extract quantities at infinity, we choose $r_{\infty}=7R$ for $\ell=1,2$ and $r_{\infty}=4R$ for $\ell=3$, with $R$ the surface of the star. They correspond to values within the range in which there is an overlap of the series expansions at infinity, \eqref{eq: ScalarSolInf} and \eqref{eq: TensorSolInf} and the numerical solutions, see Fig.~\ref{fig:OverlapPlot} for an example configuration with a percent difference of at most $1.6\times10^{-2}\%$ for the tidal moments and $1.5\times10^{-5}\%$ for the multipole moments. To further check the robustness of these choices we also computed results when dropping five orders in the series solution, which yielded no noticeable changes in the Love numbers. Varying $r_\infty$ between $2$ and $10R$ led to sub-percent level changes in the tidal deformabilities (e.g. at most $0.6\%$ for the SLy EoS for both choices of $\beta$ and different kinds of Love numbers), see Fig.~\ref{fig:TidalDefVsrinf}. We also note that higher multipolar orders require extracting quantities at an $r_{\infty}$ that is closer to the star's surface in order to capture the increasingly smaller contributions. 

When numerically extracting the dipolar Love number, we observed a very sensitive dependence on the initial conditions. During the initial computations, we obtained outliers (around $1-10$) that we attribute to numerical problems hidden within \texttt{Mathematica} built-in functions. A way to mitigate this is by repeating the analysis with a shift in the central densities and combining the different data points, getting rid of the rogue results. Also, we tried different combinations for (\texttt{PrecisionGoal}, \texttt{AccuracyGoal} and \texttt{WorkingPrecision}) options of the function \texttt{NDSolve} of \texttt{Mathematica}. We found the best choice to be $(10,10,20)$ for the background solutions, and the automatic setting for the dipolar perturbations for $\beta=-4.5$.

We note that all quantities shown in the plots below correspond to the their respective frame. However, we omitted asterisks and subindices for clarity. The \texttt{Mathematica} code with the analytical and numerical calculations can be found in the \texttt{GitHub} repository \cite{Creci_Tidal_properties_of_2024}.

\subsection{Mass-radius and charge-mass curves}
 Figure~\ref{fig:MassRadius} shows the mass-radius curves computed with the above methodology and setup in the Einstein and Jordan frames. The Einstein frame mass $M$ is the ADM mass \eqref{eq: ADMMassEinstein} and $R$ the radius of the star, defined as the distance from the origin at the center of the star at which $p_0=0$. For the Jordan frame mass and radius we make use of \eqref{eq: ADMMassJordan} and \eqref{eq: RadiusEinteinToJordan}. Each point in the mass-radius curve corresponds to a configuration with different central density, increasing from the right to the left of the plot. We denote with a cross and a circle the maximum mass configuration for $\beta=0,-4.5$, and $\beta=-6$, respectively.
\begin{figure*}[p]
\begin{center}
{\includegraphics[width=0.49\textwidth,clip]{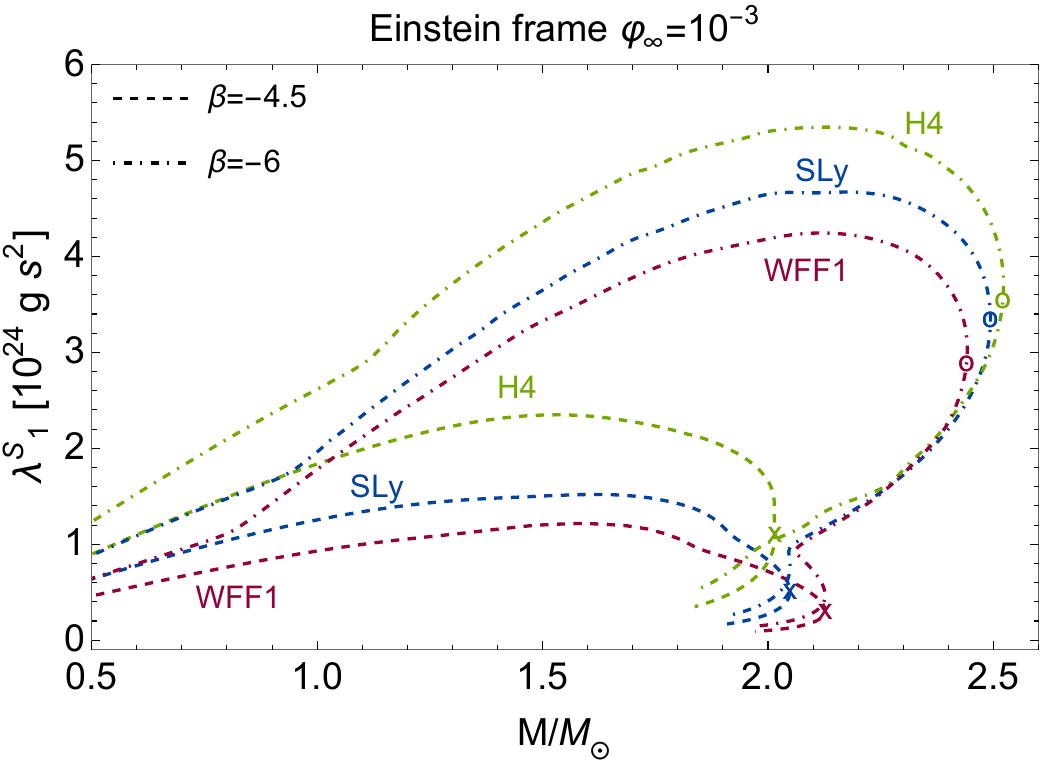}}
{\includegraphics[width=0.49\textwidth,clip]{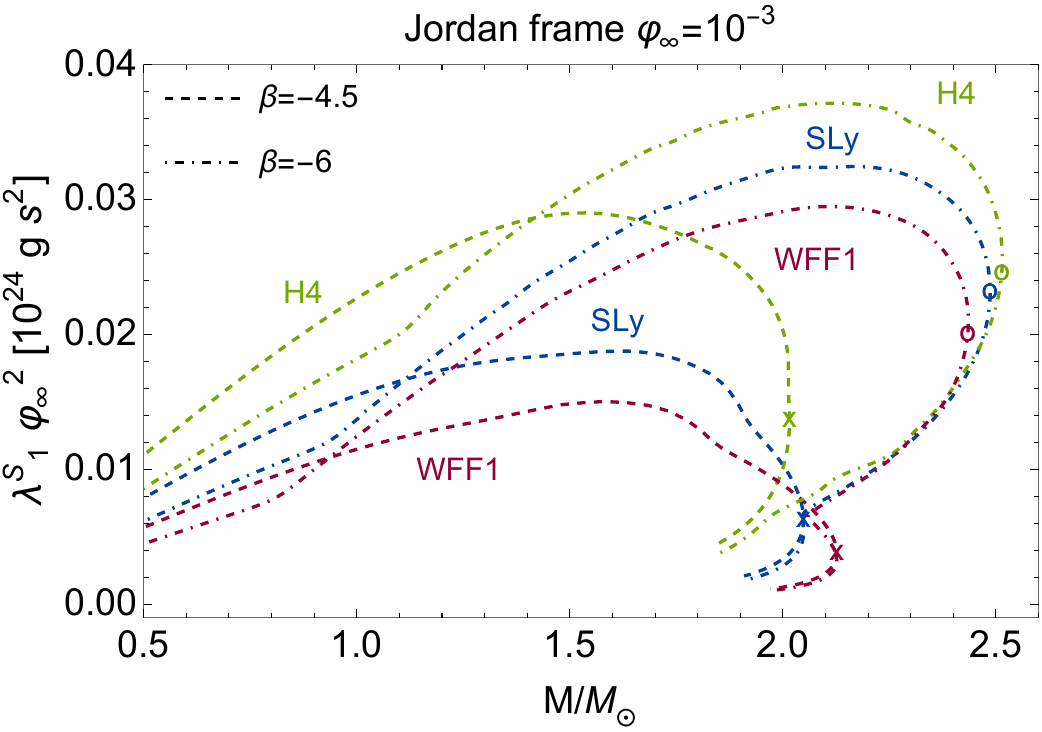}}
\caption{\textbf{\emph{Dipolar scalar tidal deformabilities}} $\lambda_1^S$ in the Einstein and Jordan frames for three equations of state (WFF1, SLy, and H4). The dashed and dot-dashed lines are the scalarized configurations with $\beta=-4.5$ and $\beta=-6$, respectively. For $\beta=-6$ we have omitted the data beyond the maximum mass configuration for better readability. The cross represents the maximum mass configuration for $\beta=0,-4.5$, and the circle for $\beta=-6$. All plots correspond to a scalar field at infinity $\varphi_{0\infty}=10^{-3}$.}\label{fig:lambdaScalarl1}
\end{center}

\begin{center}
\begin{minipage}{0.49\textwidth}
{\includegraphics[width=\textwidth,clip]{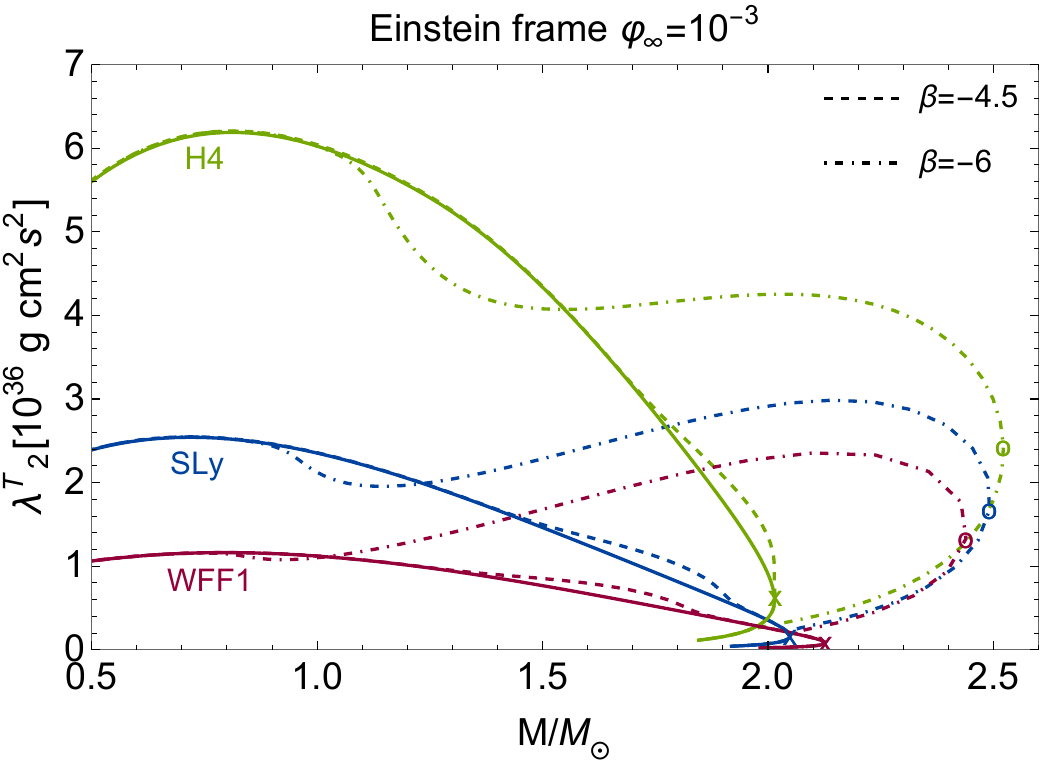}}
{ (a) Tensor}\label{fig:lambdaTensor}
\end{minipage}
\begin{minipage}{0.49\textwidth}
{\includegraphics[width=\textwidth,clip]{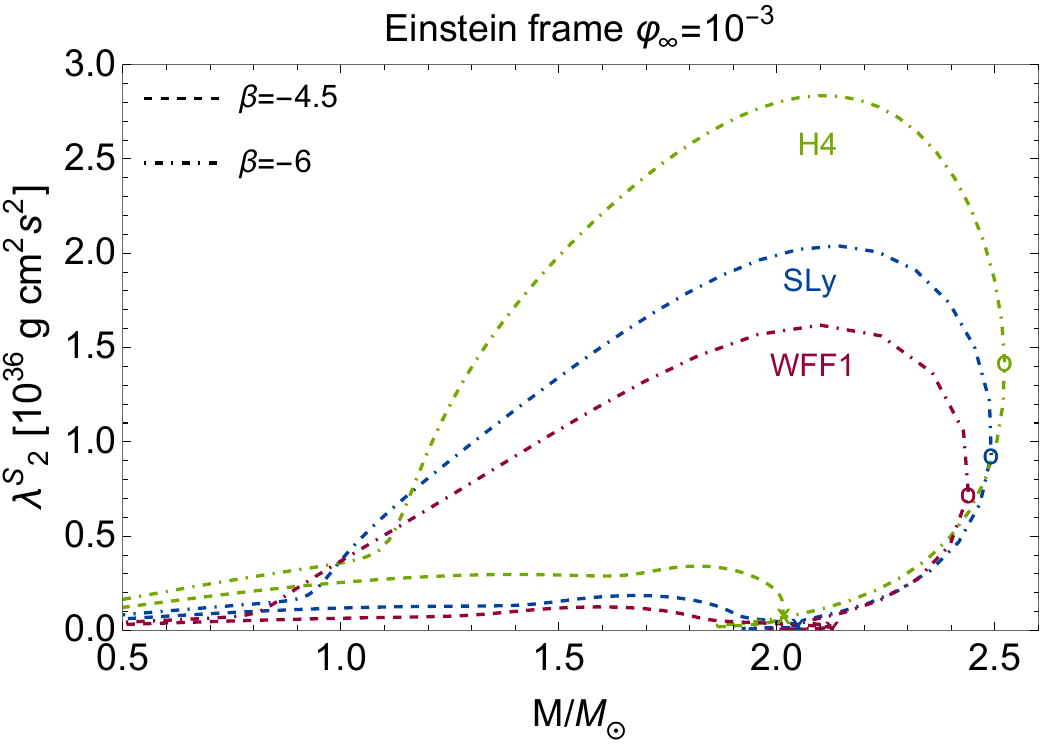}}
{(b) Scalar}\label{fig:lambdaScalar}
\end{minipage}
\begin{minipage}{0.49\textwidth}
{\includegraphics[width=\textwidth,clip]{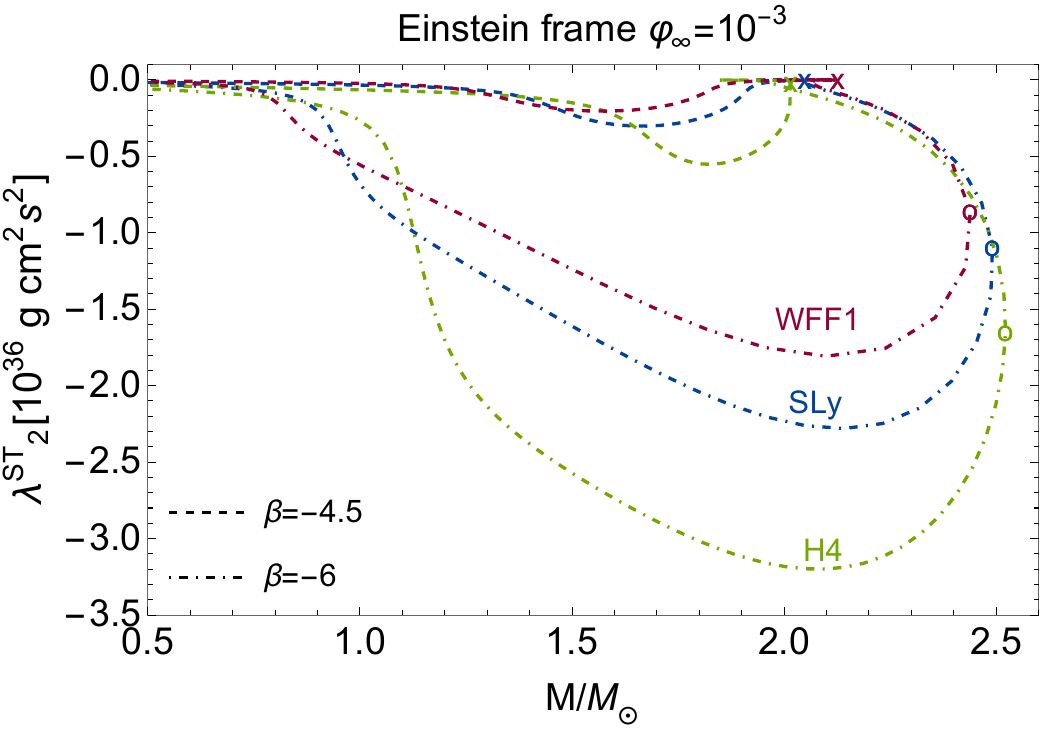}}
{(c) Scalar-tensor}\label{fig:lambdaScalarTensor}
\end{minipage}
\caption{\textbf{\emph{Quadrupolar tidal deformabilities}} $\lambda_2$ in the Einstein frame for three equations of state (WFF1, SLy, and H4). Panel (a) illustrates the tensor deformability, (b) the scalar one, and (c) the mixed scalar-tensor one. The solid lines represent the GR configurations $\beta=0$ and the dashed and dot-dashed lines are the scalarized configurations with $\beta=-4.5$ and $\beta=-6$, respectively. The plot corresponds to a scalar field at infinity $\varphi_{0\infty}=10^{-3}$. The cross represents the maximum mass configuration for $\beta=0,-4.5$, and the circle for $\beta=-6$.
The corresponding results in the Jordan frame shown in Appendix~\ref{app:Plotslambda} are qualitatively similar, the tensor Love numbers are the same in both frames [see~\eqref{eq:ScalarTidalDefTransf}], while the scalar and scalar-tensor ones are about 2 orders of magnitude smaller in the Jordan frame.}\label{fig:lambdaquadrupolar}
\end{center}
\end{figure*}

In agreement with \cite{Pani:2014jra,PhysRevLett.70.2220}, we see that certain configurations (dot-dashed lines in Fig.~\ref{fig:MassRadius}) deviate away from the GR values (solid lines) due to scalarization, as corroborated by the behavior of the scalar charge shown in Fig.~\ref{fig:ChargeRadius}. This indicates that configurations exhibit a sudden growth in the scalar field beyond a certain compactness, leading to a larger radius and higher mass than their GR counterparts. This is because the non-negligible amount of scalar field increases both the mass and pressure of the fluid, yielding more massive and bigger stars.

\subsection{Love numbers}
\begin{figure*}[htpb!]
\begin{center}
\begin{minipage}{0.49\textwidth}
{\includegraphics[width=\textwidth,clip]{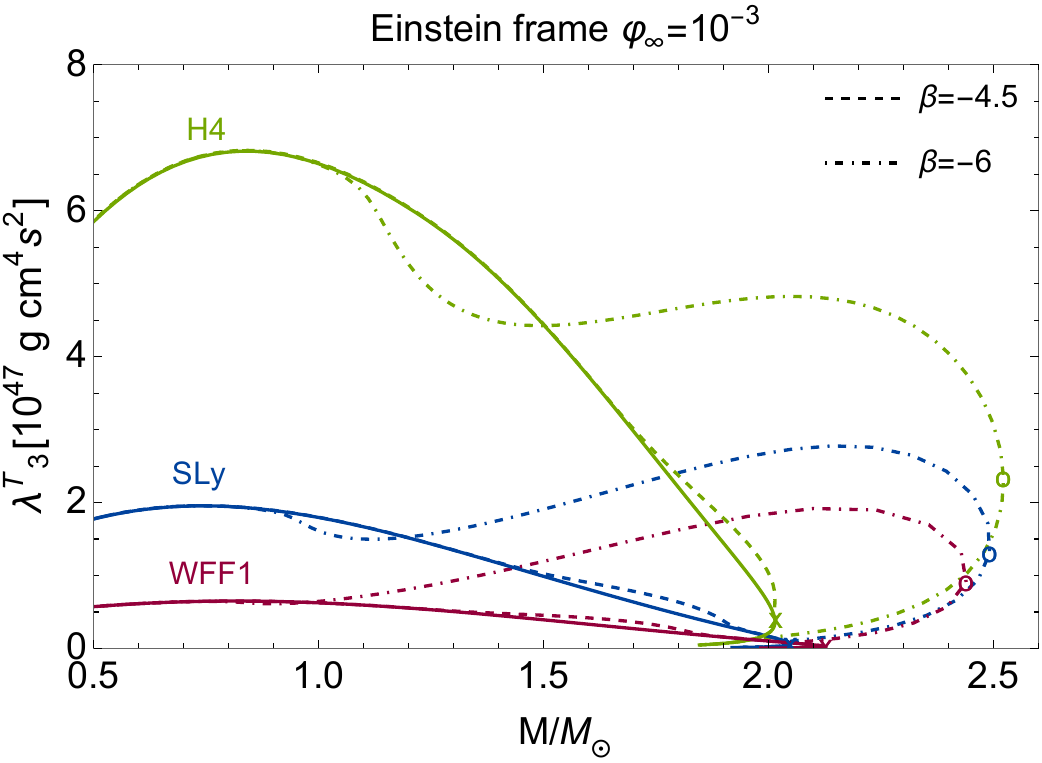}}
{(a) Tensor}
\end{minipage}
\begin{minipage}{0.49\textwidth}
{\includegraphics[width=\textwidth,clip]{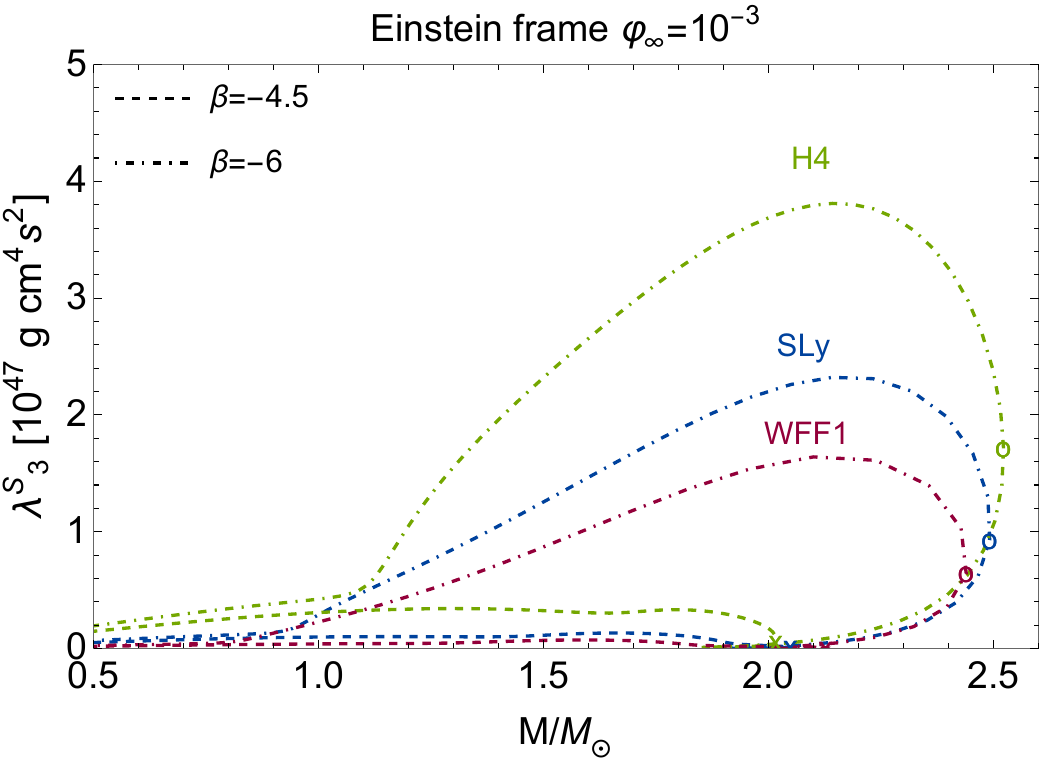}}
{(b) Scalar}
\end{minipage}
\begin{minipage}{0.49\textwidth}
{\includegraphics[width=\textwidth,clip]{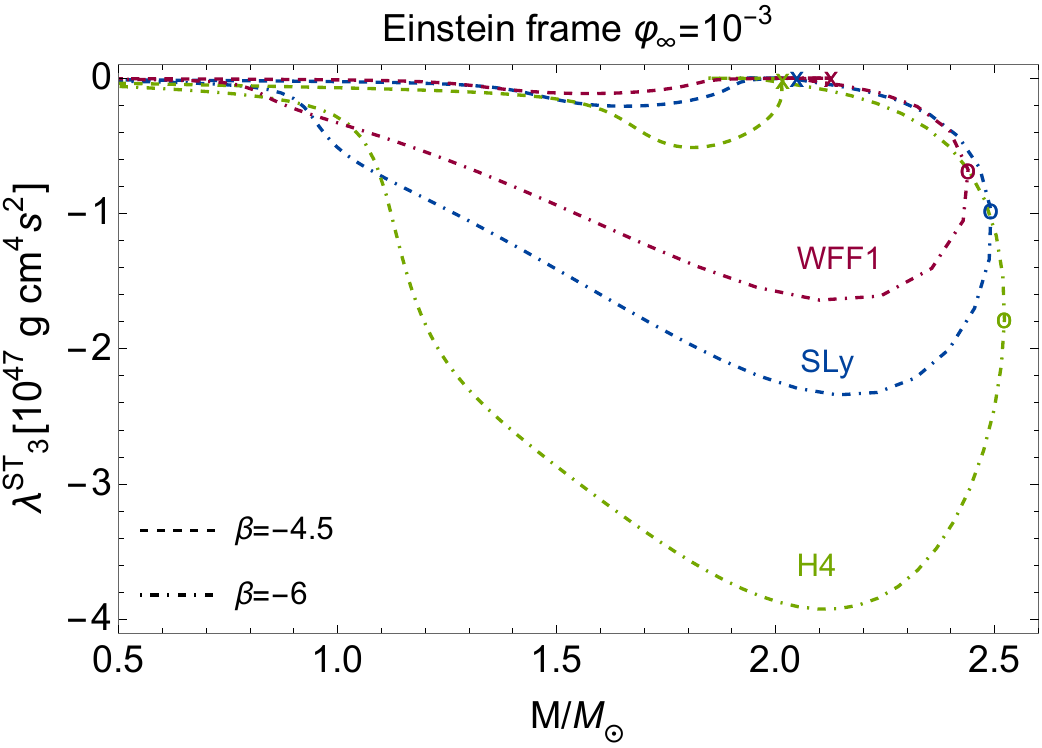}}
{(c) Scalar-tensor}
\end{minipage}
\caption{\textbf{\emph{Octupolar tidal deformabilities}} $\lambda_3$ in the Einstein frame for three equations of state (WFF1, SLy, and H4). Panel (a) illustrates the tensor deformability, (b) the scalar one, and (c) the mixed scalar-tensor one. The solid lines represent the GR configurations $\beta=0$ and the dashed and dot-dashed lines are the scalarized configurations with $\beta=-4.5$ and $\beta=-6$, respectively. The cross represents the maximum mass configuration for $\beta=0,-4.5$, and the circle for $\beta=-6$. The plot corresponds to a scalar field at infinity $\varphi_{0\infty}=10^{-3}$. 
The corresponding results in the Jordan frame shown in Appendix~\ref{app:Plotslambda} are qualitatively similar, the tensor Love numbers are the same in both frames [see~\eqref{eq:ScalarTidalDefTransf}], while the scalar and scalar-tensor ones are about 2 orders of magnitude smaller in the Jordan frame.}\label{fig:lambdaoctupolar}
\end{center}
\end{figure*}

The tidal deformabilities computed with the method described above are shown in Fig.~\ref{fig:lambdaScalarl1} for dipolar $\ell=1$ perturbations, Fig.~\ref{fig:lambdaquadrupolar} for quadrupolar $\ell=2$ perturbations and Fig.~\ref{fig:lambdaoctupolar} for octupolar $\ell=3$ perturbations. We plot the tidal deformabilities for different values of $\beta$, a fixed $\varphi_{0\infty}=10^{-3}$ and the three considered EoS: WFF1, SLy, and H4 in both the Einstein and Jordan frames.

As explained in Sec.~\ref{sec: ComputingLoveNumbers}, the tensor dipolar perturbations can be made to vanish by a gauge transformation and consequently, there are no dipolar tensor nor mixed scalar-tensor tidal deformabilities as these are pure gauge quantities. The dipolar scalar tidal deformability is shown in Fig.~\ref{fig:lambdaScalarl1}. The shape of these curves as functions of mass changes significantly depending on the value of $\beta$, however, the order of magnitude remains similar. We also observe structures in the curves which, based on further analysis, we attribute to consequences of the charge-dependent shift in the scalar dipole moment when choosing the center-of-mass gauge, c.f.~\eqref{eq:DipoleShift}. Additionally, for small masses in the Einstein frame (left panel of Fig.~\ref{fig:lambdaScalarl1}), we observe an overlap between curves corresponding to a 'stiffer' equation of state (with generally larger $\lambda$ for a given mass) and a large value of the scalar coupling $|\beta|$ and the results for a 'softer' equation of state and smaller value of $|\beta|$. This exemplifies how degeneracies can appear between the equation of state and spacetime and therefore highlights the importance of accurately modeling both.

Figure~\ref{fig:lambdaquadrupolar}(a) shows the quadrupolar tensor tidal deformability  curves. Similar to the mass-radius curves, deviations from the GR case appear for scalarized configurations. These deviations are similar to those computed in \cite{Pani:2014jra} and \cite{Brown:2022kbw} and are smaller for smaller $|\beta|$. It is interesting to note that the regions of masses in which the tidal deformabilities are above or below their GR counterparts do not coincide with those of the mass-radius curves. The behavior of the  scalar tidal deformability shown in Fig.~\ref{fig:lambdaquadrupolar}(b) shows a much greater sensitivity to changes in $\beta$ (dashed versus dot-dashed curves). These differences for the choices of $\beta$ considered here are one order of magnitude, demonstrating the sensitive dependence of the scalar tidal deformability on the theory parameters.  As there are no (scalar) charged compact objects in GR and thus no mechanism to produce a scalar tidal field we lack any GR benchmarks in this case, though we note that within the scalar-tensor theories there exist the GR-like, marginally stable equilibrium configurations discussed in Sec.~\ref{ssec:ExtractingMult}. In taking the limit $\varphi_{0\infty}\rightarrow0$ these would correspond to connecting curves underneath the bumps exhibited by the curves shown in Fig.\ref{fig:lambdaquadrupolar}(b), similar to the GR curves in the tensor tidal deformability in Fig.~\ref{fig:lambdaquadrupolar}(a). 

Finally, we show the novel scalar-tensor tidal deformability in Fig.~\ref{fig:lambdaquadrupolar}(c). As we can see, they are negative throughout most of the parameter space. Similar to the scalar tidal deformability, they are also more sensitive to $\beta$ than the pure tensor tidal deformability. Furthermore, it follows a similar behavior as the scalar charge, i.e., it is nonzero for the scalarized states and zero for the unscalarized states. As shown in Appendix~\ref{app:Plotslambda}, both scalar and scalar-tensor tidal deformabilities have different orders of magnitude in the Jordan frame, and scale differently with $\varphi_{0\infty}$. This is a consequence of the relation between the Einstein and Jordan frame scalar fields.

The octupolar tidal deformabilities, shown in Figure~\ref{fig:lambdaoctupolar} exhibit qualitatively very similar trends over the parameter space considered as the quadrupolar ones. A difference is that the adimensional Love numbers $k_3$ shown in Appendix~\ref{app:PlotsLambdak} are 1 order of magnitude smaller than the quadrupolar counterparts. As seen in Fig.~\ref{fig:LambdaScalarTensor3}, the scalar-tensor adimensional tidal deformability $\Lambda_3^{ST}$ is also negative for most of the configurations but can become noticeably positive for higher-mass systems with large negative $\beta$ and soft equations of state.

\subsubsection{Investigating the parity invariance of $\lambda_\ell^{ST}$}
\begin{figure}[H]
\begin{center}
{\includegraphics[width=0.49\textwidth,clip]{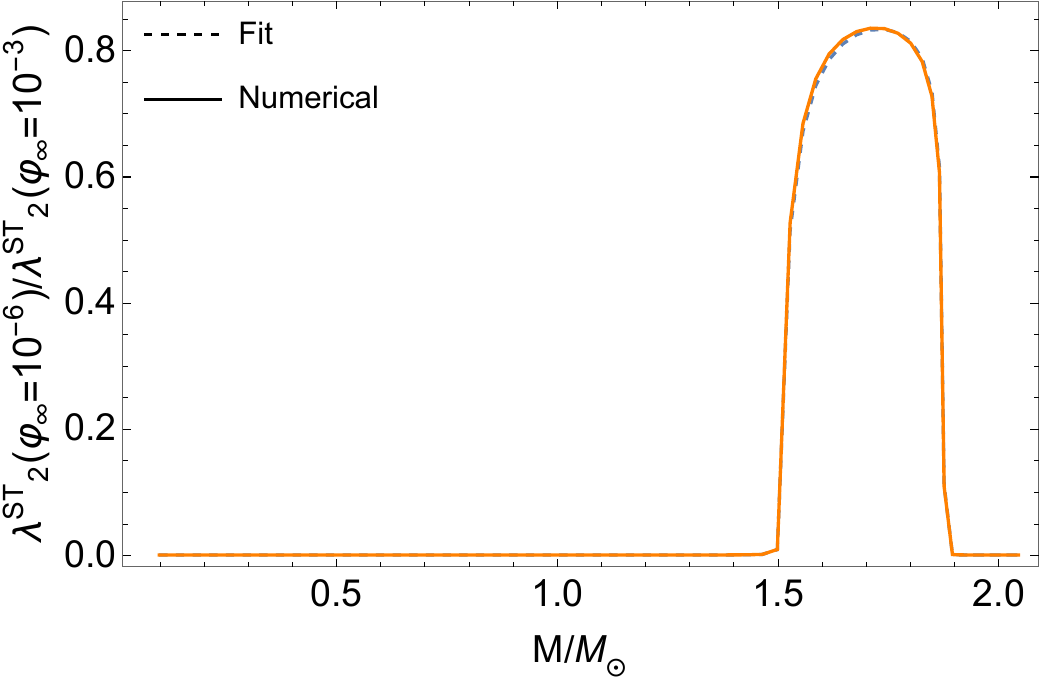}}
\caption{Ratio of scalar-tensor tidal deformabilities computed for $\varphi_\infty=10^{-6}$ and $\varphi_\infty=10^{-3}$. The solid line is the numerical result and the dashed line is the ratio using the linear fit to the polynomial $w(\varphi_\infty,q)$.}\label{fig:wPolynomialFit}
\end{center}
\end{figure}

\begin{figure}[H]
\begin{center}
{\includegraphics[width=0.49\textwidth,clip]{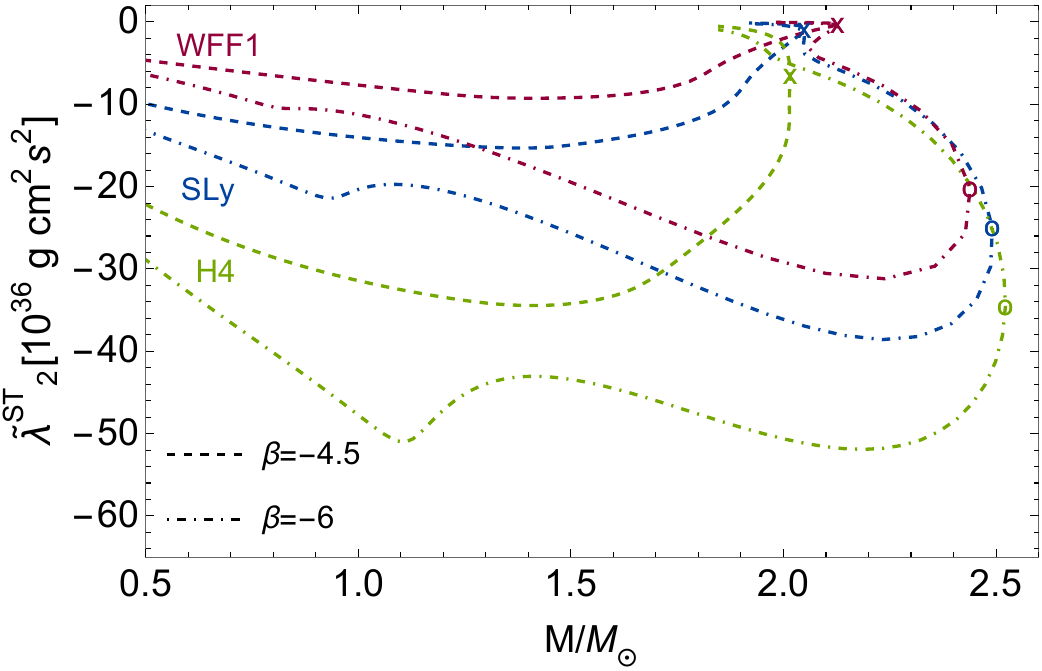}}
\caption{\textbf{\emph{Quadrupolar, scalar-field-independent scalar-tensor tidal deformabilities}} $\tilde{\lambda}_2^{ST}$ in the Einstein frame for three equations of state (WFF1, SLy, and H4). The dashed and dot-dashed lines are the scalarized configurations with $\beta=-4.5$ and $\beta=-6$, respectively. The cross represents the maximum mass configuration for $\beta=0,-4.5$, and the circle for $\beta=-6$.}\label{fig:TildeTidalDefST}
\end{center}
\end{figure}

We next return to the parity considerations of the scalar-tensor deformabilities discussed in Sec.~\ref{subsect:Paritygeneral} and compute the function $w(\varphi_\infty, Q)$ for this case study. 
We compute the ratio between scalar-tensor tidal deformabilities for three values of the asymptotic scalar field, $\varphi_{\infty}=10^{-3}$, $2\times10^{-3}$, and $10^{-6}$ and fit the data with an 
informed guess. In particular,
\begin{align}
\label{eq:wansatz}
    w(\varphi_\infty,q)&=\sum_{p=0}^{\infty}c_{p}\varphi_{\infty}^{2p+1}+\sum_{n=0}^{\infty}c_{n}q^{2n+1}\nonumber\\&\approx c_{\varphi}\sum_{p=0}^{\infty}\varphi_{\infty}^{2p+1}+c_{q}\sum_{n=0}^{\infty}q^{2n+1}\nonumber\\&\approx c_{\varphi}\frac{\varphi_\infty}{1-\varphi_\infty^2}+c_q \frac{q}{1-q^2}\nonumber\\&\approx c_\varphi\varphi_{\infty}+ c_q q~,
\end{align}
where we use the charge per unit mass $q$ instead of the charge $Q$ to make the expression nondimensional. In the second equality, we have assumed that the coefficients are the same for all terms in the sum, then summed the series and used that $\varphi_\infty^2\ll1$ and $q^2\ll1$. In the regime of vanishing scalar charge, we find that the deformabilities are directly proportional to the field. Thus, we set 
\be
c_\varphi=1.
\ee 
The best fit for the remaining coefficient $c_q$ in~\eqref{eq:wansatz} is 
\be
c_q=-0.0957
\ee 
with a fractional difference of at most $1.6\%$ and $2.6\%$ for $\varphi_{\infty}=(2\times10^{-3},10^{-6})$, respectively. In Fig.~\ref{fig:wPolynomialFit} we show the ratio of scalar-tensor tidal deformabilities computed for $\varphi_\infty=10^{-6}$ and $\varphi_\infty=10^{-3}$, respectively, and the corresponding ratio between the fitted polynomials $w(\varphi_\infty,q)$.
With this result we can compute the scalar-field-independent scalar-tensor tidal deformability $\tilde{\lambda}_\ell^{ST}$. In Figure~\ref{fig:TildeTidalDefST} we present the results for the quadrupolar case. Even though this tidal deformability is not relevant for gravitational wave observables, it is interesting to see how the field dependence suppresses the order of magnitude compared to Figure~\ref{fig:lambdaquadrupolar}c.

\section{Summary and Discussion}\label{sec:SummaryDiscussion}
\subsection{Effective action}
The tidal deformability parameters $\lambda_\ell$, or Love numbers, are useful GW observables that contain information about the fundamental physics of matter and spacetime. These quantities are computed from detailed calculations of the response of a relativistic compact object configuration to a perturbing tidal field and must be related to coefficients characterizing the resulting signatures in GWs at the orbital and radiation scales much larger than the size of the bodies. We establish this connection using an effective field theory description, where the bodies are described as worldline skeletons, i.e. a central worldline augmented with multipole moments. We demonstrated that in modified theories of gravity containing an additional scalar degree of freedom such as scalar-tensor theories, the number of tidal deformabilities needed to fully characterize the multipolar structure of the bodies is enhanced and in fact requires three kinds of Love numbers: a tensor (T), scalar (S), and mixed scalar-tensor (ST) parameter. The effective action describing adiabatic tidal effects in the binary dynamics is thus given by 
\begin{align}
S_{\rm tidal}=&\sum_{\ell} \int d\sigma~z~g^{LP}\times\nonumber\\&\left(\frac{\lambda^T_\ell}{2\ell!}E_L^{T}E_P^{T}+\frac{\lambda^S_\ell}{2\ell!}E_L^{S}E_P^{S}+\frac{\lambda^{ST}_\ell}{\ell!}E_L^{T}E_P^{S}\right),
\end{align}
with $\sigma$ and $z$ defined in \eqref{eq:Spp}. The latter is a novel term that characterizes the scalar/tensor multipole moment $Q_L$ induced by a tensor/scalar tidal field $E_L$ as a consequence of the coupling between tensor and scalar perturbations of the body,
\begin{align}
    Q^{S}_L=&-\lambda^{ S}_{\ell}E_L^{ S}-\lambda^{ ST}_{\ell}E_L^{ T}~,\\
    Q^{ T}_L=&-\lambda^{ T}_{\ell}E_L^{ T}-\lambda^{ ST}_{\ell}E_L^{ S}~.
\end{align}

\subsection{Frame transformations}
We focus on scalar-tensor theories, which are originally formulated in the Jordan frame, see \eqref{eq: STaction}. However, calculations are simpler in a conformally related Einstein frame. To transform results back to the Jordan frame, 
we derived the mapping of tidal deformabilities between these frames based on the action and obtained
\begin{align}
\lambda_\ell^T&=\lambda_\ell^{\ast T}~,\\ 
\lambda_\ell^S&=\left(\frac{A_\infty^{2}{F'}_\infty}{2\alpha_\infty}\right)^2\lambda_\ell^{\ast S}~,\\ \lambda_\ell^{ST}&=\frac{A_\infty^{2}F'_\infty}{2\alpha_\infty}\lambda_\ell^{\ast ST}~,
\end{align}
where the asterisks denote quantities in the Einstein frame and with $F$, $A$ and $\alpha$ defined in \eqref{eq: STaction}, \eqref{eq:MetricConformalTransf} and \eqref{eq:STalpha}, respectively.

\subsection{Numerical extraction of tidal deformabilities}
To numerically extract each tidal deformability parameter from the coupled system of equations, we developed a generic framework to disentangle multipolar and tidal fields. In particular, we demonstrated the use of different boundary conditions in the interior and exterior of the NS to construct the most generic solution as a linear combination of particular solutions. The different solutions are summarized in Table~\ref{tab: BCs}. Then, we proceeded by
\begin{itemize}
    \item[1.] matching the interior and exterior solutions and their derivatives at the star's surface, and
    \item[2.] fixing an arbitrary normalization.
\end{itemize}
This determines the solution up to a free constant that can be chosen to set either the tensor $E^{\ast T}_L$ or scalar $E^{\ast S}_L$ tidal fields to zero. With this, the tidal deformabilities can be computed from
\begin{subequations}
\begin{align}
\lambda^{\ast T}_\ell&=-\left.\frac{Q^{\ast T}_L}{E^{\ast T}_L}\right|_{E^{\ast S}_L=0}~,\\
\lambda^{\ast S}_\ell&=-\left.\frac{Q^{\ast S}_L}{E^{\ast S}_L}\right|_{E^{\ast T}_L=0}~,\\
\lambda^{\ast ST}_\ell&=-\left.\frac{Q^{\ast T}_L}{E^{\ast S}_L}\right|_{E^{\ast T}_L=0}=-\left.\frac{Q^{\ast S}_L}{E^{\ast T}_L}\right|_{E^{\ast S}_L=0}~.
\end{align}
\end{subequations}
\subsection{Numerical results: scalarized neutron stars}
For case studies, we calculated dipolar, quadrupolar and octupolar perturbations of NSs for a family of coupling functions that can yield scalarized configurations. In particular, depending on the theory parameters and EoS, noticeable deviations from GR may emerge in the tensor tidal deformability, as seen in  Figs.~\ref{fig:lambdaquadrupolar}(a) and \ref{fig:lambdaoctupolar}(a). The scalar tidal deformability, shown in Figs.~\ref{fig:lambdaScalarl1}, \ref{fig:lambdaquadrupolar}(b) and \ref{fig:lambdaoctupolar}(b) for the dipolar, quadrupolar and octupolar case, respectively, can have similar orders of magnitude as the tensor tidal deformability in the Einstein frame, although in the Jordan frame it scales with the square of the inverse of the cosmological value of the scalar field, as shown in Appendix~\ref{app:Plotslambda}. Regarding the scalar-tensor tidal deformability, shown in Figs.~\ref{fig:lambdaquadrupolar}(c) and \ref{fig:lambdaoctupolar}(c), it has negative values and similar order of magnitude as the scalar tidal deformability, also scaling with the inverse of the cosmological value of the scalar field in the Jordan frame.

\subsection{Comparison with previous literature}
Tidal deformabilities in scalar-tensor gravity have previously been computed in Refs.~\cite{Pani:2014jra,Brown:2022kbw}. Qualitatively, our conclusions for the features of the scalar and tensor Love numbers are similar, but a few details differ. These differences arise for the following reasons.  

In both~\cite{Pani:2014jra,Brown:2022kbw}, the mapping between quantities in the two frames is obtained by computing the transformation properties of the multipole moments and tidal fields as dictated by the transformation of the metric functions in relativistic perturbation theory. The differences to the mappings used here are that (i) as explained in Sec.~\ref{sec:EFT}, we include here the mixed scalar-tensor tidal deformability, which adds an additional contribution to the multipole moments, and (ii) we compute the transformations at the level of the effective action, where the tidal parameters appear as coupling coefficients, which are in turn directly related to GW observables. 

Another source of discrepancies between the results of \cite{Pani:2014jra,Brown:2022kbw} and those presented in Sec.~\ref{sec:CaseStudy} are calculational details. Specifically, Pani and Berti~\cite{Pani:2014jra} define the quadrupolar scalar tidal deformability as
\begin{align}
    {\lambda^S_{2}}^{\rm Pani-Berti}=-\frac{{Q_2^\ast}^S}{{E_0^\ast}^S}~,
\end{align}
with ${Q_2^\ast}^S$ the scalar quadrupole moment and ${E_0^\ast}^S$ the coefficient associated with the power $r^0$ in the asymptotic solution of the scalar field perturbation (see Sec.~\ref{ssec:ExtractingMult}). This is because they specialize to a tensor tidal field, setting the quadrupolar scalar tidal field ${E_2^\ast}^S$ to zero [which corresponds to $\delta\varphi^{(\ell=2)}=0$ in \eqref{eq: ScalarSolInf}]. 

The differences with the results of Brown~\cite{Brown:2022kbw} are a consequence of the fact that~\cite{Brown:2022kbw} sets to zero the source term responsible for the mixed, scalar-tensor tidal deformability. In particular, the functions $f_s$ and $g_s$ in \eqref{eq:H0PertEq} and \eqref{eq:PhiPertEq} are omitted therein. Comparing our results, which include these functions, against the results of~\cite{Brown:2022kbw} shows that omitting these terms leads to smaller values for the scalar and tensor tidal deformabilities, with differences of up to $18\%$ for $\beta=-4.5$ and up to $35\%$ for $\beta=-6$ for the scalar Love numbers and smaller differences for the tensor ones; see Appendix~\ref{app:ComparisonBrown} for a more detailed comparison.  

\section{Conclusions and Outlook}\label{sec:Conclusions}
In this paper, we studied the tidal deformability in scalar-tensor theories using an effective field theory approach connected with calculations based on relativistic perturbation theory. In addition to the tensor and scalar tidal deformabilities considered in the literature, our analysis revealed the need for a third kind of tidal deformability characterizing the tensor/scalar multipole moment induced by a scalar/tensor tidal field. This additional parameter introduces subtleties in the calculations due to couplings between tensor and scalar sectors. We developed a framework to decouple the different contributions and extract the three tidal deformabilities from detailed calculations of the perturbed neutron star and scalar field configurations. The mixed  scalar-tensor Love numbers have definite parity properties, which can be used to scale out the dependencies on asymptotic scalar characteristics. We performed most of the calculations in the Einstein frame, where the scalar field is minimally coupled to gravity, and derived the transformation properties of the tidal deformabilities to obtain results in the Jordan frame, where the theory is originally formulated. 

As an application of the method, we considered case studies in scalar-tensor theories with a choice of coupling function that can give rise to scalarized neutron stars. We demonstrated the feasibility of numerically extracting the various tidal deformability parameters for examples with different equations of state, scalar coupling strengths, and asymptotic values of the scalar field. For the examples considered, the tensor deformabilities become larger than the GR values for high-mass neutron stars and the scalar Love numbers are of the same order of magnitude as the tensor ones in the Einstein frame. Interestingly, the mixed scalar-tensor deformabilities are negative and also of the same order of magnitude as the others in the Einstein frame. 

Calculating the consequences of these tidal properties for GW signals is the subject of ongoing work \cite{STWaveformInPrep}. The general methodology developed here can also be applied for scalarized compact objects in other theories of gravity and can be extended to include the full dynamical tidal response~\cite{Creci:2021rkz,Ivanov:2022qqt}. This would allow studies from an EFT perspective of dynamical scalar tidal effects, like the monopolar dynamical scalarization~\cite{Palenzuela:2013hsa} that happens close to the critical point of scalarization. In this context, considering an expansion of the EFT around the marginally stable solutions (Sec.~\ref{ssec:StealthSol}) would be interesting, since it is expected to be of comparable accuracy as an expansion around the stable branch close to the critical point but does not introduce discontinuities into the model. Our work provides important inputs for future tests of GR with GWs and multimessenger observations and for understanding and assessing possible degeneracies with changes to the NS EoS or the presence of dark matter. 

\section*{Acknowledgments}
We thank Laura Bernard, Daniela Doneva and Stephanie Brown for insightful discussions, Frank Visser for providing the code for the piecewise polytropes and Dáire Scully for valuable comments on the early version of the manuscript. G.C. and T.H. acknowledge funding from the Nederlandse Organisatie voor Wetenschappelijk Onderzoek
(NWO) sectorplan. TH also acknowledges Cost Action CA18108. 
\clearpage
\appendix
\section{Frame transformations: Jordan to Einstein}\label{app:JordantoEinsteinAction}
\subsection{Action}
In this section, following \cite{Fujii_2003,Faraoni:2004pi}, we will rederive the steps necessary to go from the Jordan to the Einstein frame via a conformal transformation. We start with the scalar-tensor action in the Jordan frame,
\begin{align}\label{eq:STactionApp}
S_{\rm ST}=\int_{\mathcal{M}}d^4x \sqrt{-g}\left[K_R F(\phi) R-K_{\phi}\frac{\omega(\phi)}{\phi}\p^\mu\phi\p_\mu\phi\right]~,
\end{align}
where $\phi$ is a scalar field with self-coupling $\omega(\phi)$, coupled to the Ricci scalar $R$ via a scalar-field-dependent function $F(\phi)$, and $K_R$ and $K_{\phi}$ are the normalization constants of $R$ and $\phi$, respectively. We start by defining our (local) conformal transformation
\begin{align}
g^\ast_{\mu\nu}=\Omega(x)^2 g_{\mu\nu}~.
\end{align}
From here on we will write $\Omega(x)=\Omega$ for simplicity. Using the ansatz $g_\ast^{\mu\nu}=C g^{\mu\nu}$ and the invariance of the Kronecker delta ${\delta^\ast}_\mu^\nu=\delta_\mu^\nu$ yields $C=\Omega^{-2}$,
\begin{align}\label{eq:InverseConformalMetric}
g_\ast^{\mu\nu}=\frac{1}{\Omega^2} g^{\mu\nu}~.
\end{align}
Using that $\det(c \mathcal{M})=c^n \det(\mathcal{M})$ for an $n\times n$ matrix $\mathcal{M}$, it follows that
\begin{align}
g^\ast&=\Omega^8 g,\\
\sqrt{-g^\ast}&=\Omega^4\sqrt{-g}\label{eq:SqrtConformal}~,
\end{align}
in $n=4$ dimensions. The Christoffel symbol
\begin{align}
{\Gamma^{\mu}}_{\nu\lambda}=\frac{1}{2}g^{\mu\rho}\left(\p_\nu g_{\rho\lambda}+\p_\lambda g_{\rho\nu}-\p_\rho g_{\nu\lambda}\right)~,
\end{align}
changes as
\begin{align}\label{eq: ChristoffelToEinstein}
{\Gamma^{\mu}}_{\nu\lambda}={{\Gamma^\ast}^{\mu}}_{\nu\lambda}-\left(\delta^\mu_\lambda f_\nu+\delta^\mu_\nu f_\lambda-g^\ast_{\nu\lambda} f_\ast^\mu\right)~,
\end{align}
where
\begin{align}
f\equiv\log\Omega,\qquad f_\alpha\equiv\p_\alpha f,\qquad f_\ast^\alpha\equiv\p_\ast^\alpha f=g_\ast^{\alpha\beta}\p_\beta f~.
\end{align}
The Ricci scalar will therefore change according to
\begin{align}\label{eq:ConformalR}
R=\Omega^2\left(R_\ast+6\Box_\ast f-6 g_\ast^{\mu\nu}f_\mu f_\nu\right)~,
\end{align}
where $R_\ast$ is the Ricci scalar of the metric $g^\ast_{\mu\nu}$ and 
\begin{align}\label{eq:Boxf}
\Box_\ast f = \frac{1}{\sqrt{-g^\ast}}\p_\mu\left(\sqrt{-g^\ast}g_\ast^{\mu\nu}\p_\nu f\right)
\end{align}
is the Laplace-Beltrami operator associated with the metric $g^\ast_{\mu\nu}$. We will now split the Lagrangian in \eqref{eq:STactionApp} into two pieces,
\begin{align}
L_R&=\sqrt{-g}K_R F(\phi)R,\\
L_\phi&=\sqrt{-g}K_{\phi}\frac{\omega(\phi)}{\phi}\p^\mu\phi\p_\mu\phi~.
\end{align}
We will start with the first piece. Inverting \eqref{eq:InverseConformalMetric} and using \eqref{eq:SqrtConformal} and \eqref{eq:ConformalR} yields
\begin{align}\label{eq:LagrangianR}
L_R=\sqrt{-g^\ast} F(\phi) \Omega^{-2} K_R \left[R_\ast+6\Box_\ast f - 6 g_\ast^{\mu\nu}f_\mu f_\nu\right]~.
\end{align}
If we want to obtain a minimally coupled Lagrangian we must choose
\begin{align}\label{eq:ConformalChoice}
F(\phi)\Omega^{-2}=1~,
\end{align}
such that we recover the Einstein-Hilbert term\footnote{Notice that, although cumbersome, if we wish to have a different coefficient for the Ricci-scalar term in the Einstein frame, the right hand side should read $K_{R_\ast}/K_R$, with $K_{R_\ast}$ the new normalisation of the Ricci scalar in the minimally coupled action.}. The second term, \eqref{eq:Boxf}, is a boundary term and therefore will not contribute to the Lagrangian. Now, given that we made the choice $\eqref{eq:ConformalChoice}$, we have
\begin{align}
f_\alpha=\p_{\alpha} f=\frac{\p_\alpha\Omega}{\Omega}=\frac{1}{2}\frac{\p_\alpha F}{F}=\frac{1}{2}\frac{F'}{F}\p_\alpha\phi~,
\end{align}
where $F'\equiv dF/d\phi$, and therefore the third term in \eqref{eq:LagrangianR} reads
\begin{align}
6 g_\ast^{\mu\nu}f_\mu f_\nu&=6 g_\ast^{\mu\nu}\frac{1}{4}\left(\frac{F'}{F}\right)^2\p_\mu\phi\p_\nu\phi\nonumber\\&=\frac{3}{2}\left(\frac{F'}{F}\right)^2g_\ast^{\mu\nu}\p_\mu\phi\p_\nu\phi~.
\end{align}
We now focus on the second term,
\begin{align}
L_\phi&=\sqrt{-g}K_{\phi}\frac{\omega(\phi)}{\phi}g^{\mu\nu}\p_\mu\phi\p_\nu\phi\nonumber\\&=(\sqrt{-g^\ast}\Omega^{-4})K_{\phi}\frac{\omega(\phi)}{\phi}(\Omega^2g_\ast^{\mu\nu})\p_\mu\phi\p_\nu\phi\\&
=\sqrt{-g^\ast}F^{-1}K_{\phi}\frac{\omega(\phi)}{\phi}g_\ast^{\mu\nu}\p_\mu\phi\p_\nu\phi~.
\end{align}
Adding the two pieces together we obtain
\begin{align}
L_{ST}=&\sqrt{-g^\ast}\Bigg\{K_R R_\ast\nonumber\\&- \left[\frac{3 K_R}{2}\left(\frac{F'}{F}\right)^2+K_\phi\frac{\omega(\phi)}{\phi F}\right]g_\ast^{\mu\nu}\p_\mu\phi\p_\nu\phi\Bigg\}~.
\end{align}
Introducing a new field by
\begin{align}
	&\frac{d\varphi}{d\phi}=\sqrt{\Delta}~,\label{eq:ConformalField}\\
	&\p_\alpha\phi=\frac{1}{\sqrt{\Delta}}\p_\alpha\varphi~,\label{eq:ConformalPartial}	
\end{align}
with
\begin{align}
\Delta\equiv\frac{3}{2}\frac{K_R}{K_{\varphi}}\left(\frac{F'}{F}\right)^2+\frac{K_\phi}{K_{\varphi}}\frac{\omega(\phi)}{\phi F}~,
\end{align}
yields the Einstein frame Lagrangian
\begin{align}
L_{ST}=\sqrt{-g^\ast}\left\{K_R R_\ast- K_{\varphi} g_\ast^{\mu\nu}\p_\mu\varphi\p_\nu\varphi\right\}~.
\end{align}
We can now define a new coupling $A(\varphi)$ by $A=\Omega^{-1}$. The reason for this particular choice is that, if one adds a matter action, the matter Lagrangian in the Einstein frame will contain a metric $A(\varphi)^2g^\ast_{\mu\nu}$, and therefore can be seen as a coupling of the scalar field to matter. Hence, with this new parameter we have
\begin{align}\label{eq: MetricConformalTransfA}
g_{\mu\nu}=A^2 g^\ast_{\mu\nu}~.
\end{align}
We can relate $A$ to the new field using \eqref{eq:ConformalChoice},
\begin{align}\label{eq: AtoOmegatoF}
A=\Omega^{-1}=F^{-1/2}&=\exp\left(-\frac{1}{2}\log F\right)\nonumber\\&=\exp\left(-\int dF \frac{1}{2F}\right)~,
\end{align}
and \eqref{eq:ConformalField}
\begin{align}
\frac{d\varphi}{dF}F'=\sqrt{\Delta}~,
\end{align}
to obtain
\begin{align}
A(\varphi)=\exp\left(-\int d\varphi \frac{F'}{2F\sqrt{\Delta}}\right)~.
\end{align}
Analogously, we define
\begin{align}\label{eq:STalphaApp}
\alpha(\varphi)=-\frac{1}{A}\frac{dA}{d\varphi}=\frac{F'}{2F\sqrt{\Delta}}~,
\end{align}
where all the quantities are understood as a function of the new field $\varphi$. 

In order to transform the skeletonized/EFT action, it is useful to transform the following quantities
\begin{subequations}\label{eq:DifferentialTransf}
\begin{align}
d\sigma^2&=-ds^2=-\frac{1}{\Omega^2}ds_\ast^2=\frac{1}{\Omega^2} d\sigma_\ast^2~,\\
u^\mu&=\frac{dx^\mu}{d\sigma}=\Omega\frac{dx^\mu}{d\sigma^\ast}=\Omega~ u_\ast^\mu~,\\
u_\mu&=g_{\mu\nu}u^\nu=\frac{1}{\Omega^2}g^\ast_{\mu\nu}\Omega~u_\ast^\nu=\frac{1}{\Omega}u^\ast_\mu~,\\
u_\mu u^\mu&=u^\ast_\mu u_\ast^\mu~.
\end{align}
\end{subequations}
\subsection{Covariant derivatives}
The covariant derivatives acting on the scalar and tensor fields will yield higher-order contributions in the EFT. In this section, we will show that this is the case by analyzing the transformation of the covariant derivatives. Then, as an explicit example, we will fix the multipolar order for the scalar and tensor cases and show the explicit terms yielding higher-order contributions. Additionally, we show a recurrence formula to transform any number of covariant derivatives between Jordan and Einstein frames. An important point to notice is that the quantities appearing in the EFT are symmetric and trace-free (STF). In this section, we do not project the covariant derivatives onto their STF part, but that does not change the rationale of the calculations.

We start with the transformation of the covariant derivative. Using 
\begin{align}\label{eq: fToAlpha}
f_\mu=\frac{1}{2}\frac{F'}{F}\p_\mu\phi=\frac{1}{2}\frac{F'}{F}\frac{1}{\sqrt{\Delta}}\p_\mu\varphi=\alpha\p_\mu\varphi=\alpha E^S_\mu~,
\end{align}
we can express \eqref{eq: ChristoffelToEinstein} in terms of the scalar dipolar tidal field as
\begin{align}
{\Gamma^{\mu}}_{\nu\lambda}={{\Gamma^\ast}^{\mu}}_{\nu\lambda}-\alpha\left(2\delta^\mu_{(\lambda} E^S_{\nu)}-g^\ast_{\nu\lambda}{E_S}^\mu\right)~,
\end{align}
where indices between parentheses are symmetrized. Therefore, when transforming a covariant derivative we will have, schematically,
\begin{align}
\nabla\rightarrow\nabla^\ast\pm\alpha\left(\delta~E^S- g E_S\right)~.
\end{align}
As the tidal action involves terms quadratic in covariant derivatives we use 
\begin{align}
    (\nabla E)^2\rightarrow(\nabla^\ast E)^2+\alpha^2 c_{EE} E^2{E_S}^2\pm\alpha c_{EE_S}(\nabla^\ast E)EE_S~,
\end{align}
with $c_{EE}$ and $c_{E E_S}$ coefficients containing Dirac deltas and metrics, and where $E$ is a generic, i.e. scalar or tensor, tidal field. Given that each tidal field $E_L$ scales as powers of $1/r$, the terms proportional to $\alpha$ will always be of higher order than the first term at large separations, and are therefore suppressed in the EFT. In the following subsections we will see how this is the case explicitly for both the scalar and tensor tidal fields.
\subsubsection{Scalar field}
We start with the scalar field. Since for $\ell=1$ the covariant derivative reduces to a partial derivative we will start with the case $\ell=2$,
\begin{align}
    \nabla_{\mu\nu}\phi=\nabla_\mu\p_\nu\phi=\p_\mu\p_\nu\phi-\Gamma^\gamma_{\mu\nu}\p_\gamma\phi~.
\end{align}
Using \eqref{eq: ChristoffelToEinstein} and \eqref{eq:ConformalPartial}, the second term reads
\begin{align}
\Gamma^\gamma_{\mu\nu}\p_\gamma\phi=\frac{1}{\sqrt{\Delta}}\left[{\Gamma^\ast}^\gamma_{\mu\nu}-\left(2\delta^\gamma_{(\mu} f^{}_{\nu)}-g^\ast_{\mu\nu} f_\ast^\gamma\right)\right]\p_\gamma\varphi~.
\end{align}
Additionally, using \eqref{eq: fToAlpha} we obtain
\begin{align}
\Gamma^\gamma_{\mu\nu}\p_\gamma\phi=\frac{1}{\sqrt{\Delta}}\left[{\Gamma^\ast}^\gamma_{\mu\nu}-\alpha\left(2\delta^\gamma_{(\mu} \p^{}_{\nu)}\varphi-g^\ast_{\mu\nu}g_\ast^{\gamma\kappa}\p_\kappa\varphi\right)\right]\p_\gamma\varphi~.
\end{align}
On the other hand, the partial derivative will transform as follows
\begin{align}
\p_{\mu\nu}\phi&=\frac{1}{\sqrt{\Delta}}\p_{\mu\nu}\varphi+\p_\mu\varphi\p_\nu\left(\frac{1}{\sqrt{\Delta}}\right)\nonumber\\&=\frac{1}{\sqrt{\Delta}}\left[\p_{\mu\nu}\varphi-\frac{\Delta'}{2\Delta^{3/2}}\p_\mu\varphi\p_\nu\varphi\right]~,
\end{align}
where we have used that
\begin{align}
\p_\mu\left(\frac{1}{\sqrt{\Delta}}\right)=\p_\mu\phi\frac{d}{d\phi}\left(\frac{1}{\sqrt{\Delta}}\right)=-\frac{\Delta'}{2\Delta^{3/2}\sqrt{\Delta}}\p_\mu\varphi~.
\end{align}
Putting all together, the $\ell=2$ covariant derivative acting on the scalar field will transform as
\begin{align}
\nabla_{\mu\nu}\phi=\frac{1}{\sqrt{\Delta}}\Bigg[&\nabla^\ast_{\mu\nu}\varphi+\alpha\left(2\delta^\gamma_{(\mu} \p^{}_{\nu)}\varphi-g^\ast_{\mu\nu}g_\ast^{\gamma\kappa}\p_\kappa\varphi\right)\p_\gamma\varphi\nonumber\\&-\frac{\Delta'}{2\Delta^{3/2}}\p_\mu\varphi\p_\nu\varphi\Bigg]~.
\end{align}
However, given that the second and third terms are proportional to ${E_S}^2=(\p\varphi)^2$, they will yield higher-order terms in the action. This is because the tidal action will read, schematically
\begin{align}
(\nabla\phi)^2=&\frac{1}{\Delta}\Big[(\nabla^\ast\varphi)^2+c_{E E_S}~(\nabla^\ast\varphi)^{\mu\nu}E^S_{\mu}E^S_{\nu}\nonumber\\&+c_{E_SE_S}~E^S_{\mu}E^S_{\nu}E_S^{\mu}E_S^{\nu}\Big]\nonumber\\=&\frac{1}{\Delta}(\nabla^\ast\varphi)^2+\mathcal{O}\left({E_S}^3\right)~,
\end{align}
with $c_{E E_S}$ and $c_{E_SE_S}$ coefficients containing factors of $\alpha$ and $\Delta$. Therefore, given that we only have to consider the first term, for generic multipolar order we will have
\begin{align}\label{eq:TidalScalarTransf}
\nabla_L\phi=\frac{1}{\sqrt{\Delta}}\nabla^\ast_L\varphi+\dots~,
\end{align}
where "$\dots$" denotes high-order terms in the EFT that are therefore omitted in the Einstein frame.
\subsubsection{Tensor field}
For the tensor (gravitational) tidal field, we will start with the case $\ell=2$,
\begin{align}
E_{\mu\nu}=\frac{1}{z^2}C_{\mu\alpha\nu\beta}u^\alpha u^\beta~,
\end{align}
with $z=\sqrt{-u_\mu u^\mu}$ and   $C_{\mu\alpha\nu\beta}$ the Weyl tensor. Using that the (3,1) Weyl tensor is invariant under conformal transformations,
\begin{align}
C_{\mu\alpha\nu\beta}=g_{\beta\gamma}C^{~}_{\mu\alpha\nu}{}^{\gamma}=\frac{1}{\Omega^2}g^\ast_{\beta\gamma}C^\ast_{\mu\alpha\nu}{}^\gamma=\frac{1}{\Omega^2}C^\ast_{\mu\alpha\nu\beta}
\end{align}
we obtain
\begin{align}\label{eq:TensorTidaltoEinstein}
E_{\mu\nu}=\frac{1}{\Omega^2 z_\ast^2}C^\ast_{\mu\alpha\nu\beta}\Omega^2u_\ast^\alpha u_\ast^\beta=E^\ast_{\mu\nu}~.
\end{align}
Therefore, the $\ell=2$ tidal tensor is invariant under conformal transformations. We now consider the case $\ell=3$,
\begin{align}
E_{\gamma\mu\nu}=\nabla_{\gamma}E_{\mu\nu}~.
\end{align}
Similar to the scalar case, the covariant derivative reads
\begin{align}
\nabla_{\gamma}E_{\mu\nu}=\p_\gamma E_{\mu\nu}-{\Gamma^\delta}_{\gamma\mu}E_{\delta\nu}-{\Gamma^\omega}_{\gamma\nu}E_{\omega\mu}
\end{align}
Using \eqref{eq: ChristoffelToEinstein} and \eqref{eq:TensorTidaltoEinstein} we obtain
\begin{align}
\nabla_{\gamma}E_{\mu\nu}=&\nabla^\ast_{\gamma}E^\ast_{\mu\nu}\nonumber\\&+3\alpha\Big[E^\ast_
{(\alpha\beta}E^S_{\gamma)}-g^\ast_{(\alpha\beta}E^\ast_{\gamma)\xi}g_\ast^{\xi\delta}E^S_\delta\Big]\nonumber\\&+\alpha\Big[E^\ast_
{\alpha\beta}E^S_{\gamma}+g^\ast_{\alpha\beta}E^\ast_{\gamma\xi}g_\ast^{\xi\delta}E^S_\delta\Big]~.
\end{align}
In analogy with the scalar case above, all terms except the first one will yield higher-order contributions in the action that are suppressed in the EFT,
\begin{align}\label{eq:TidalTensorTransf}
E_L=E_L^\ast+\dots~,
\end{align}
where, as above, "$\dots$" denotes high-order terms suppressed in the EFT.
\subsubsection{Recurrence relation for generic $\ell$}
In general, for any multipolar order $L$, we can find a recurrence relation for computing the $L$th covariant derivative,
\begin{align}
\nabla_L E_{\alpha\beta}=&\nabla^\ast_\mu E_{L-1 \alpha\beta}+\ell\alpha\nonumber\\&\times\left[E_{(L-1 \alpha \beta}\p_{\mu)}\varphi-\frac{(\ell-1)}{2}g^\ast_{(\alpha\beta}E_{\mu L-1)\xi}g_\ast^{\xi\kappa}\p_\kappa\varphi\right]\nonumber\\&+\alpha\left(\ell-2\right)\Big[E_{L-1 \alpha \beta}\p_{\mu}\varphi\nonumber\\&+\frac{(\ell-1)}{2}g^\ast_{(\alpha\beta}E_{L-1)\mu\xi}g_\ast^{\xi\kappa}\p_\kappa\varphi\Big]~,
\end{align}
 where $\nabla_\mu E_{-1 \alpha\beta}=E_{\alpha\beta}$ and $E_{-1\alpha\beta}=0$. This expression is valid for any symmetric tensor $E_{\alpha\beta}$ and can be used for the scalar tidal field as well, given that $E^S_{-1\alpha\beta}=E^S_{\alpha}$. Note that the first term does not contain an asterisk. This would be the case for $\ell=0,1$ since $E_{\mu\nu}=E^\ast_{\mu\nu}$. For the other cases one has to substitute the expression for the $(\ell-1)$th tidal field. This will yield derivatives of the second term, which will eventually be expressed as a combination of the $\ell$th tidal field in the Einstein frame and its derivatives. With this generic expression we can also reason, similar to all the cases above, that 
\begin{align}
    E_L=E^\ast_L+\dots
\end{align}
where "$\dots$" denote terms that give higher-order contributions to the EFT. This recurrence relation can be used in order to speed up the computation of any number of covariant derivatives using e.g. \texttt{Mathematica}.

\subsection{Tidal deformabilities}\label{app:TidalDefTransf}
We start with the tidal action in the Jordan frame given in~\eqref{eq:TidalactionJordan}. In the full theory describing an isolated body, we use coordinates such that the background spacetime is asymptotically flat. Therefore, we will adapt these coordinates in the EFT for consistency. As explained in Appendix~\ref{app: ADMMass}, demanding an asymptotically flat spacetime in the Jordan frame requires a rescaling of the coordinates such that
\begin{align}
\label{eq:rescaledcoords}
d\tilde{x}^\mu=A(\varphi_\infty)dx^\mu~,
\end{align}
which implies
\begin{align}
\nabla_L\rightarrow{A(\varphi_\infty)^{\ell}\tilde{\nabla}_L}~.
\end{align}
For the rest of this appendix we will adopt these coordinates and drop the tilde. Transforming the line element, four-velocities, and tidal fields using \eqref{eq:DifferentialTransf}, \eqref{eq:TidalScalarTransf}, and \eqref{eq:TidalTensorTransf}, we obtain
\begin{align}
 S_{\rm tidal}=&\sum_{\ell} \int d\sigma^\ast~z^\ast~\, g_\ast^{LP}({A/A_\infty})^{1-2\ell}\times\nonumber\\&\left(\frac{\lambda^T_\ell}{2\ell!}E_L^{\ast T}E_P^{\ast T}+\frac{\lambda^S_\ell}{2\ell!\Delta}E_L^{\ast S}E_P^{\ast S}+\frac{\lambda^{ST}_\ell}{\ell!\sqrt{\Delta}}E_L^{\ast T}E_P^{\ast S}\right)~,
\end{align}
where 
\be
\label{eq:Ainftydef}
A_\infty\equiv{A(\varphi_\infty)},
\ee
and we use a similar notation for any other function of $\varphi$ evaluated at infinity. 
From~\eqref{eq:gLP} with~\eqref{eq:MetricConformalTransf} we have
\begin{align}\label{eq: gLPtogstarLP}
g^{L P}=\prod_{n=1}^{\ell} g^{l_n p_n}=A^{-2\ell}\prod_{n=1}^{\ell} g_{\ast}^{l_n p_n}=A^{-2\ell}g_\ast^{L P}~.
\end{align}
Comparing with the tidal action in the Einstein frame in~\eqref{eq:TidalactionEinstein} we read off that the coefficients of the bilinears in the tidal fields are related by
 \begin{align}
    \lambda_\ell^T&=\bar{A}^{2\ell-1}\lambda_\ell^{\ast T}~,\\
    \lambda_\ell^S&=\bar{A}^{2\ell-1}\Delta\lambda_\ell^{\ast S}~,\\\lambda_\ell^{ST}&=\bar{A}^{2\ell-1}\sqrt{\Delta}\lambda_\ell^{\ast ST}~,
 \end{align}
with $\bar{A}=A/A_\infty$ and
\begin{align}
\Delta\equiv\frac{3}{2}\frac{K_R}{K_{\varphi}}\left(\frac{F'}{F}\right)^2+\frac{K_\phi}{K_{\varphi}}\frac{\omega(\phi)}{\phi F}~.
\end{align}
We can also rewrite these expressions in terms of $\alpha$. Using \eqref{eq:STalpha} and assuming $\alpha\neq0$,
\begin{align}
\Delta=\left(\frac{F'}{2F\alpha}\right)^2=\left(\frac{A^2F'}{2\alpha}\right)^2~,
\end{align}
and therefore
\begin{align}
    \lambda_\ell^T&=\bar{A}^{2\ell-1}\lambda_\ell^{\ast T}~,\\
\lambda_\ell^S&=\frac{\bar{A}^{2\ell-1}A^4F'^2}{4\alpha^2}\lambda_\ell^{\ast S}~,\\
    \lambda_\ell^{ST}&=\frac{\bar{A}^{2\ell-1}A^2F'}{2\alpha}\lambda_\ell^{\ast ST}~,
 \end{align}
where $F'$ can be expressed in terms of $A(\varphi)$ for a specific $F(\phi)$. Note that coefficients in front of $\lambda_\ell^\ast$ have to be evaluated at infinity. This is because the tidal and multipole moments or, equivalently, the tidal deformability are extracted at infinity \eqref{eq:EFTEoMEinstein}. This leads to the relations
\begin{subequations}
\label{eq:lambdatransformApp}
\begin{align}
\lambda_\ell^T&=\lambda_\ell^{\ast T}~,\label{eq:ScalarTidalDefTransfApp}\\
\lambda_\ell^S&=\left(\frac{A_\infty^{2}{F'}_\infty}{2\alpha_\infty}\right)^2\lambda_\ell^{\ast S}~,\\
\lambda_\ell^{ST}&=\frac{A_\infty^{2}F'_\infty}{2\alpha_\infty}\lambda_\ell^{\ast ST}~.
\end{align}
\end{subequations}

\subsubsection{Explicit example}
To give an explicit example of the application of the transformations~\eqref{eq:lambdatransformApp} we consider the coupling function $F(\phi)=\phi^n$. 
In this case we have 
\begin{align}
F'=n \phi^{n-1}=n F^{\frac{n-1}{n}}=n A^{-2\frac{n-1}{n}}~,
\end{align}
and substituting into~\eqref{eq:lambdatransformApp} leads to the result for general choices of $n$. For the case $n=1$, corresponding to Jordan-Brans-Dicke gravity, and additionally choosing $A(\varphi_\infty)=e^{\frac{1}{2}\beta\varphi_\infty^2}$, and hence $\alpha(\varphi_\infty)=-\beta\varphi_\infty$, relevant for spontaneous scalarization, we obtain
\begin{align}
\lambda_\ell^T=&\lambda_\ell^{\ast T}~,\label{eq:LambdaTEinsteinLambdaTJordanApp}\\  
\lambda_\ell^S=&\frac{e^{2\beta\varphi_\infty^2}}{4\beta^2\varphi_\infty^2}\lambda_\ell^{\ast S}~,\\
\lambda_\ell^{ST}=&-\frac{e^{\beta\varphi_\infty^2}}{2\beta\varphi_\infty}\lambda_\ell^{\ast ST}~.
\end{align}

\section{Just coordinate system}\label{app:JustCoordApp}
In this appendix we review the Just coordinate system and provide some explicit derivations missing in the literature. This coordinate system was introduced by Kurt Just \cite{Just+1959+751+751} and later used by Damour and Esposito-Farèse \cite{Damour_1992,Damour:1996ke,Damour:1998jk}. The Just coordinate system is useful because it provides a closed-form solution for the background scalar field. This solution is given in terms of two constants, which correspond to physical quantities such as the mass or the charge. We will derive the vacuum TOV equations and explicitly perform the matching to the surface of the star, which will allow us to compute the value of the scalar field at infinity and its scalar charge. 
\subsection{Metric functions and scalar field}
The metric in the Just coordinate system reads
\begin{align}
ds^2=-e^{\nu} dt^2 + e^{-\nu} d\rho^2 + e^{\mu-\nu}d\Omega^2~.
\end{align}
In vacuum, the Einstein Field equations read
\begin{subequations}
\begin{align}
G_{tt}&=\mu''-\mu'\nu'-e^{-\mu}-\nu''+\frac{{\nu'}^2+3{\mu'}^2}{4}+{\varphi_0'}^2=0~,\\
G_{\rho\rho}&=e^{-\mu}+\frac{{\nu'}^2-{\mu'}^2}{4}+{\varphi_0'}^2=0,\\
G_{\theta\theta}&=2\mu''+{\mu'}^2+{\nu'}^2+4{\varphi_0'}^2=0~.\label{eq: JustTheta}
\end{align}
\end{subequations}
We can now take combinations of the different components in order to solve for the metric functions. In particular,
\begin{subequations}
\begin{align}
\frac{G_{\theta\theta}}{2}-G_{tt}-G_{\rho\rho}=\nu''+\mu'\nu'=0~,\label{eq: Justnueq}\\
\frac{G_{\theta\theta}}{2}-2G_{\rho\rho}=-2e^{-\mu}+{\mu'}^2+\mu''=0~.\label{eq: Justgammaeq}
\end{align}
\end{subequations}
From \eqref{eq: Justgammaeq} we obtain
\begin{align}
\mu=\log\left[-\frac{c_1}{4}+(c_2+\rho)^2\right]=\log\left[\rho^2\left(1-\frac{a}{\rho}\right)\right]~,\label{eq: Justgammasol}
\end{align}
where in the last step we choose the integration constant $c_1=4{c_2}^2$ and redefine $c_2=-a/2$. Plugging this solution into \eqref{eq: Justnueq} and solving for $\nu$ yields
\begin{align}\label{eq: Justnusol}
\nu=k_1+\frac{k_2}{a}\log\left(1-\frac{a}{\rho}\right)=\frac{b}{a}\log\left(1-\frac{a}{\rho}\right)~,
\end{align}
where we have redefined $k_2=b$, and set $k_1=0$ by demanding asymptotic flatness.

Now that we solved for the metric functions we can solve the background scalar field. Its vacuum equation of motion in the Just coordinate system reads
\begin{align}
\varphi_0''+\mu'\varphi_0'=e^{-\mu}\left(\varphi_0'e^{\mu}\right)'=0~.
\end{align}
Substituting \eqref{eq: Justgammasol} and integrating yields
\begin{align}\label{eq: JustScalarField}
\varphi_0=b_1+\frac{b_2}{a}\log\left(1-\frac{a}{\rho}\right)={\varphi_{0\infty}}+\frac{d}{a}\log\left(1-\frac{a}{\rho}\right)~,
\end{align}
where we have redefined $b_2=d$ and $b_1={\varphi_{0\infty}}$ is the value of the scalar field background at infinity. Finally, we can substitute all the solutions into \eqref{eq: JustTheta} and find that the constants obey
\begin{align}\label{eq: JustConstants}
    a^2=b^2+4d^2~.
\end{align}
\subsection{Relating the constants to physical quantities}
In order to relate the constants to some known physical quantities we can start by relating the Just coordinate system to the standard Schwarzschild coordinates,
\begin{align}
ds^2=-e^\nu dt^2 + e^\gamma dr^2 + r^2 d\Omega^2~.
\end{align}
By comparing the metric functions we obtain the following relations:
\begin{subequations}
\begin{align}
r^2&=e^{\mu-\nu}=\rho^2\left(1-\frac{a}{\rho}\right)^{1-b/a}~,\label{eq: JustRadialRelation}\\
e^\gamma&=e^{-\nu}\left(\frac{d\rho}{dr}\right)^2=\left(1-\frac{a}{\rho}\right)\left(1-\frac{a+b}{2\rho}\right)^{-2}\label{eq: JustMassMetric}~.
\end{align}  
\end{subequations}
Given that, at infinity,
\begin{align*}
    e^\gamma&=1+\frac{2M}{r}+\mathcal{O}\left(\frac{1}{r^2}\right)~,\\
    \left(1-\frac{a}{\rho}\right)\left(1-\frac{a+b}{2\rho}\right)^{-2}&=1+\frac{b}{\rho}+\mathcal{O}\left(\frac{1}{\rho^2}\right)~,\\
    r&=\rho+\mathcal{O}\left(\rho^{1/2},\frac{1}{\rho}\right)~,
\end{align*}
it follows that $b=2M$, with $M$ the point-particle or ADM mass. Similarly, comparing the scalar field at infinity
\begin{align*}
\varphi_0(r)&=\varphi_{0\infty}-\frac{q M}{r}+\mathcal{O}\left(\frac{1}{r^2}\right)~,\\
\varphi_0(\rho)&=\varphi_{0\infty}-\frac{d}{\rho}+\mathcal{O}\left(\frac{1}{\rho^2}\right)~,
\end{align*}
implies $d=q M=q b/2$, where $q=-Q/M$ is (minus) the scalar charge $Q$ per unit mass. Using \eqref{eq: JustConstants} we obtain $a=b\sqrt{1+q^2}$. To summarize, the relations between the constants and the physical quantities are
\begin{subequations}
\begin{align}
b&=2M~,\\
\frac{d}{b}&=\frac{q}{2}~,\label{eq: JustConstAlpha}\\
\frac{a}{b}&=\sqrt{1+q^2}~.
\end{align}
\end{subequations}
\subsection{Obtaining the constants by matching at the surface.}
We can use the relation between the constants and the physical quantities, together with the relations between the coordinate systems, in order to extract the scalar charge per unit mass, the value of the scalar field at infinity, and the ADM mass from the metric components evaluated at the surface of the star. 

We start with the charge. Using \eqref{eq: Justnusol} we can write \eqref{eq: JustScalarField} as
\begin{align*}
\varphi(\rho)=\varphi_{0\infty}+\frac{d}{b}\nu(\rho)~.
\end{align*}
Taking a derivative and using \eqref{eq: JustConstAlpha} yields
\begin{align}
    q=\frac{2 \varphi_0'(\rho)}{\nu'(\rho)}=\frac{2 \varphi_0'(r)}{\nu'(r)}~,
\end{align}
where we have changed coordinates in the last step. In order to compute the ADM mass in terms of surface quantities we first have to relate the metric function $\nu$ and the surface radius in the two coordinate systems. In order to obtain the relation between the radial coordinate at the surface we can take a derivative of $\nu$ with respect to the radial coordinate $r$. For that, it may be useful to rewrite $\nu$ using \eqref{eq: JustRadialRelation},
\begin{align*}
\nu(\rho)=\frac{2b}{a-b}\log\left(\frac{r}{\rho}\right)~.
\end{align*}
Taking an $r$ derivative and evaluating at the surface $r=R$ yields
\begin{align*}
\nu'_S=\frac{2b}{R(2\rho_S-a-b)}~,
\end{align*}
with $\nu'\equiv d\nu/dr$ and the subscript $S$ denotes the quantity evaluated at the surface. Hence, the relation between radial coordinates at the surface reads
\begin{align}\label{eq: JustSurfaceRadial}
    \rho_S=\frac{a+b}{2}+\frac{b}{R \nu'_S}~.
\end{align}
We can now use this equation in order to compute the relation between $\nu$ evaluated at the surface in the two coordinate systems,
\begin{align}
\nu_S&=\nu(\rho_S)=\frac{b}{a}\log\left(1-\frac{a}{\rho_S}\right)=-\frac{b}{a}\log\left(\frac{1+x}{1-x}\right)\nonumber\\&=-\frac{2b}{a}{\text{arctanh}}(x)=-\frac{2b}{a}{\text{arctanh}}\left(\frac{a/b}{1+\frac{2}{R\nu'_S}}\right)~,
\end{align}
where in the second equality we have rewritten the argument in the logarithm in terms of $x=\frac{a R \nu'_S}{b(2+R\nu'_S)}$ and used a trigonometric identity in the third equality. Next, we use equation \eqref{eq: JustMassMetric} with [see \eqref{eq: BackgroundGamma}]
\begin{align*}
    e^{\gamma(r)}=\left(1-\frac{2m(r)}{r}\right)^{-1}~
\end{align*}
evaluated at the surface in order to derive
\begin{align*}
    a^2=\frac{b}{Q_2 R^2 {\nu'_S}^2}\left[b Q_2 (R\nu'_S-2)^2-4 e^{\nu_S/2}R^2\nu'_S\right]~,
\end{align*}
where, following \cite{Damour:1998jk}, we define $Q_2=\sqrt{1-2m_S/R}$, with $m_S$ the mass at the surface of the star.
Substituting the expression for $a$ above into \eqref{eq: JustRadialRelation} evaluated at the surface
\begin{align*}
R=e^{-\nu_S/2}\rho_S\left(1-\frac{a}{\rho_S}\right)^{1/2}~,
\end{align*}
together with \eqref{eq: JustSurfaceRadial}, yields
 \begin{align}
     b=2M=e^{\nu_S/2}Q_2R^2\nu'_S~,
\end{align}    
 and hence,
 \begin{align}
 d&=q\frac{b}{2}=\frac{1}{2}e^{\nu_S/2}Q_2R^2\varphi_0'(R)~,\\
a&=b\sqrt{1+q^2}=e^{\nu_S/2}Q_2R^2\nu'_S\sqrt{1+\frac{4\varphi_0'(R)^2}{\nu'_S{}^2}}~.
 \end{align}

To the best of our knowledge the relations between $d$ and $a$ and the star's quantities at the surface have not yet appeared in the literature. We can use these results to relate the value of the scalar field at infinity with its value at the surface,
\begin{align}
\varphi_{0\infty}=\varphi_0(R)-\frac{\varphi_0'(R)}{\nu'_S}\nu_S~.
\end{align}
To sum up, the charge per unit mass, the scalar field at infinity and the ADM mass are related to the surface quantities in the Schwarzschild coordinates by
\begin{align}
q&=\frac{2 \varphi_0'(R)}{\nu'_S}~,\\
\varphi_{0\infty}&=\varphi_0(R)-\frac{\varphi_0'(R)}{\nu'_S}\nu_S~,\\
M&=\frac{1}{2}e^{\nu_S/2}Q_2R^2\nu'_S~,\label{eq: ADMMassEinstein}
\end{align}
with 
\begin{align}
\nu_S=-\frac{2}{\sqrt{1+q^2}}{\text{arctanh}}\left(\frac{\sqrt{1+q^2}}{1+\frac{2}{R\nu'_S}}\right)~.
\end{align}
\section{Relating physical quantities between frames}\label{app: ADMMass}
In this appendix, we rederive the transformations of the ADM mass and charge between Jordan and Einstein frames presented in \cite{Pani:2014jra,Damour_1992}. From \cite{Fierz:1956zz,Jordan1959ZumGS,PhysRev.124.925,1970ApJ...161.1059N,Zhang:2023nil}, and assuming the common normalization $K_\phi=K_R$, the $\phi$-dependent gravitational constant measured in experiments $\tilde{G}(\phi_{\infty})$, such as those by Cavendish, is related to the gravitational constant $G$ appearing in the normalisation coefficients by
\begin{align}
    \tilde{G}(\phi_\infty)=\frac{1}{F(\phi_\infty)}\left(\frac{4 F'(\phi_\infty)^2+2\frac{F(\phi_\infty)}{\phi_\infty}\omega_\infty}{3 F'(\phi_\infty)^2+2 \frac{F(\phi_\infty)}{\phi_\infty}\omega_\infty}\right)G~,
\end{align}
where the subscript $\infty$ means evaluation at infinity. Using \eqref{eq: AtoOmegatoF} and \eqref{eq:STalpha} we can rewrite it as 
\begin{align}
    \tilde{G}(\varphi_\infty)=A(\varphi_\infty)^2\left(1+\frac{2K_R}{K_\varphi}\alpha_\infty^2\right)G~,
\end{align}
hence generalizing the different normalizations in the literature. With the normalizations and coupling chosen in Sec. \ref{sec: ComputingLoveNumbers} we have
\begin{align}
\tilde{G}(\varphi_\infty)=e^{\beta \varphi_\infty^2}
\left(1+\beta^2\varphi_\infty^2\right)G~.
\end{align}
\subsection{ADM mass}
In order to relate the ADM mass between frames we can compare the ${r-r}$ component of the metrics at infinity. However, from \eqref{eq: MetricConformalTransfA}, we see that, at infinity
\begin{align}
ds^2=A^2(\varphi_\infty)\eta^\ast_{\mu\nu}dx^\mu dx^\nu~,
\end{align}
given that we demand asymptotic flatness in the Einstein frame (see Section \ref{sec: ComputingLoveNumbers}).
Therefore, in order to obtain a Minkowski spacetime at infinity in the Jordan frame we must rescale our Jordan frame coordinates by a constant
\begin{align}
d\tilde{x}^\mu=A(\varphi_\infty)dx^\mu~,
\end{align}
such that $\tilde{r}=A(\varphi_\infty)r$. With these rescaled coordinates we can write the Jordan metric at infinity in terms of the Einstein metric at infinity as 
\begin{align}
g_{rr}&=A(\varphi)^2g^\ast_{rr}=\frac{A(\varphi)^2}{A(\varphi_\infty)^2}\frac{1}{1-\frac{2A(\varphi_\infty)G m(\tilde{r})}{\tilde{r}}}\nonumber\\&=1+\frac{2A(\varphi_\infty)G\left(M-Q\alpha_\infty\right)}{\tilde{r}}+\mathcal{O}\left(\frac{1}{\tilde{r}^2}\right)~,
\end{align}
with $M$ the ADM mass \eqref{eq: ADMMassEinstein} and $Q$ the scalar charge in the Einstein frame. Comparing with the Jordan frame metric at infinity
\begin{align}
g_{rr}&=1+\frac{2\tilde{G}M_J}{\tilde{r}}+\mathcal{O}\left(\frac{1}{\tilde{r}^2}\right)~,
\end{align}
with $M_J$ the ADM mass in the Jordan frame, yields
\begin{align}\label{eq: ADMMassJordanGeneric}
    M_J=\frac{M - Q \alpha_\infty}{A(\varphi_\infty) \left(1+\frac{2K_R}{K_\varphi}\alpha_\infty^2\right)}=\frac{M\left(1 + q \alpha_\infty\right)}{A(\varphi_\infty) \left(1+\frac{2K_R}{K_\varphi}\alpha_\infty^2\right)}~,
\end{align}
with $q=-Q/M$ (minus) the charge per unit mass. Again, using the normalizations and coupling function of the main text we obtain
\begin{align}\label{eq: ADMMassJordan}
    M_J=Me^{-\frac{1}{2}\beta\varphi_\infty^2}\frac{1-q\beta\varphi_\infty}{ 1+\beta^2\varphi_\infty^2}~,
\end{align}
in agreement with \cite{Damour_1992}. Ignoring the terms quadratic in $\varphi_\infty$ we recover the result of \cite{Pani:2014jra},
\begin{align}
    M_J=e^{-\frac{1}{2}\beta\varphi_\infty^2}\left( M + \beta\varphi_\infty Q \right)~.
\end{align}
\subsection{Scalar charge}
In order to transform the scalar charge we can use \eqref{eq:ConformalPartial} and \eqref{eq:STalphaApp} in order to write
\begin{align*}
    \p_\mu\phi=\frac{1}{\sqrt{\Delta}}\p_\mu\varphi=\frac{F'}{2 F \alpha}\p_\mu\varphi~.
\end{align*}
Using that, at infinity,
\begin{subequations}\label{eq: ScalarAtInf}
\begin{align}
\phi=&\phi_\infty+\frac{\tilde{G} Q_J}{\tilde{r}}+\mathcal{O}\left(\frac{1}{\tilde{r}^2}\right)~,\\
\varphi_0=&\varphi_{0\infty}+\frac{G Q}{r}+\mathcal{O}\left(\frac{1}{r^2}\right)~,
\end{align}
\end{subequations}
we have, at zeroth order in $1/r$,
\begin{align}
    \frac{G Q}{r^2}=\frac{F_\infty'}{2 F_\infty \alpha_\infty}\frac{\tilde{G} Q_J}{\tilde{r}^2}\frac{d\tilde{r}}{dr}~,
\end{align}
and therefore
\begin{align}
    Q_J=\frac{2\alpha_\infty}{A(\varphi_\infty)}\frac{G}{\tilde{G}}Q=\frac{2\alpha_\infty}{A(\varphi_\infty)^3\left(1+\frac{2K_R}{K_\varphi}\alpha_\infty^2\right)}Q~.
\end{align}
With our normalizations we have
\begin{align}
    Q_J=-\frac{2\beta\varphi_\infty }{1+\beta^2\varphi_\infty^2}e^{-\frac{3}{2}\beta\varphi_\infty^2}Q~.
\end{align}
Note that here we define the coefficient in front of $1/r$ in \eqref{eq: ScalarAtInf} with the explicit factors of $G$ and $\tilde{G}$. However, in the literature this might vary, but then one has to take into account the proper dimensionality of $\alpha_\infty$, which with our normalizations is dimensionless. If we ignore the factors of $G$, then we recover the same expression as in \cite{Pani:2014jra},
\begin{align}
      \tilde{G}Q_J=-2\beta\varphi_\infty e^{-\frac{1}{2}\beta\varphi_\infty^2}GQ~.
\end{align}

\begin{figure*}
\section{Additional plots of $\lambda_\ell$}\label{app:Plotslambda}
\subsection{Quadrupolar $\ell=2$}
\begin{justify}
Figure \ref{fig:lambdaScalarApp} shows the results for the scalar quadrupolar tidal deformabilities for different equations of state (different colors), coupling coefficients (different dashings), and in both the Einstein (left panels) and Jordan frame (right panels). The panels in the bottom row are zoomed in on the features for small-$\lambda$ regime relevant for the smaller coupling. Figure \ref{fig:lambdaScalarTensorApp} shows analogous results but for the mixed scalar-tensor deformability.
\end{justify}
\begin{center}
{\includegraphics[width=0.49\textwidth,clip]{figs/ScalarTidalDefEinsteinMinus3.pdf}}
{\includegraphics[width=0.49\textwidth,clip]{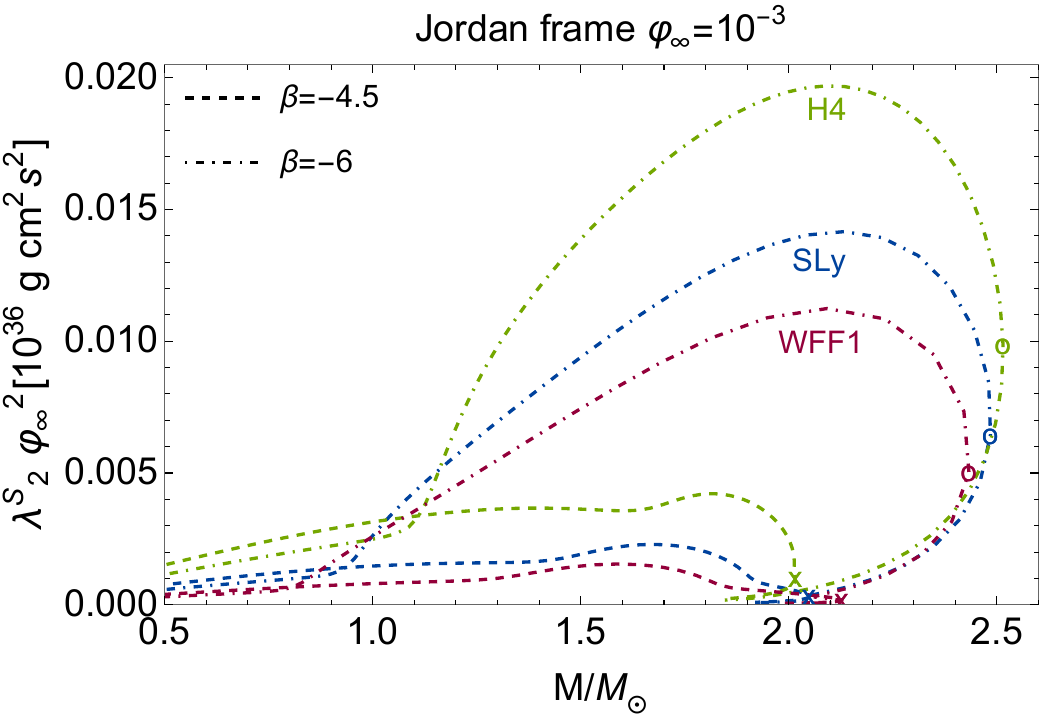}}
{\includegraphics[width=0.49\textwidth,clip]{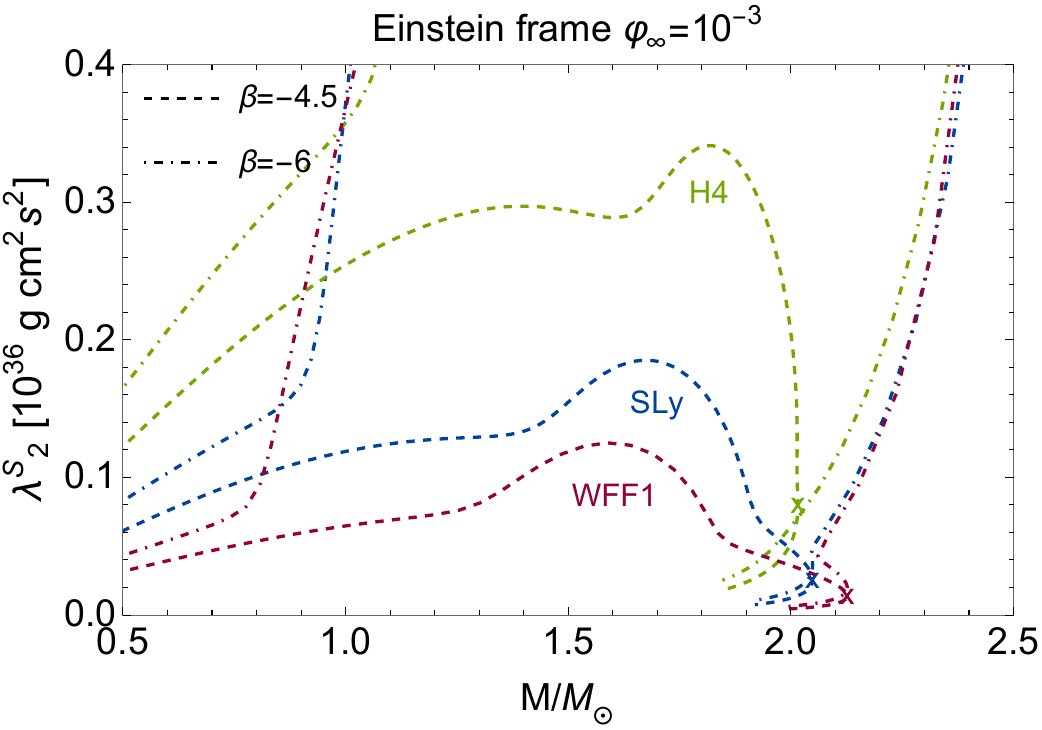}}
{\includegraphics[width=0.49\textwidth,clip]{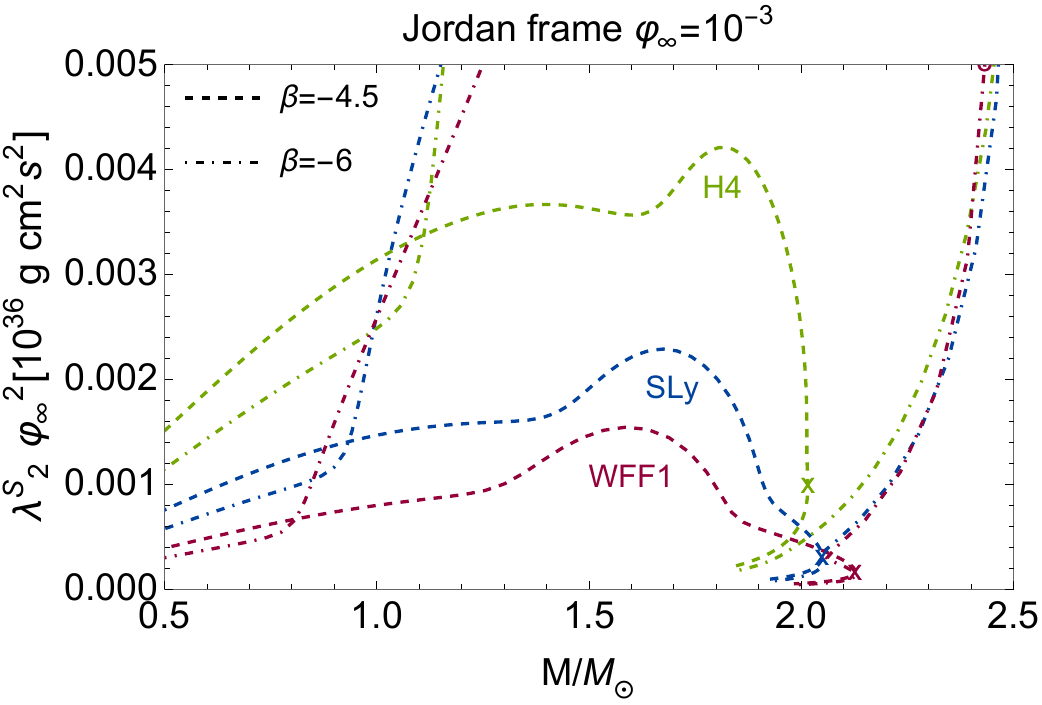}}
\caption{\textbf{\emph{Quadrupolar scalar tidal deformabilities}} $\lambda_2^S$ in the Einstein and Jordan frames for three equations of state (WFF1, SLy, and H4). The Jordan frame plots use the Jordan frame mass. The dashed and dot-dashed lines are the scalarized configurations with $\beta=-4.5$, and $\beta=-6$, respectively, and the plots in the row are enlarged with respect to their counterparts in the top row. The cross represents the maximum mass configuration for $\beta=0,-4.5$, and the circle for $\beta=-6$. All plots correspond to a scalar field at infinity $\varphi_{0\infty}=10^{-3}$.}\label{fig:lambdaScalarApp}
\end{center}
\end{figure*}
\begin{figure*}[htpb!]
\begin{center}
{\includegraphics[width=0.49\textwidth,clip]{figs/ScalarTensorTidalDefEinsteinMinus3.pdf}}
{\includegraphics[width=0.49\textwidth,clip]{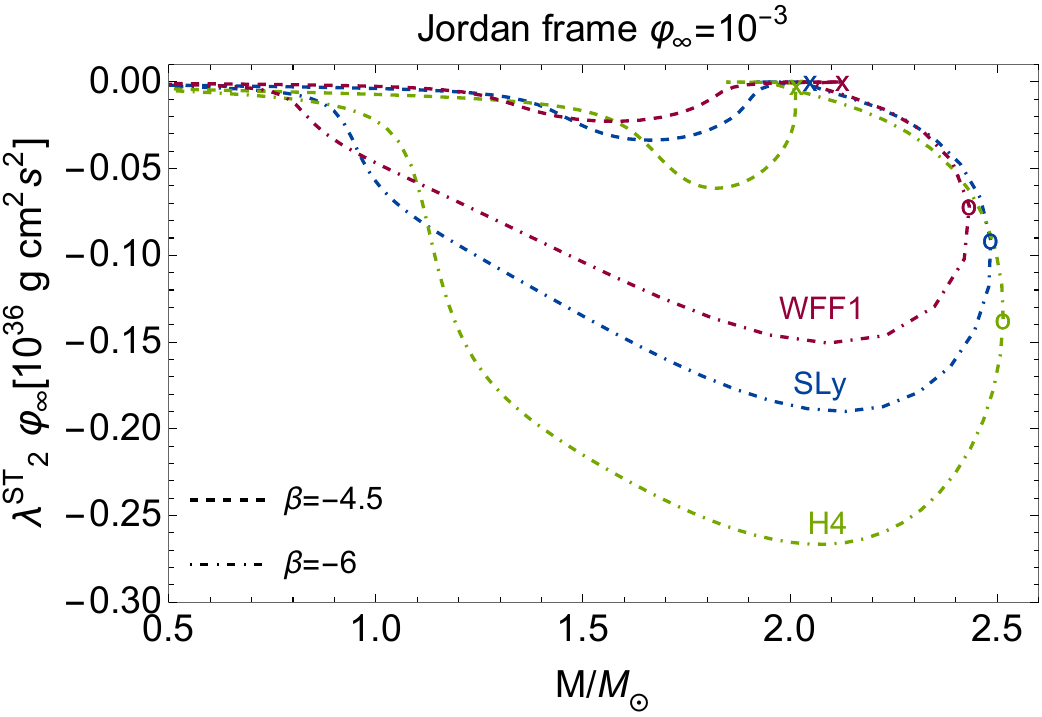}}
{\includegraphics[width=0.49\textwidth,clip]{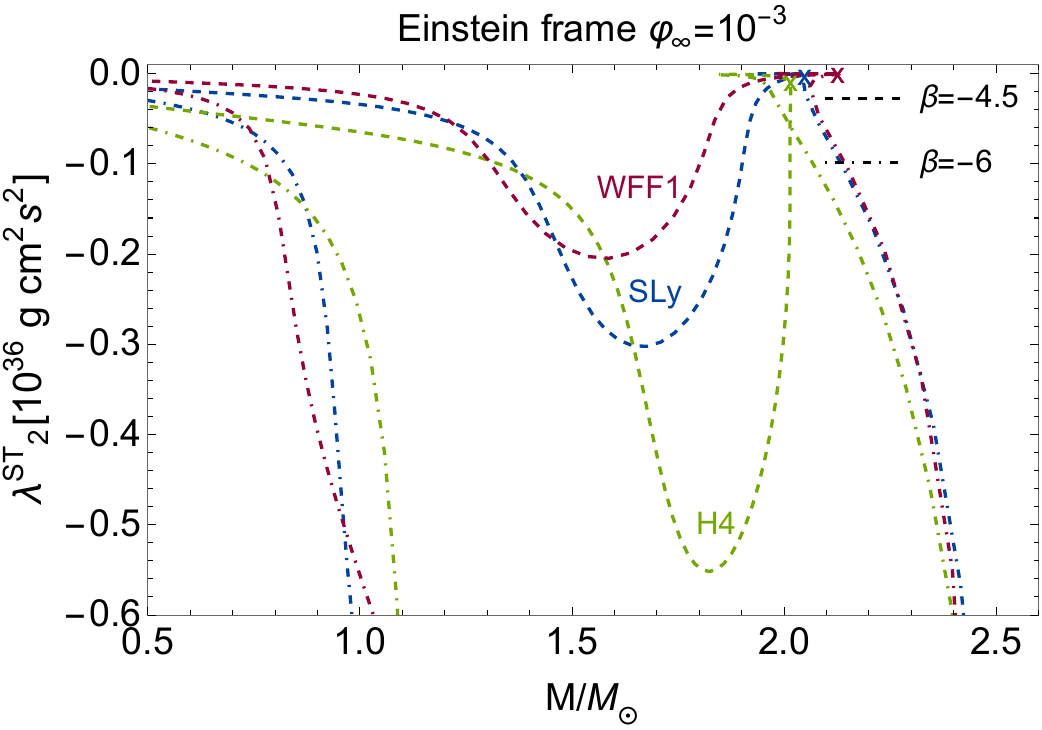}}
{\includegraphics[width=0.49\textwidth,clip]{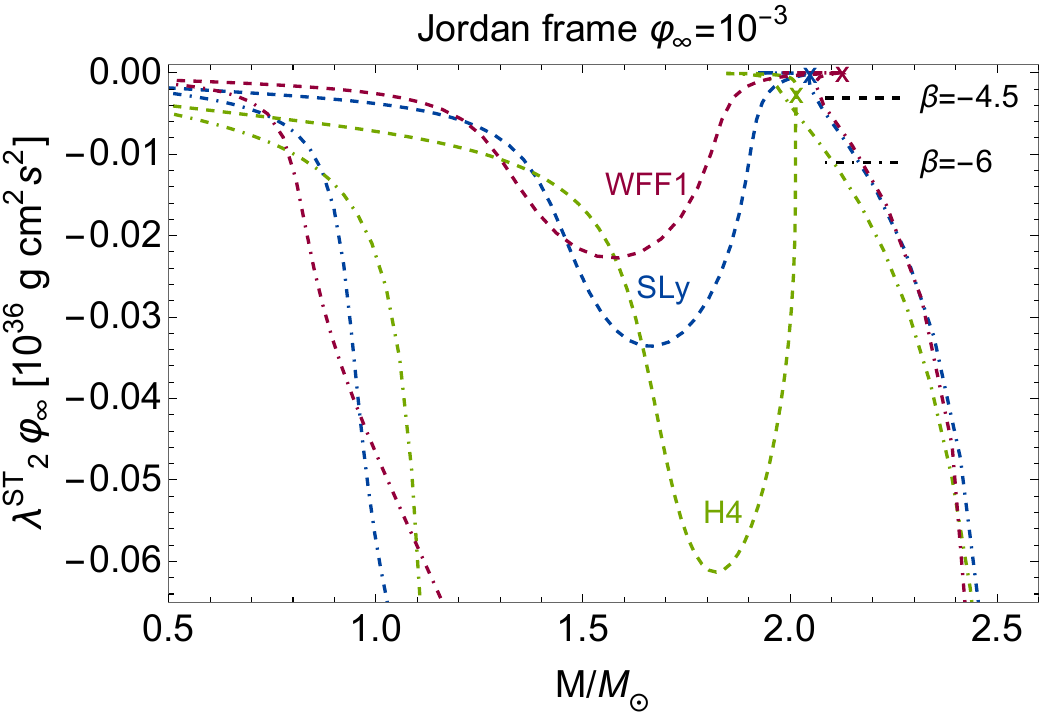}}
\caption{\textbf{\emph{Quadrupolar scalar-tensor tidal deformabilities}} $\lambda_2^{ST}$ in the Einstein and Jordan frames for three equations of state (WFF1, SLy, and H4). The Jordan frame plots use the Jordan frame mass. The dashed and dot-dashed lines are the scalarized configurations with $\beta=-4.5$, and $\beta=-6$, respectively, and the plots in the bottom row are enlarged with respect to their counterparts in the top row. The cross represents the maximum mass configuration for $\beta=0,-4.5$, and the circle for $\beta=-6$. All plots correspond to a scalar field at infinity $\varphi_{0\infty}=10^{-3}$.}\label{fig:lambdaScalarTensorApp}
\end{center}
\end{figure*}

\begin{figure*}
\subsection{Octupolar $\ell=3$}
\begin{justify}
Figure \ref{fig:lambdaScalarl3App} shows the results for the scalar octupolar tidal deformabilities for different equations of state (different colors), coupling coefficients (different dashings), and in both the Einstein (left panels) and Jordan frame (right panels). The panels in the bottom row are zoomed in on the features for small-$\lambda$ regime relevant for the smaller coupling. Figure \ref{fig:lambdaScalarTensorl3App} shows analogous results but for the mixed scalar-tensor deformability.    
\end{justify}
\begin{center}
{\includegraphics[width=0.49\textwidth,clip]{figs/ScalarTidalDefEinsteinMinus3l3.pdf}}
{\includegraphics[width=0.49\textwidth,clip]{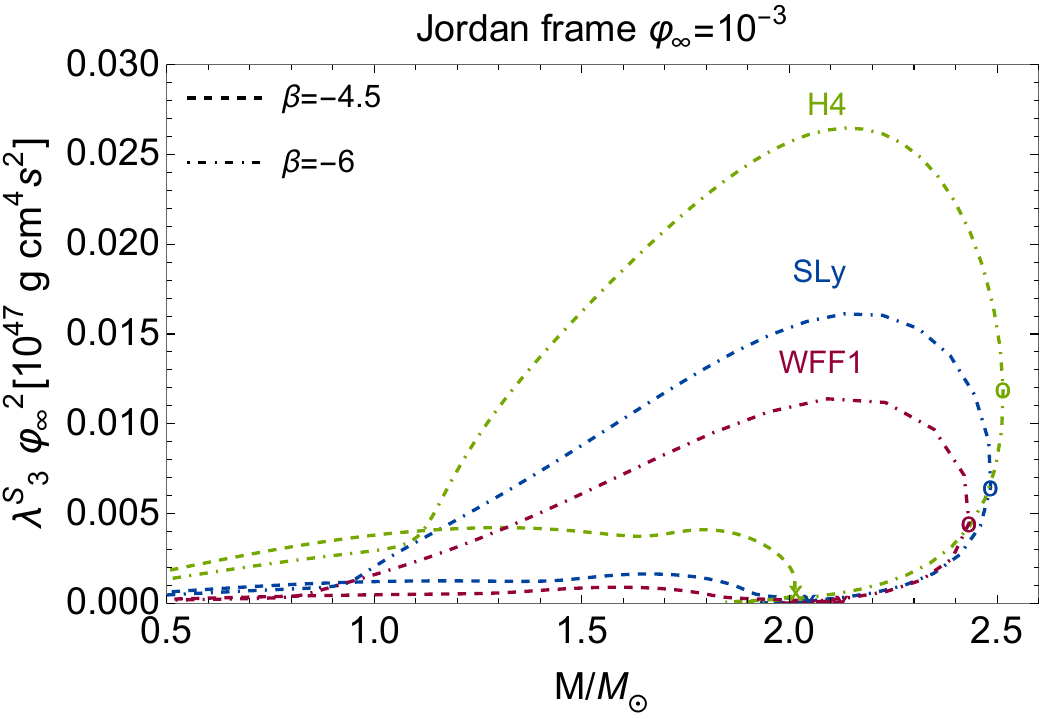}}
{\includegraphics[width=0.49\textwidth,clip]{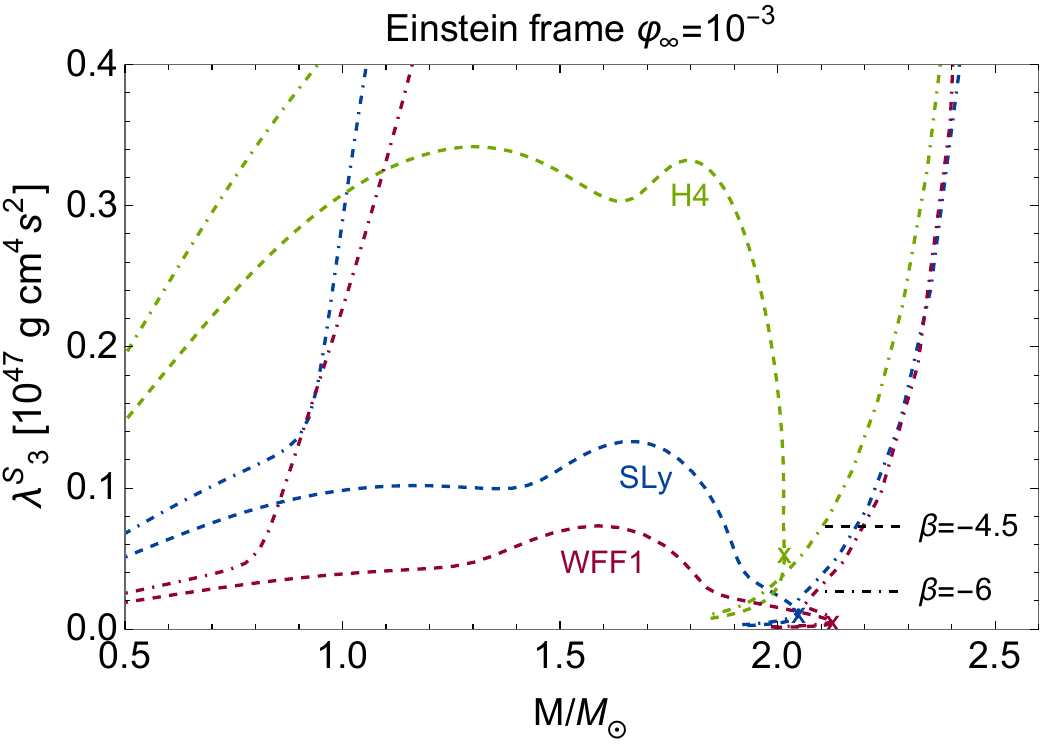}}
{\includegraphics[width=0.49\textwidth,clip]{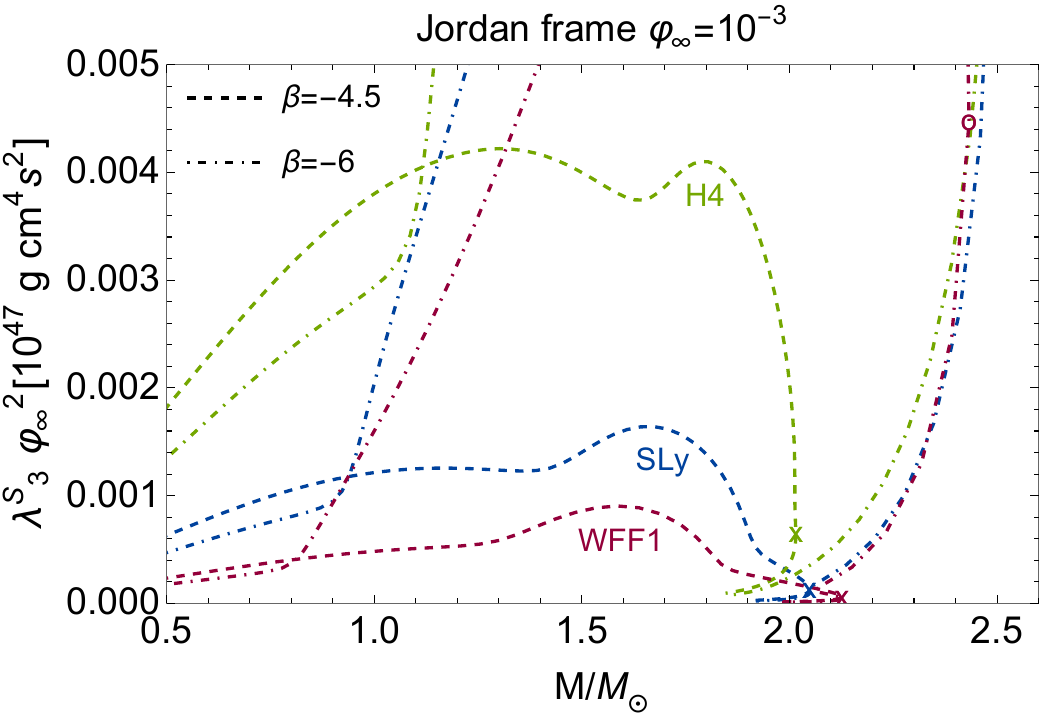}}
\caption{\textbf{\emph{Octupolar scalar tidal deformabilities}} $\lambda_3^S$ in the Einstein and Jordan frames for three equations of state (WFF1, SLy, and H4). The Jordan frame plots use the Jordan frame mass. The dashed and dot-dashed lines are the scalarized configurations with $\beta=-4.5$, and $\beta=-6$, respectively, and the plots in the bottom row are enlarged with respect to their counterparts in the top row. The cross represents the maximum mass configuration for $\beta=0,-4.5$, and the circle for $\beta=-6$. All plots correspond to a scalar field at infinity $\varphi_{0\infty}=10^{-3}$.}\label{fig:lambdaScalarl3App}
\end{center}
\end{figure*}
\begin{figure*}[htpb!]
\begin{center}
{\includegraphics[width=0.49\textwidth,clip]{figs/ScalarTensorTidalDefEinsteinMinus3l3.pdf}}
{\includegraphics[width=0.49\textwidth,clip]{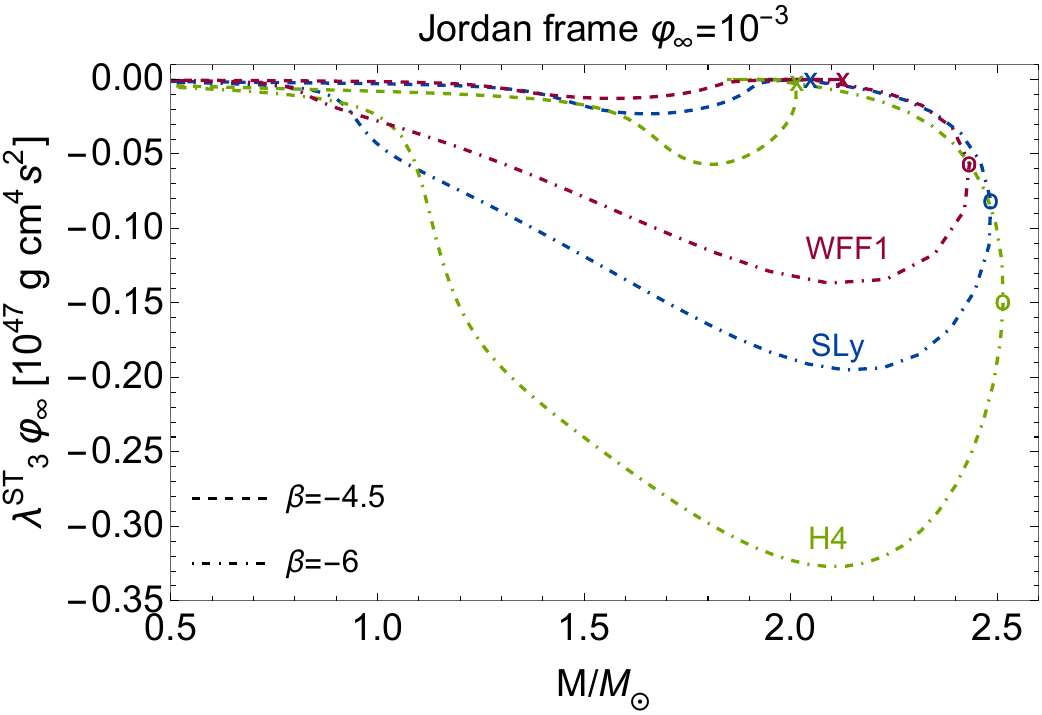}}
{\includegraphics[width=0.49\textwidth,clip]{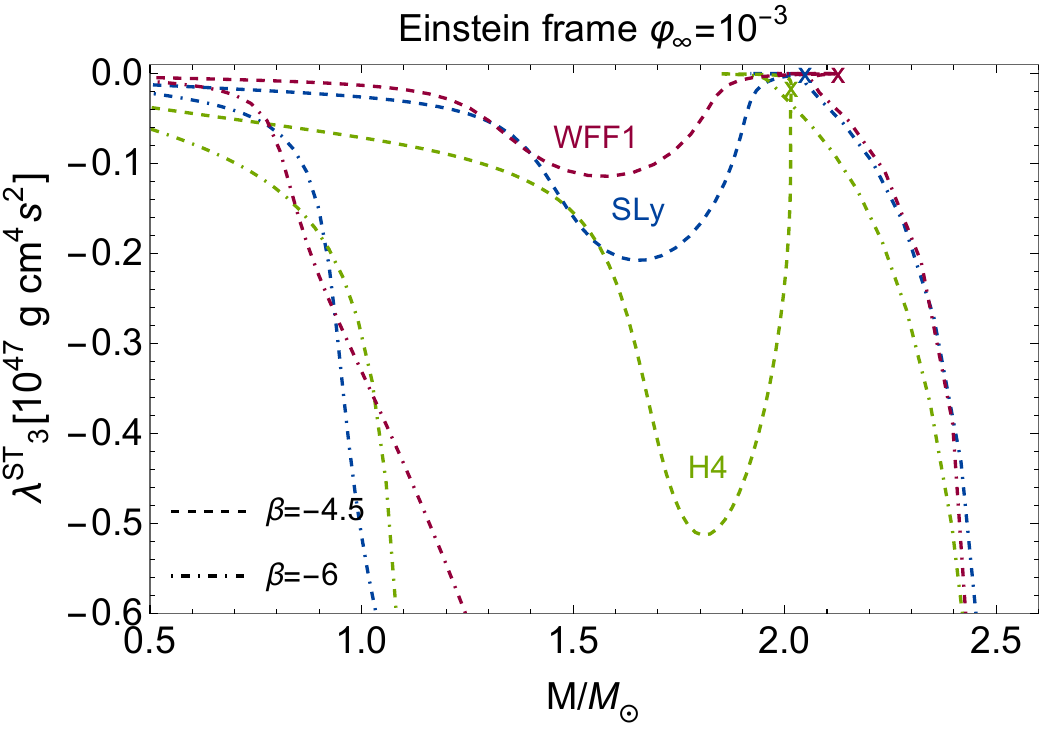}}
{\includegraphics[width=0.49\textwidth,clip]{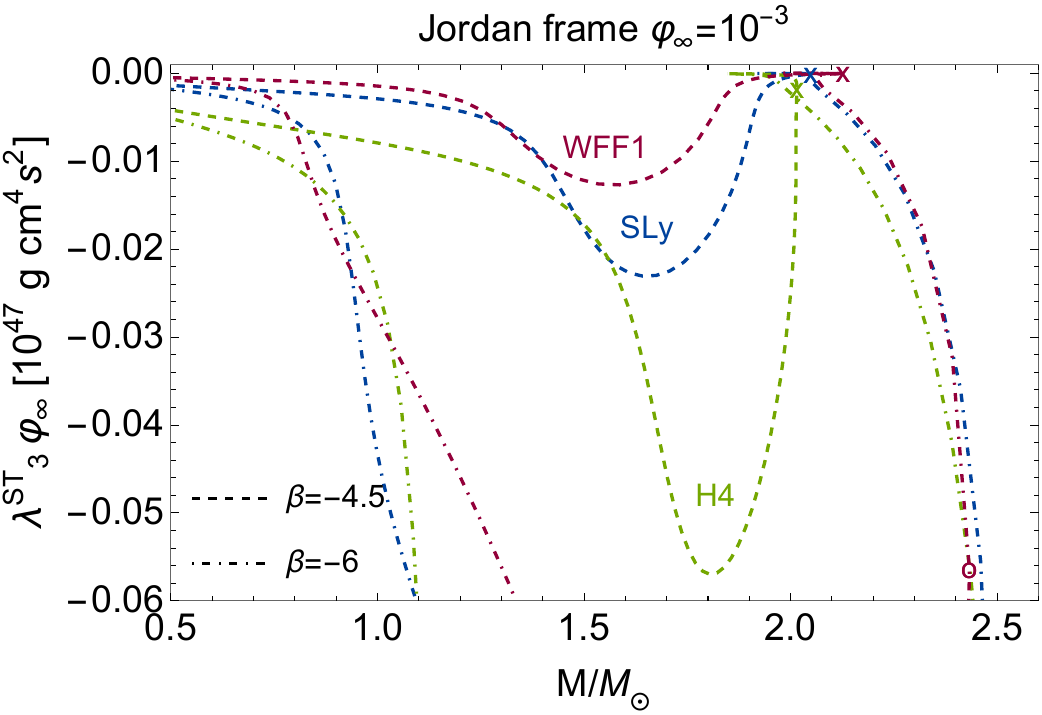}}
\caption{\textbf{\emph{Octupolar scalar-tensor tidal deformabilities}} $\lambda_3^{ST}$ in the Einstein and Jordan frames for three equations of state (WFF1, SLy, and H4). The Jordan frame plots use the Jordan frame mass. The dashed and dot-dashed lines are the scalarized configurations with $\beta=-4.5$, and $\beta=-6$, respectively, and the plots in the bottom row are enlarged with respect to their counterparts in the top row. The cross represents the maximum mass configuration for $\beta=0,-4.5$, and the circle for $\beta=-6$. All plots correspond to a scalar field at infinity $\varphi_{0\infty}=10^{-3}$.}\label{fig:lambdaScalarTensorl3App}
\end{center}
\end{figure*}

\begin{figure*}
\section{Plots of the dimensionless quantities $\Lambda_{\ell}$ and $k_{\ell}$}\label{app:PlotsLambdak}
\subsection{Dipolar $\ell=1$}
\begin{justify}
    Figure \ref{fig:DipolarScalarLambda} shows the results for the dimensionless scalar dipolar tidal deformability quantities $\Lambda_1^S$ defined in Eq.~\eqref{eq:LambdaDef} for different equations of state (different colors), coupling coefficients (upper and lower panels), and in both the Einstein (left panels) and Jordan frame (right panels). Figure \ref{fig:DipolarScalarLove} shows analogous results for the dimensionless Love number defined in Eq.~\eqref{eq:LoveNumvsTidalDef}.
\end{justify}
\begin{center}
{\includegraphics[width=0.49\textwidth,clip]{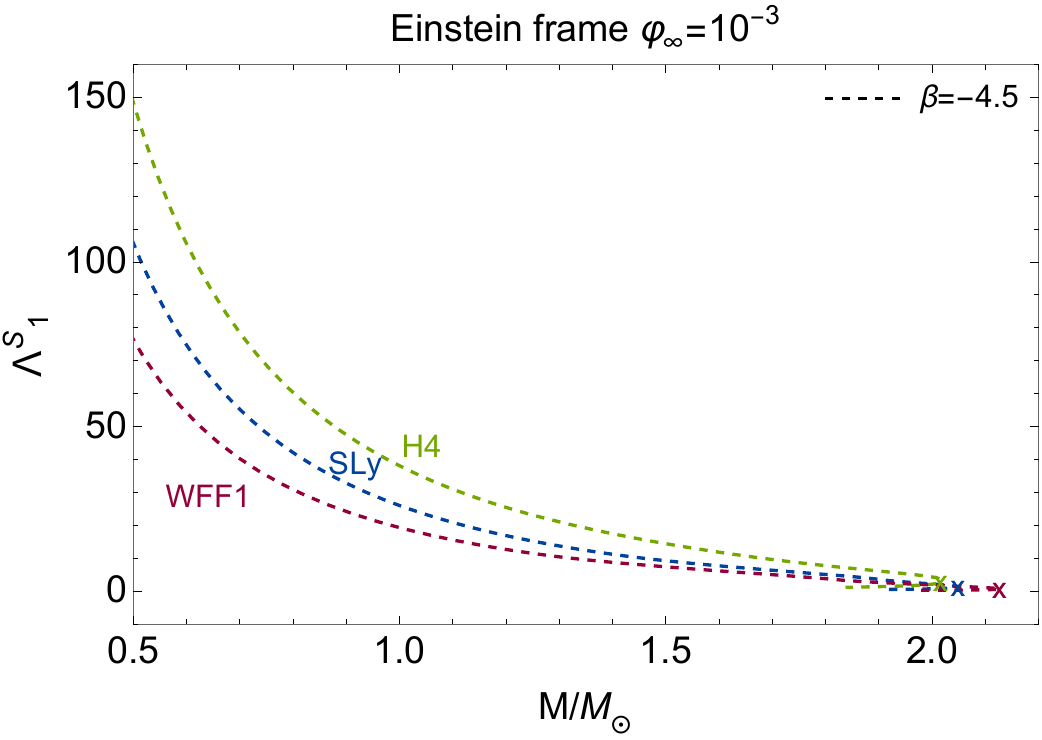}}
{\includegraphics[width=0.49\textwidth,clip]{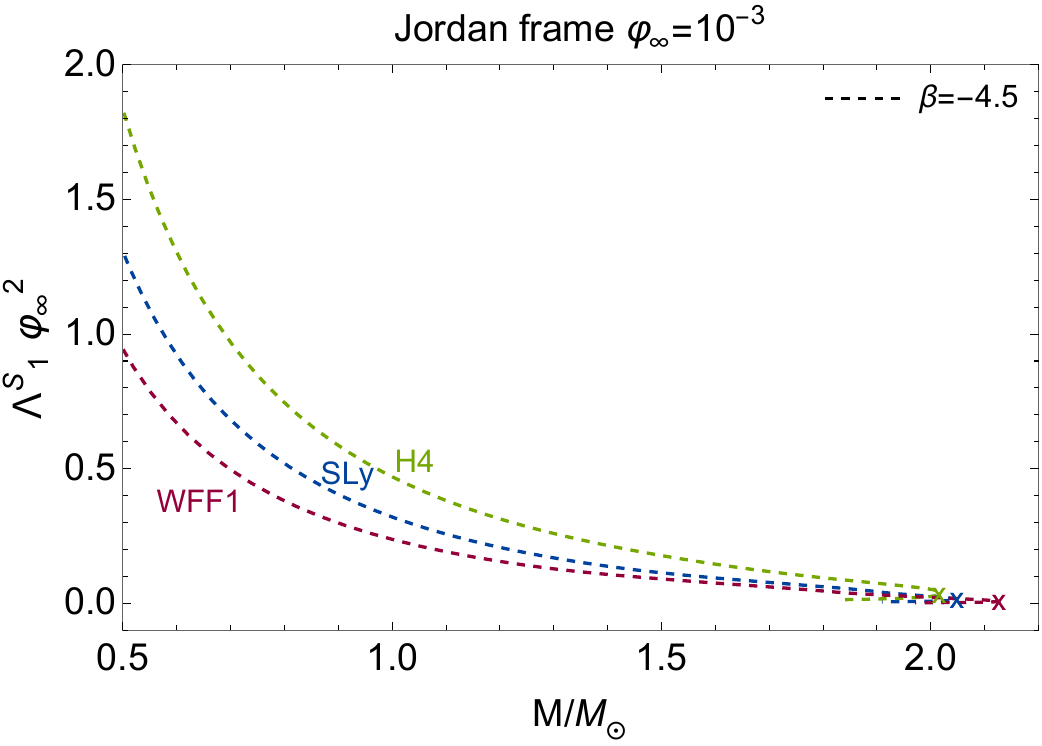}}
{\includegraphics[width=0.49\textwidth,clip]{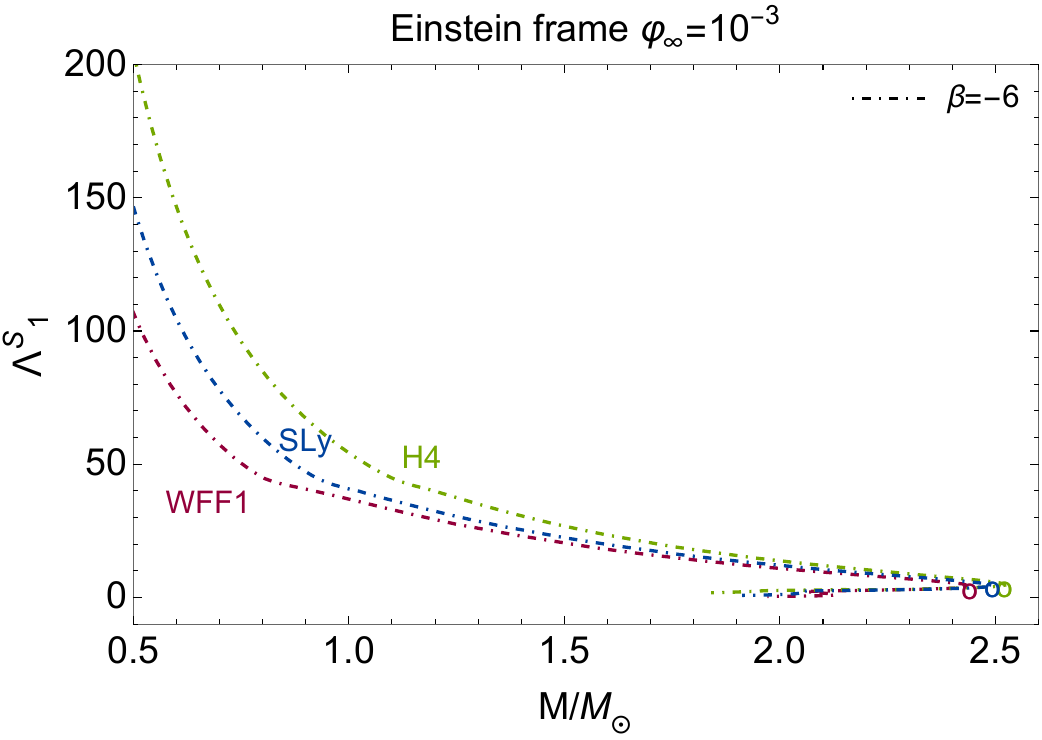}}
{\includegraphics[width=0.49\textwidth,clip]{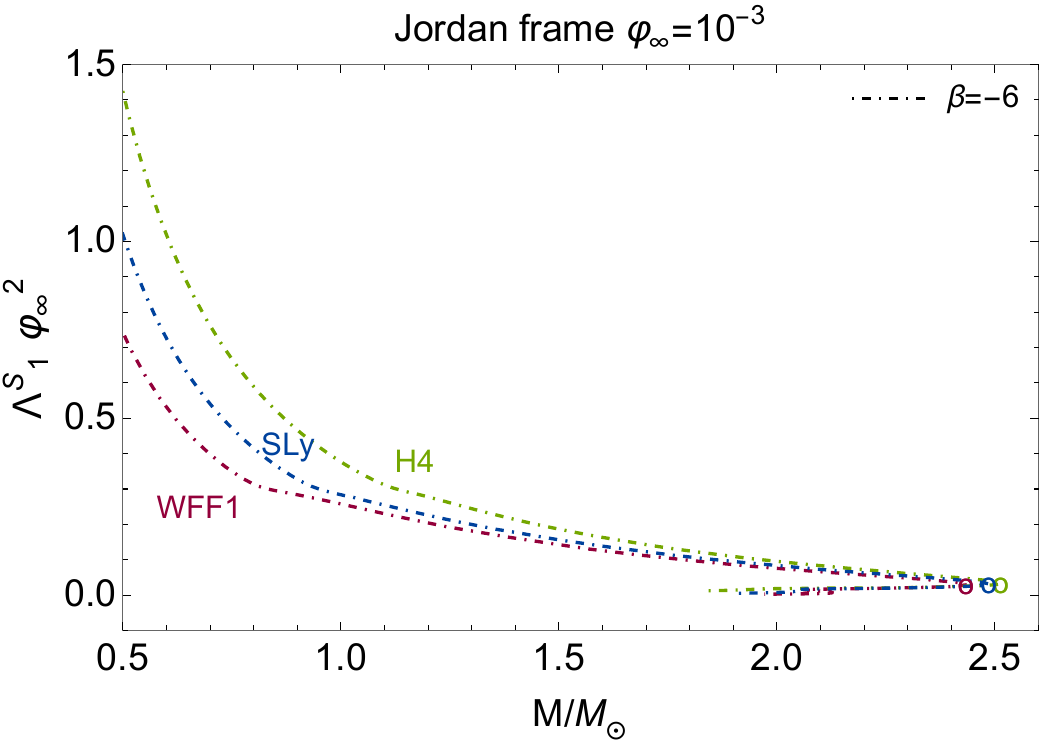}}
\caption{\textbf{\emph{Dipolar scalar adimensional tidal deformabilities}} $\Lambda_1^S$ in the Einstein and Jordan frames for three equations of state (WFF1, SLy, and H4). The Jordan frame plots use the Jordan frame mass. The dashed and dot-dashed lines are the scalarized configurations with $\beta=-4.5$ and $\beta=-6$, respectively. The cross represents the maximum mass configuration for $\beta=0,-4.5$, and the circle for $\beta=-6$.  For $\beta=-6$ we have omitted the data beyond the maximum mass configuration for better readability. All plots correspond to a scalar field at infinity $\varphi_{0\infty}=10^{-3}$.}\label{fig:DipolarScalarLambda}
\end{center}
\end{figure*}

\begin{figure*}
\begin{center}
{\includegraphics[width=0.49\textwidth,clip]{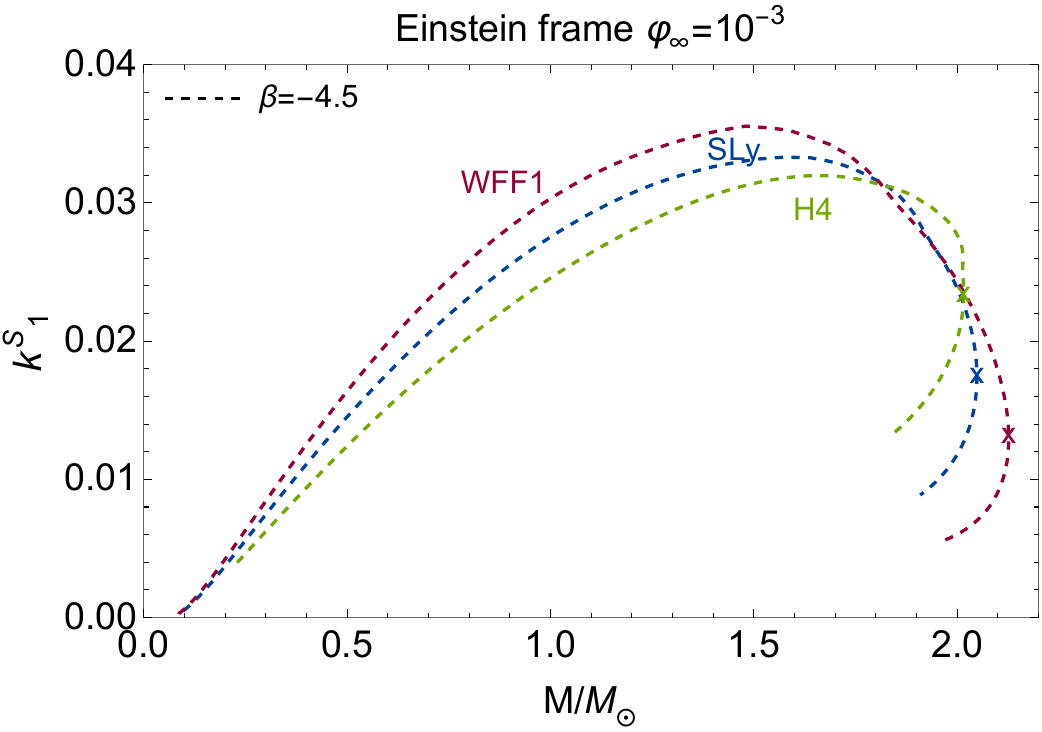}}
{\includegraphics[width=0.49\textwidth,clip]{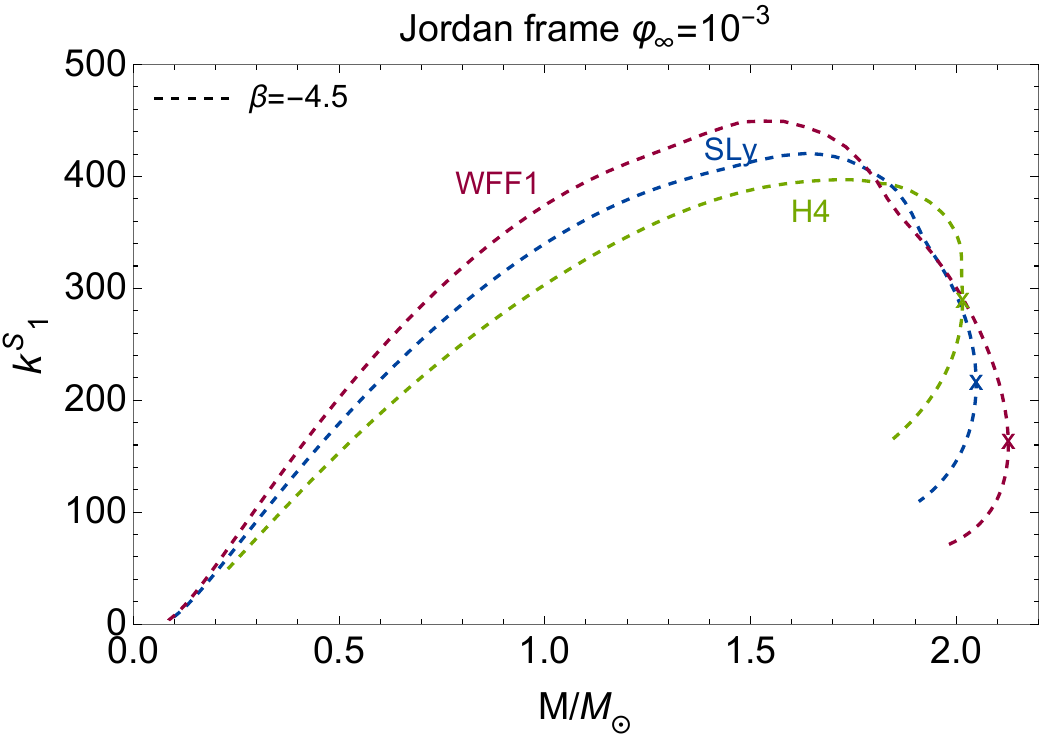}}
{\includegraphics[width=0.49\textwidth,clip]{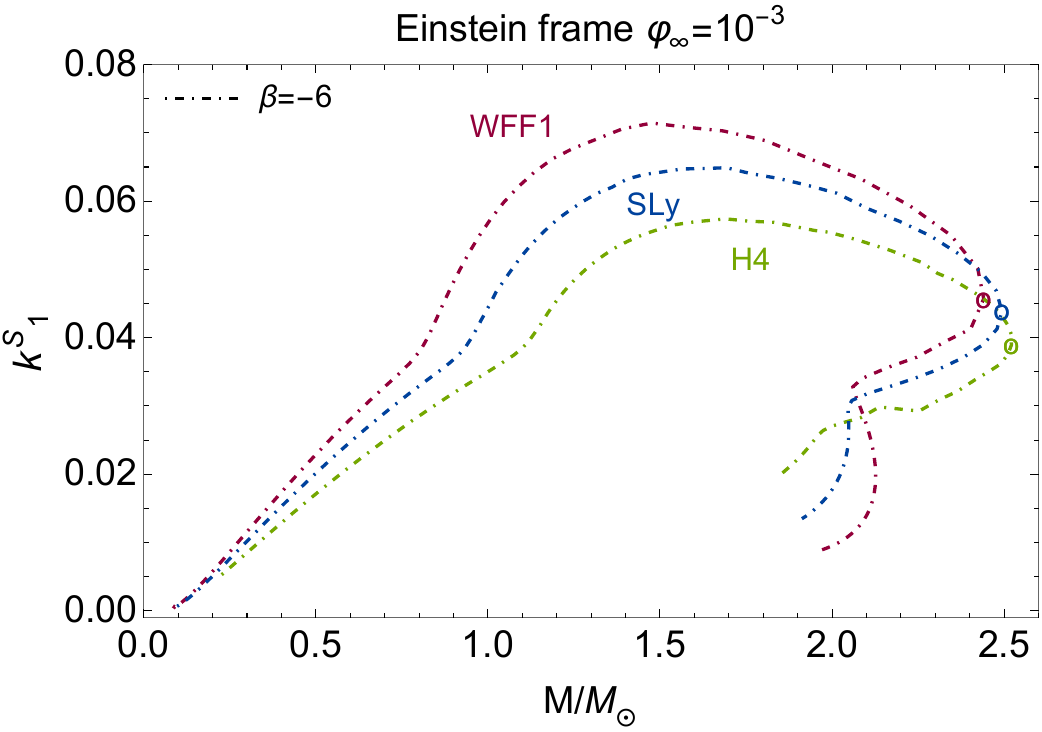}}
{\includegraphics[width=0.49\textwidth,clip]{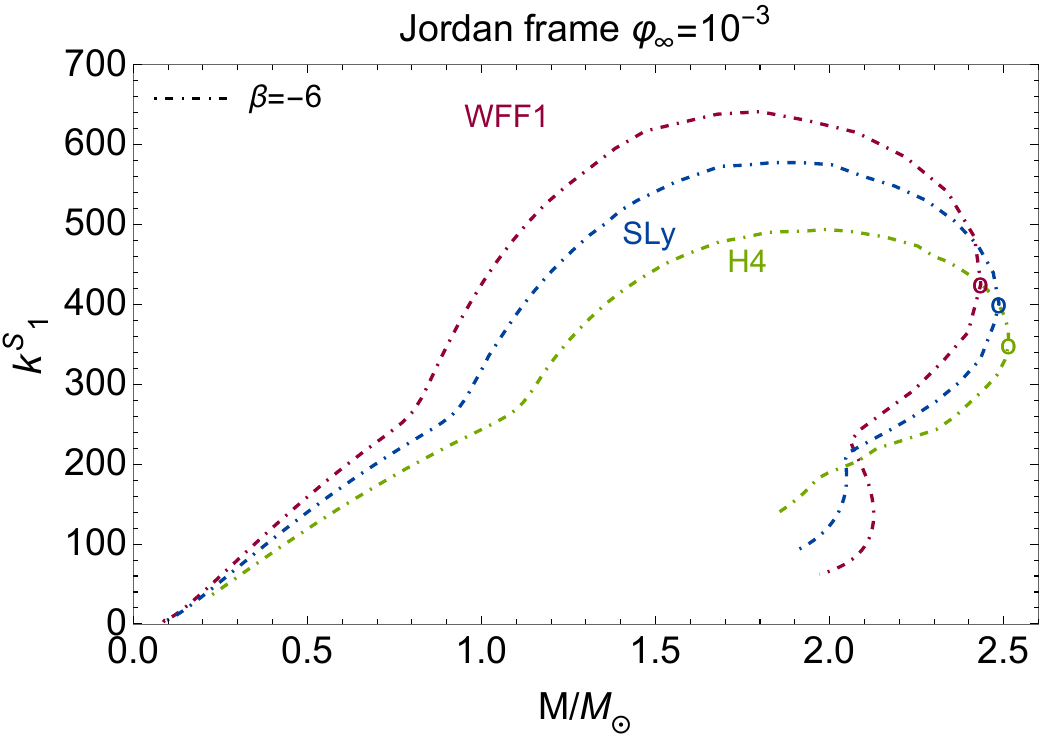}}
\caption{\textbf{\emph{Dipolar scalar Love number}} $k_1^S$ in the Einstein and Jordan frames for three equations of state (WFF1, SLy, and H4). The Jordan frame plots use the Jordan frame mass. The dashed and dot-dashed lines are the scalarized configurations with $\beta=-4.5$ and $\beta=-6$, respectively. The cross represents the maximum mass configuration for $\beta=0,-4.5$, and the circle for $\beta=-6$. For $\beta=-6$ we have omitted the data beyond the maximum mass configuration for better readability. All plots correspond to a scalar field at infinity $\varphi_{0\infty}=10^{-3}$.}\label{fig:DipolarScalarLove}
\end{center}
\end{figure*}

\begin{figure*}
\subsection{Quadrupolar $\ell=2$}
\begin{justify}
Figure \ref{fig:LambdaTensor} shows the results for the dimensionless tensor quadrupolar tidal deformability quantities $\Lambda_2^T$ defined in Eq.~\eqref{eq:LambdaDef} for different equations of state (different colors), coupling coefficients (different dashings), and in both the Einstein (left panel) and Jordan frame (right panel). Figure \ref{fig:LambdaScalar} shows analogous results for the scalar deformability, while Fig. \ref{fig:LambdaScalarTensor2} illustrates those for the mixed scalar-tensor quantity. The corresponding results for the dimensionless Love numbers $k_2$ defined in Eq.~\eqref{eq:LoveNumvsTidalDef} are shown in Fig.~\ref{fig:Love2TPlot} for the tensor, Fig.~\ref{fig:Love2SPlot} for the scalar, and Fig.~\ref{fig:Love2STPlot} for the mixed quantities
\end{justify}
\begin{center}
{\includegraphics[width=0.49\textwidth,clip]{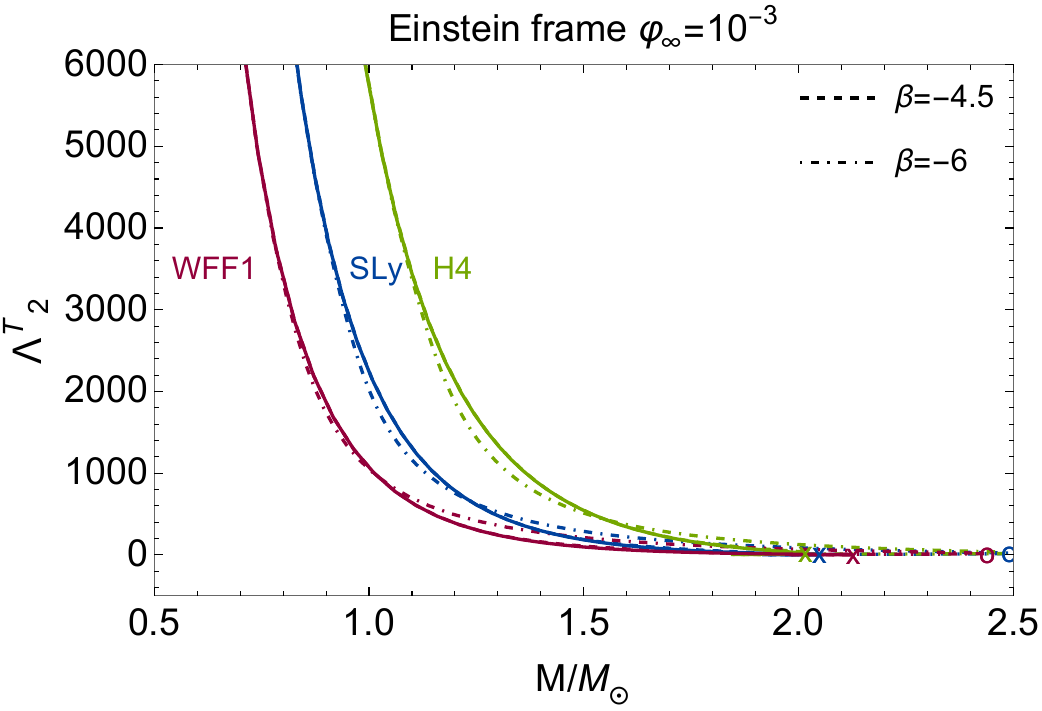}}
{\includegraphics[width=0.49\textwidth,clip]{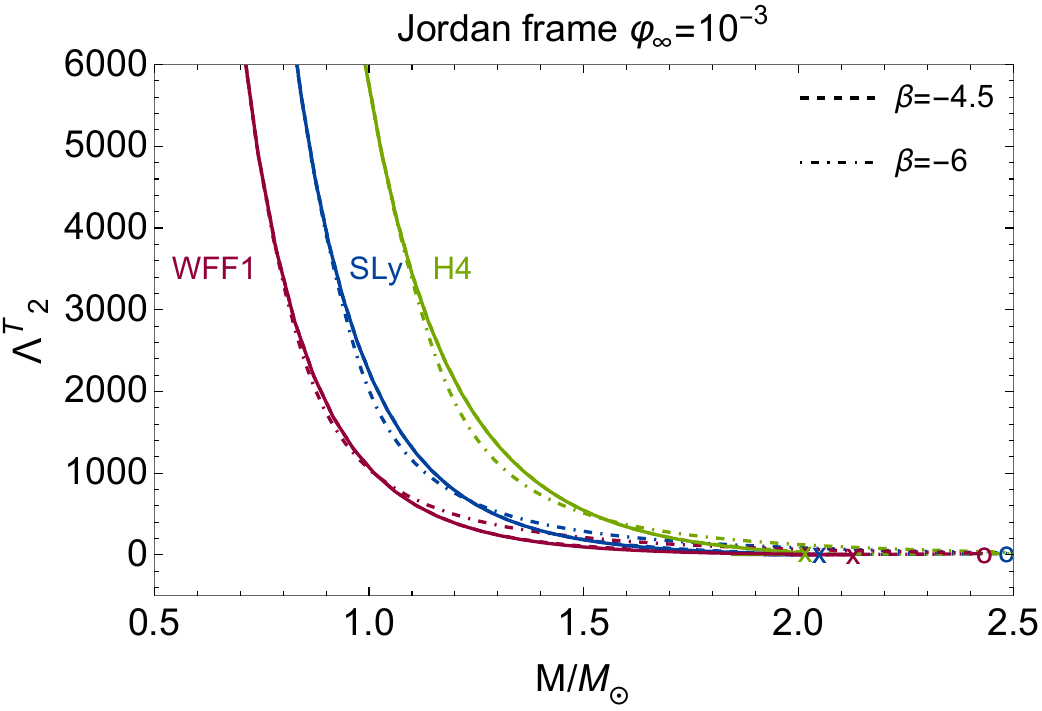}}
\caption{\textbf{\emph{Quadrupolar tensor adimensional tidal deformabilities}} $\Lambda_2^T$ in the Einstein and Jordan frames for three equations of state (WFF1, SLy, and H4). The Jordan frame plots use the Jordan frame mass. The solid lines represent the GR configurations $\beta=0$ and the dashed and dot-dashed lines are the scalarized configurations with $\beta=-4.5$ and $\beta=-6$, respectively. The cross represents the maximum mass configuration for $\beta=0,-4.5$, and the circle for $\beta=-6$. Both plots correspond to a scalar field at infinity $\varphi_{0\infty}=10^{-3}$.}\label{fig:LambdaTensor}
\end{center}
\end{figure*}
\begin{figure*}
\begin{center}
{\includegraphics[width=0.49\textwidth,clip]{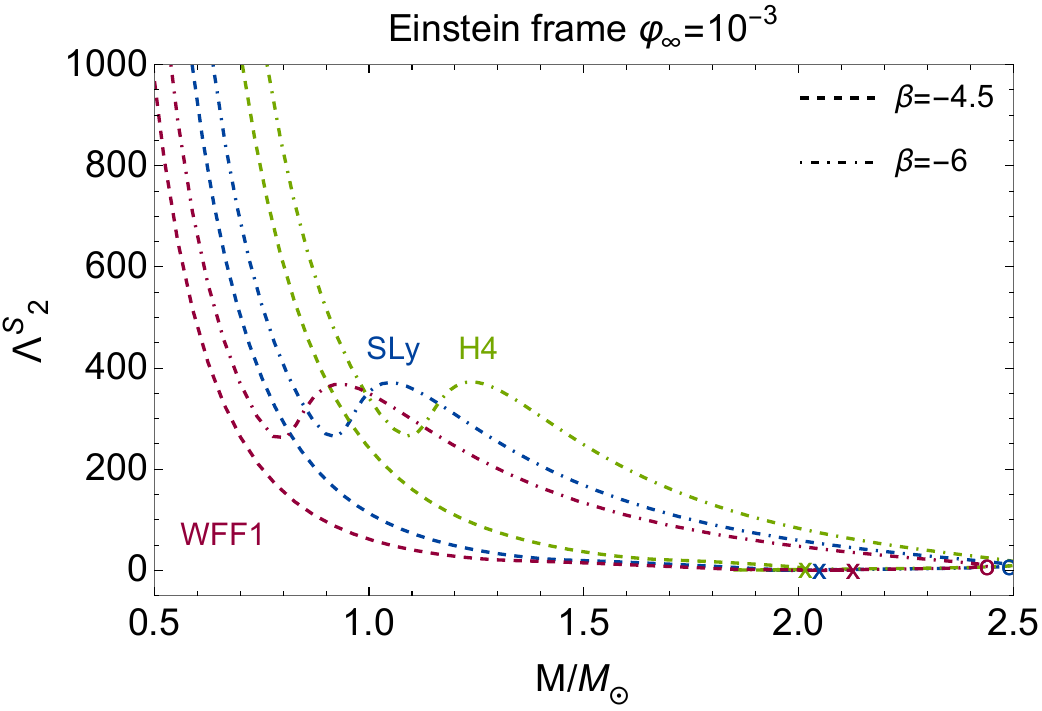}}
{\includegraphics[width=0.49\textwidth,clip]{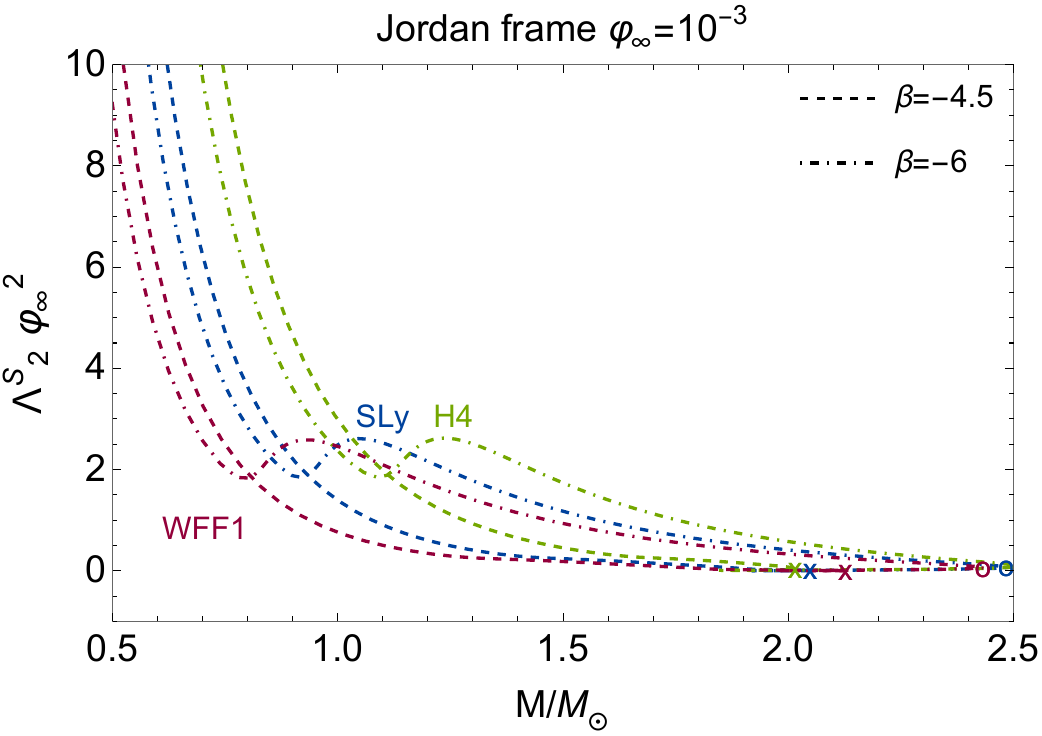}}
\caption{\textbf{\emph{Quadrupolar scalar adimensional tidal deformabilities}} $\Lambda_2^S$ in the Einstein and Jordan frames for three equations of state (WFF1, SLy, and H4). The Jordan frame plots use the Jordan frame mass. The dashed and dot-dashed lines are the scalarized configurations with $\beta=-4.5$ and $\beta=-6$, respectively. The cross represents the maximum mass configuration for $\beta=0,-4.5$, and the circle for $\beta=-6$. Both plots correspond to a scalar field at infinity $\varphi_{0\infty}=10^{-3}$.}\label{fig:LambdaScalar}
\end{center}
\end{figure*}
\begin{figure*}[p]
\begin{center}
{\includegraphics[width=0.49\textwidth,clip]{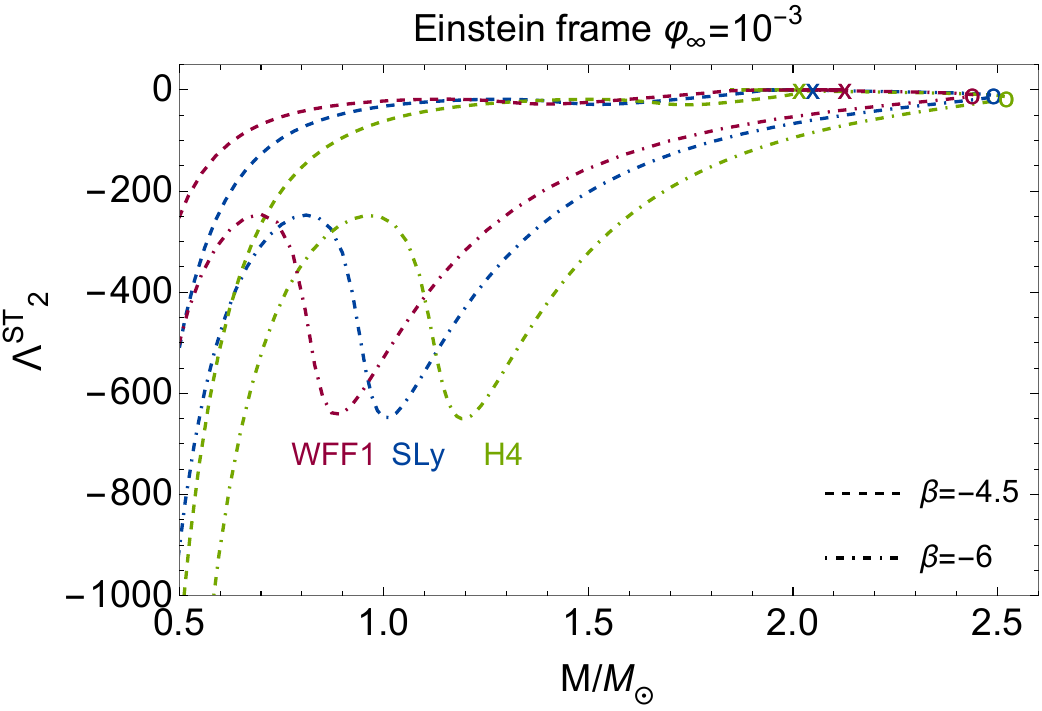}}
{\includegraphics[width=0.49\textwidth,clip]{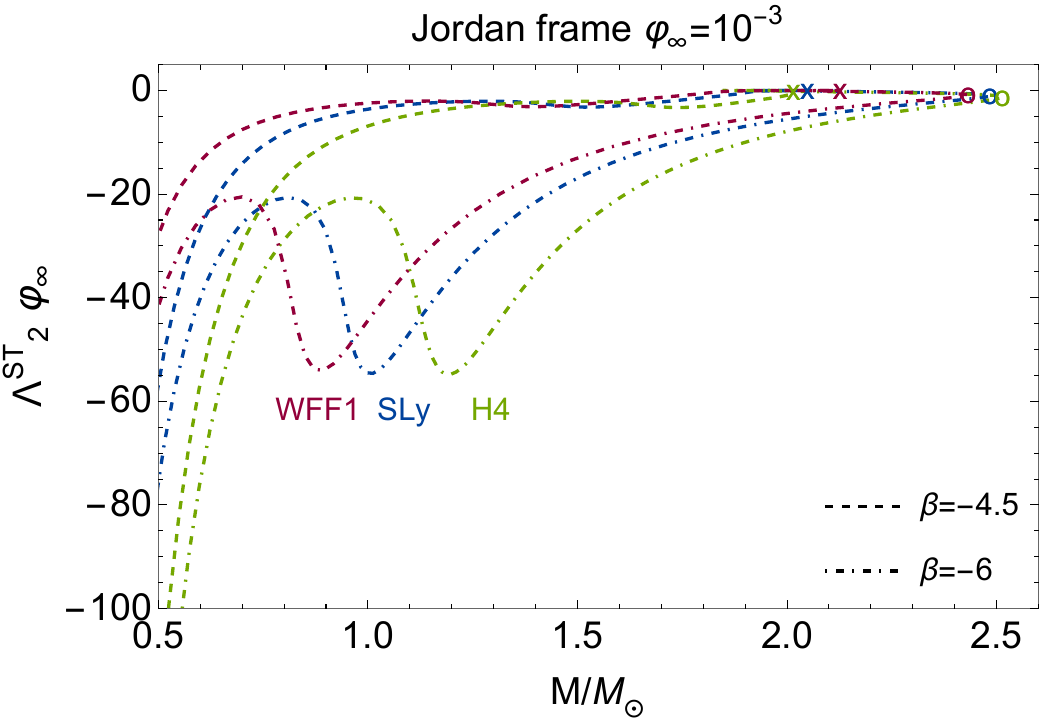}}
{\includegraphics[width=0.49\textwidth,clip]{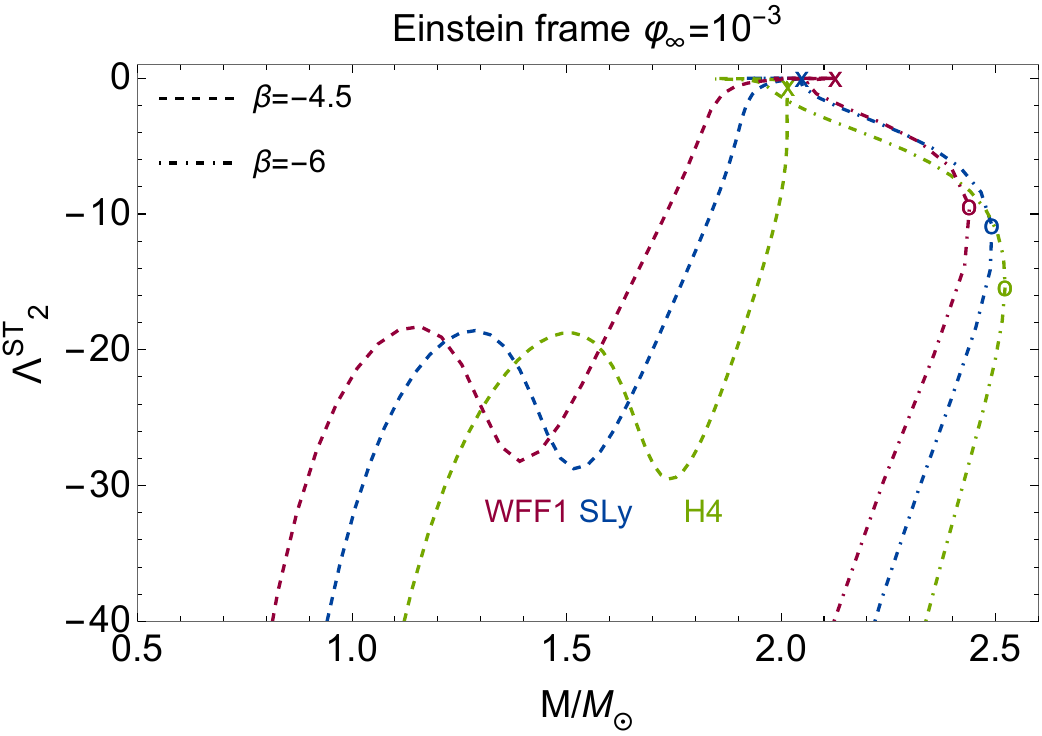}}
{\includegraphics[width=0.49\textwidth,clip]{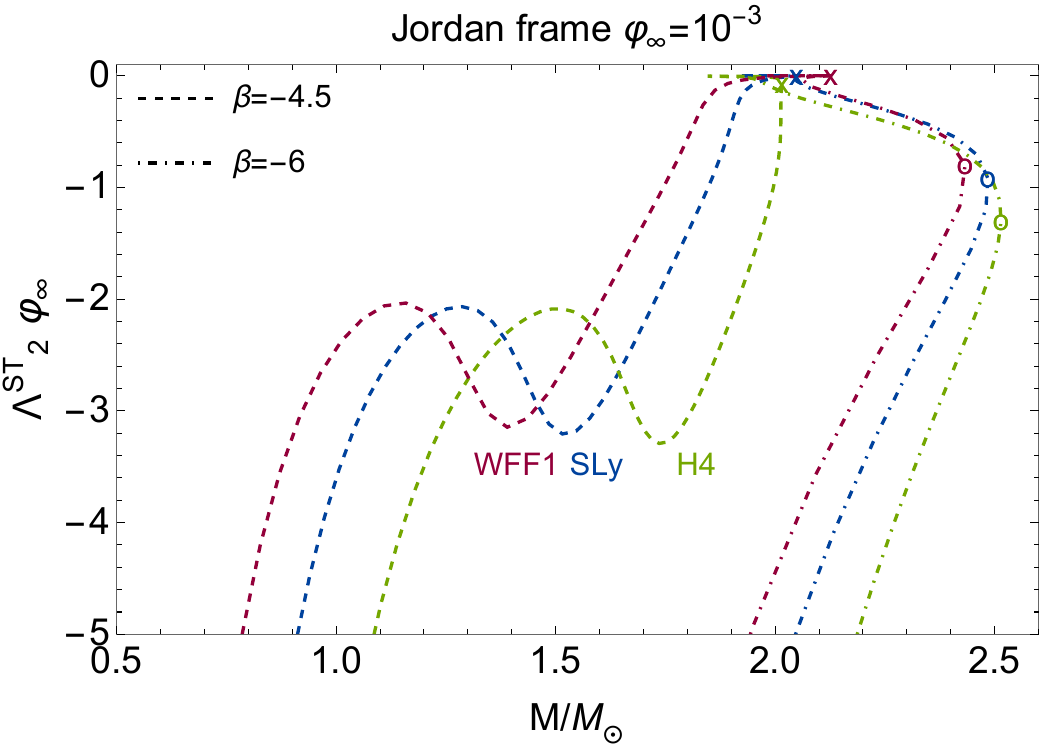}}
\caption{\textbf{\emph{Quadrupolar scalar-tensor adimensional tidal deformabilities}} $\Lambda_2^{ST}$ in the Einstein and Jordan frames for three equations of state (WFF1, SLy, and H4).  The Jordan frame plots use the Jordan frame mass. The solid lines represent the GR configurations $\beta=0$ and the dashed and dot-dashed lines are the scalarized configurations with $\beta=-4.5$ and $\beta=-6$, respectively, and the plots in the bottom row are enlarged with respect to their counterparts in the top row. The cross represents the maximum mass configuration for $\beta=0,-4.5$, and the circle for $\beta=-6$. All plots correspond to a scalar field at infinity $\varphi_{0\infty}=10^{-3}$.}\label{fig:LambdaScalarTensor2}
\end{center}
\begin{center}
{\includegraphics[width=0.49\textwidth,clip]{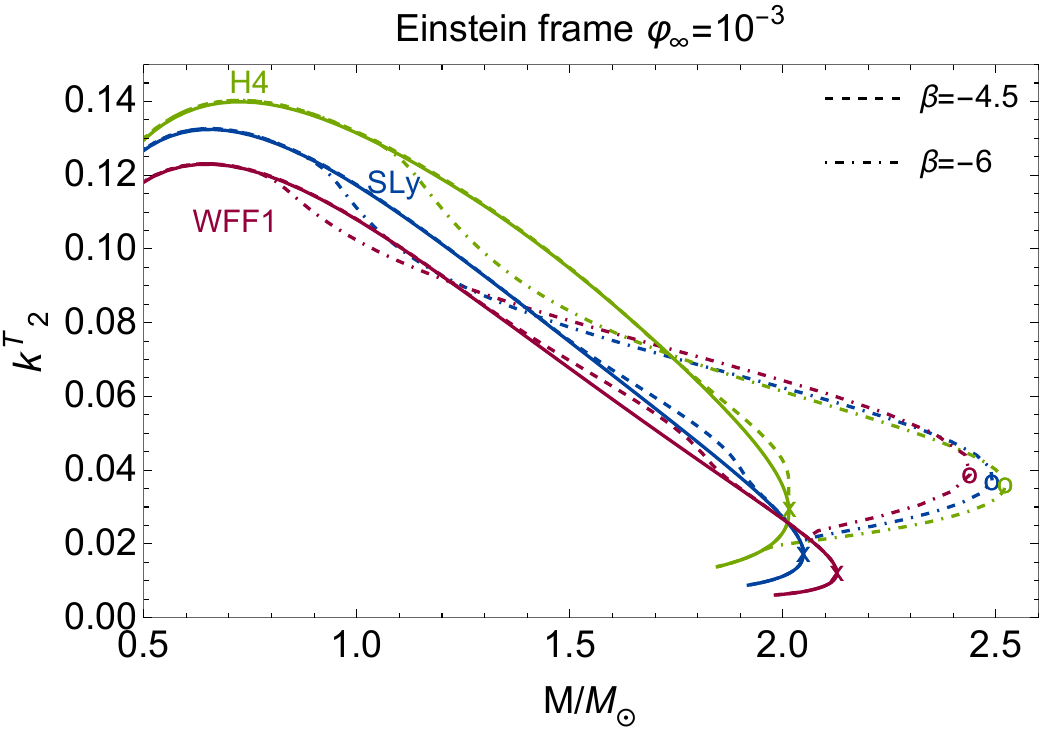}}
{\includegraphics[width=0.49\textwidth,clip]{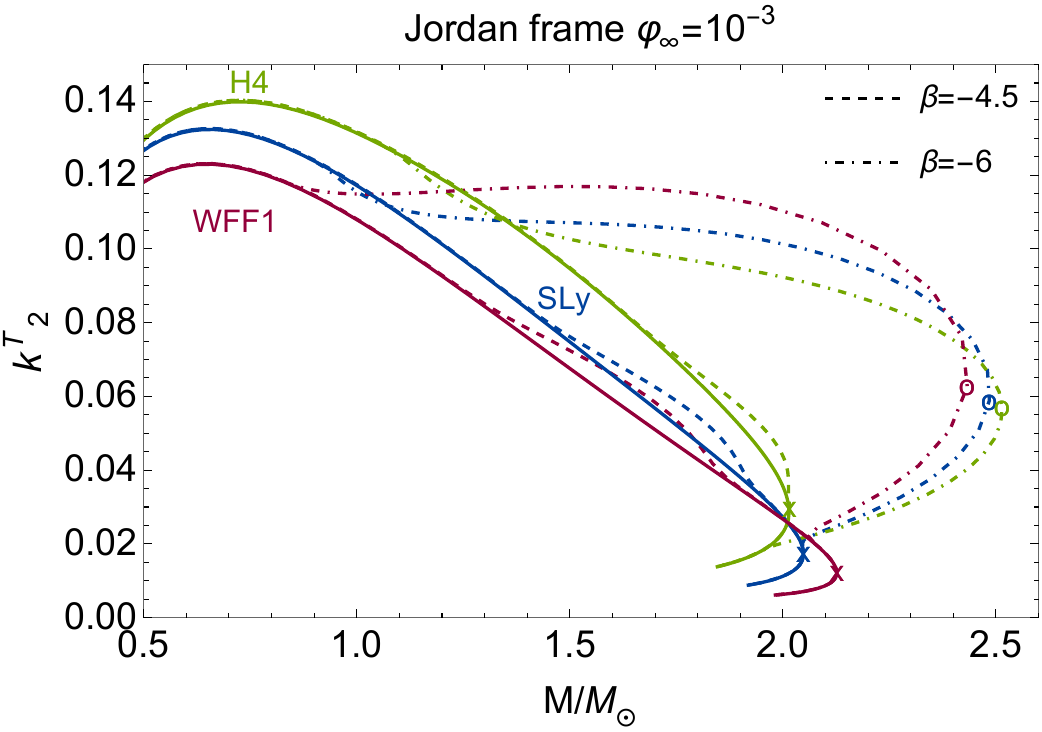}}
\caption{\textbf{\emph{Quadrupolar tensor Love numbers}} $k_2^{T}$ in the Einstein and Jordan frames for three equations of state (WFF1, SLy, and H4). The Jordan frame plots use the Jordan frame mass. The solid lines represent the GR configurations $\beta=0$ and the dashed and dot-dashed lines are the scalarized configurations with $\beta=-4.5$ and $\beta=-6$, respectively. The cross represents the maximum mass configuration for $\beta=0,-4.5$, and the circle for $\beta=-6$. Both plots correspond to a scalar field at infinity $\varphi_{0\infty}=10^{-3}$.}\label{fig:Love2TPlot}
\end{center}
\end{figure*}

\begin{figure*}[p]
\begin{center}
{\includegraphics[width=0.49\textwidth,clip]{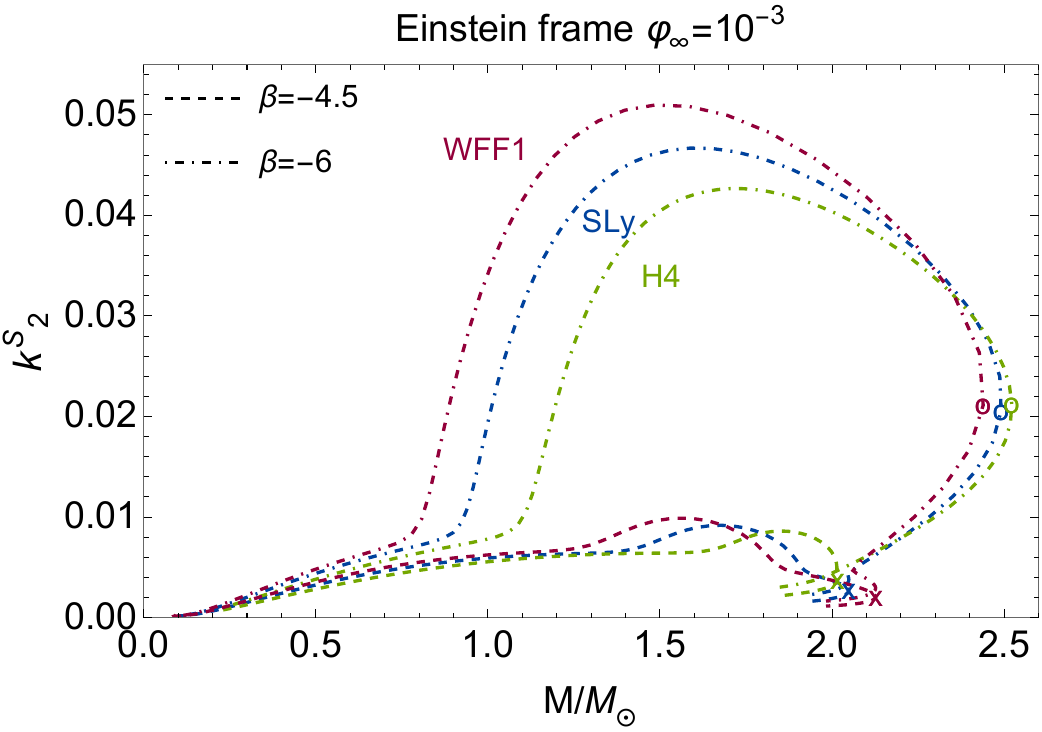}}
{\includegraphics[width=0.49\textwidth,clip]{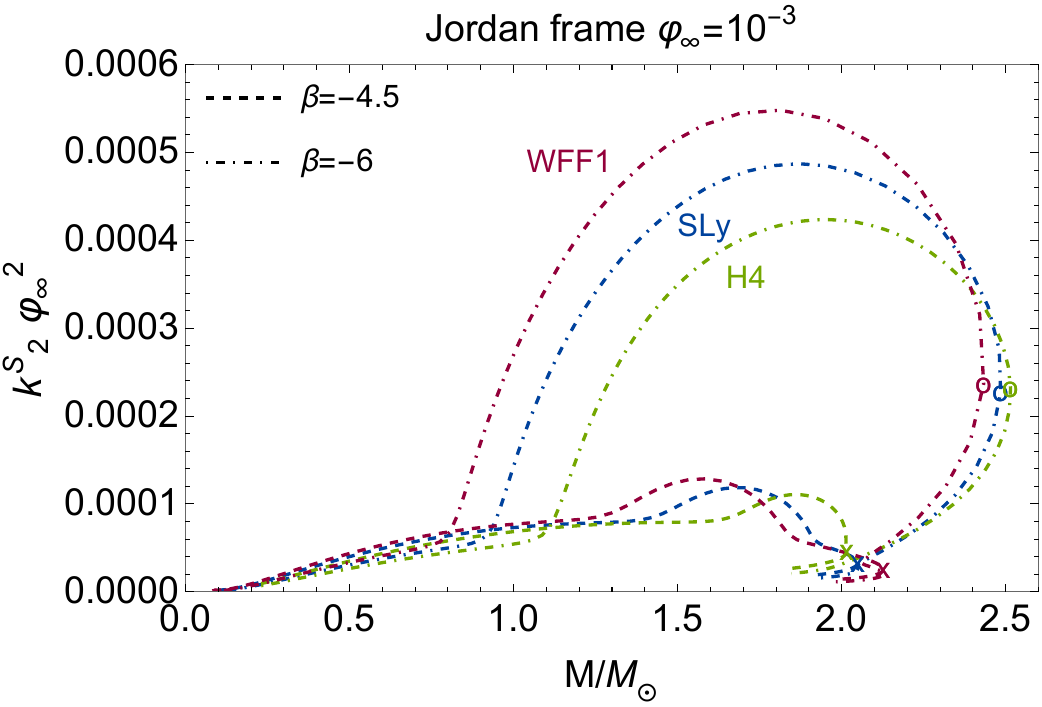}}
{\includegraphics[width=0.49\textwidth,clip]{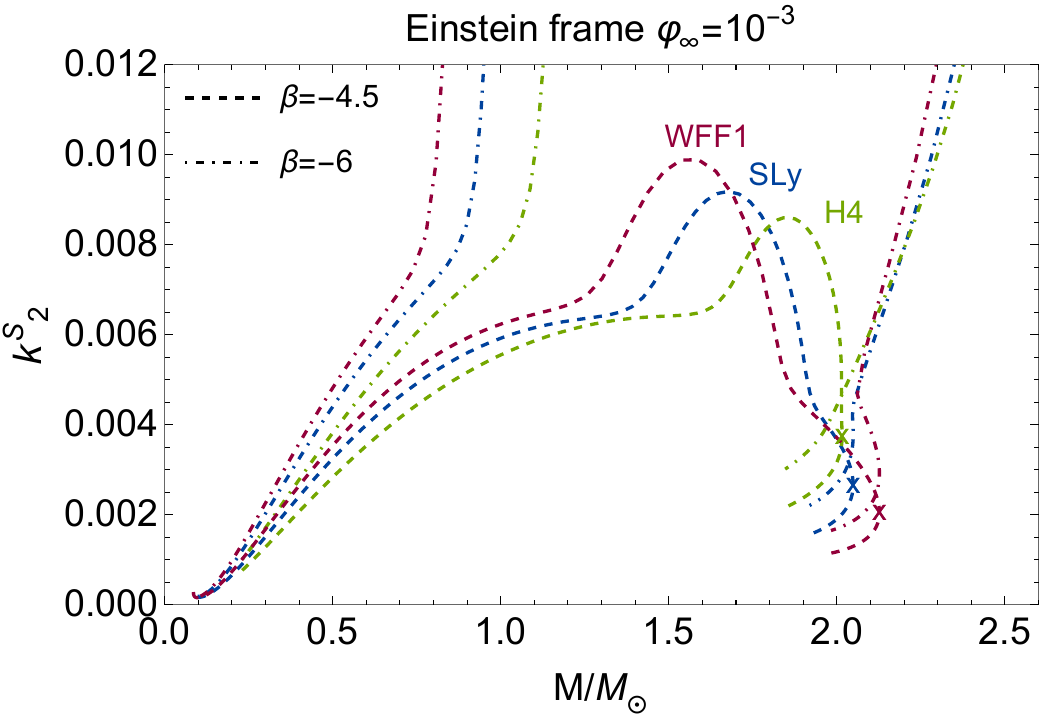}}
{\includegraphics[width=0.49\textwidth,clip]{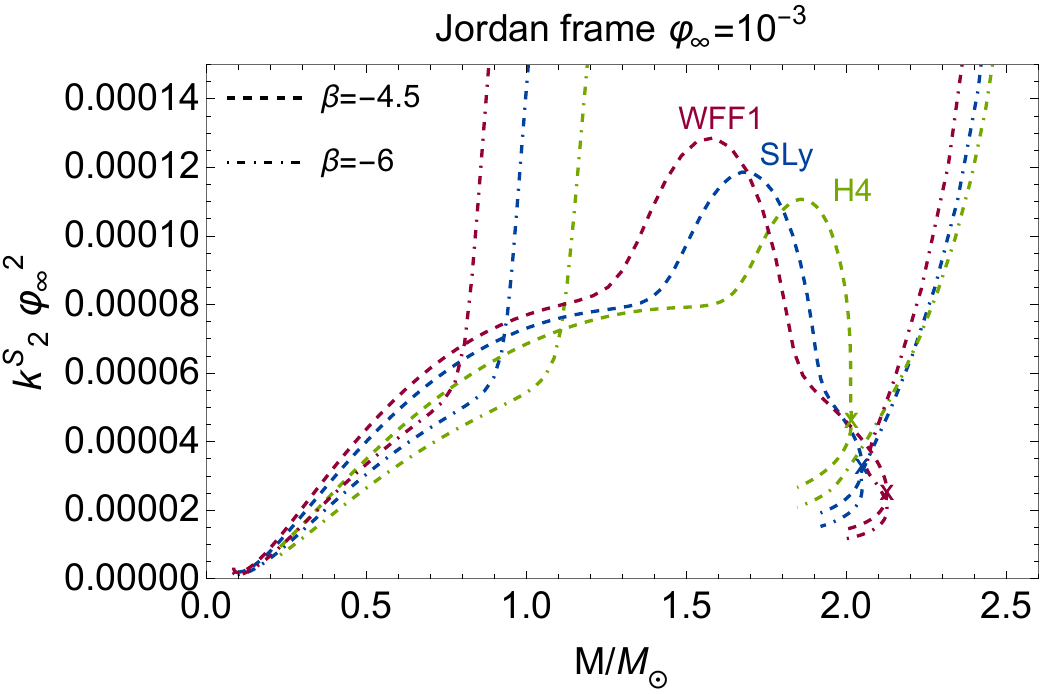}}
\caption{\textbf{\emph{Quadrupolar scalar Love numbers}} $k_2^{S}$ in the Einstein and Jordan frames for three equations of state (WFF1, SLy, and H4). The Jordan frame plots use the Jordan frame mass. The solid lines represent the GR configurations $\beta=0$ and the dashed and dot-dashed lines are the scalarized configurations with $\beta=-4.5$ and $\beta=-6$, respectively, and the plots in the bottom row are enlarged with respect to their counterparts in the top row. The cross represents the maximum mass configuration for $\beta=0,-4.5$, and the circle for $\beta=-6$. All plots correspond to a scalar field at infinity $\varphi_{0\infty}=10^{-3}$.}\label{fig:Love2SPlot}
\end{center}
\begin{center}
{\includegraphics[width=0.49\textwidth,clip]{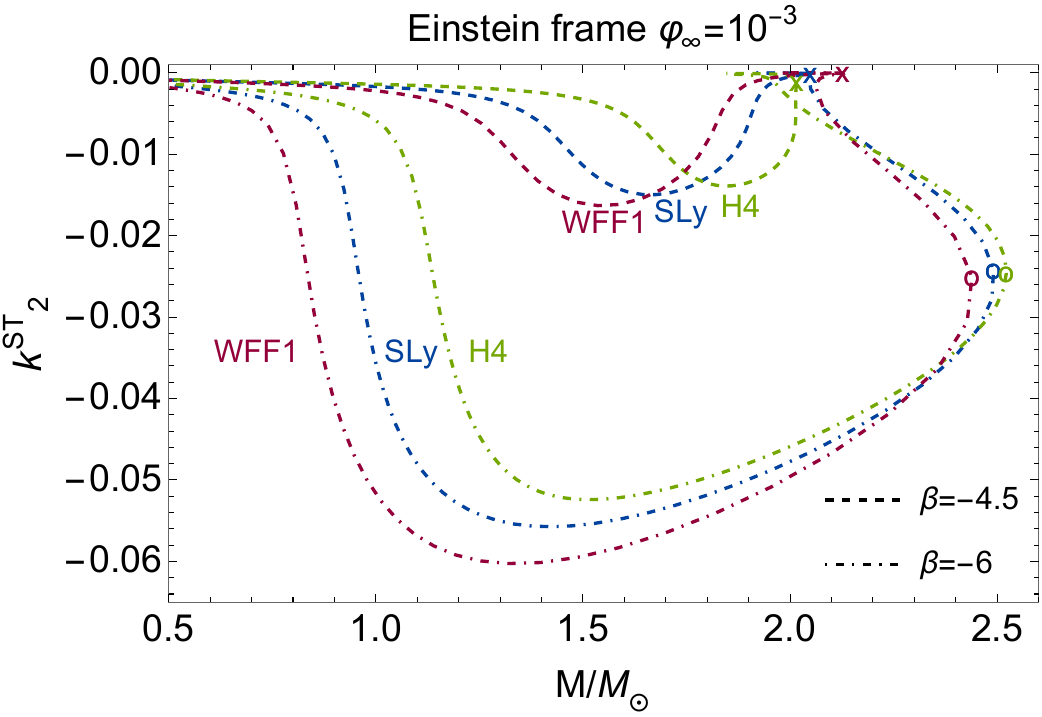}}
{\includegraphics[width=0.49\textwidth,clip]{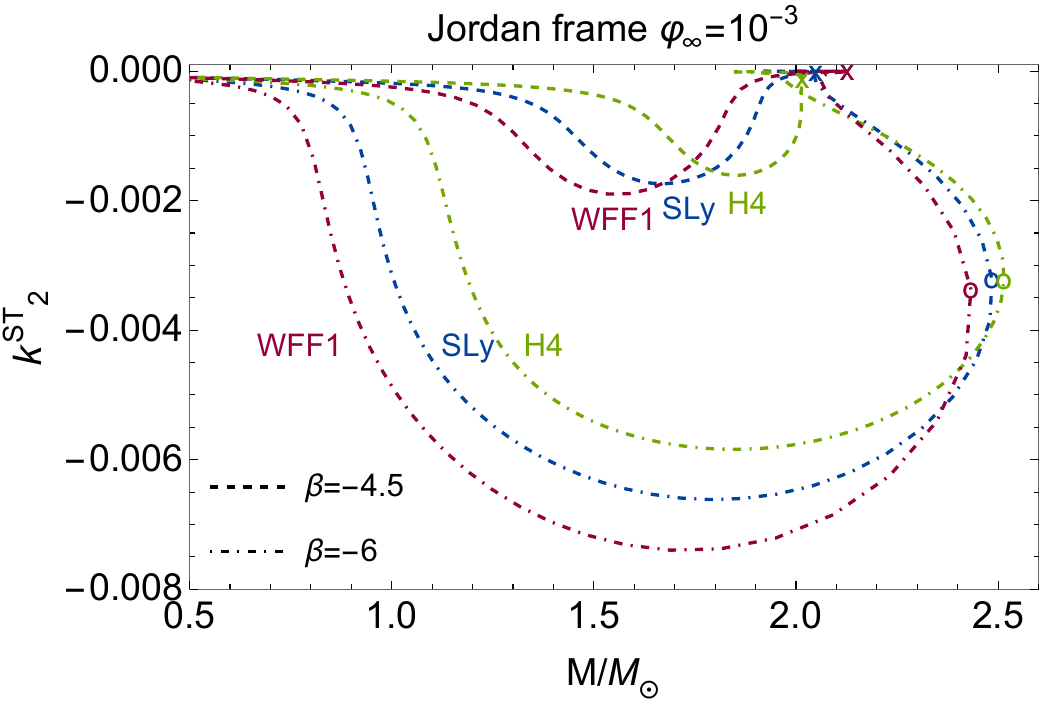}}
\caption{\textbf{\emph{Quadrupolar scalar-tensor Love numbers}} $k_2^{ST}$ in the Einstein and Jordan frames for three equations of state (WFF1, SLy, and H4). The Jordan frame plots use the Jordan frame mass. The solid lines represent the GR configurations $\beta=0$ and the dashed and dot-dashed lines are the scalarized configurations with $\beta=-4.5$ and $\beta=-6$, respectively. The cross represents the maximum mass configuration for $\beta=0,-4.5$, and the circle for $\beta=-6$. Both plots correspond to a scalar field at infinity $\varphi_{0\infty}=10^{-3}$.}\label{fig:Love2STPlot}
\end{center}
\end{figure*}

\begin{figure*}
\subsection{Octupolar $\ell=3$}
\begin{justify}
Figure \ref{fig:Lambda3Tensor} shows the results for the dimensionless tensor octupolar tidal deformability quantities $\Lambda_3^T$ defined in Eq.~\eqref{eq:LambdaDef} for different equations of state (different colors), coupling coefficients (different dashings), and in both the Einstein (left panel) and Jordan frame (right panel). Figure~\ref{fig:Lambda3Scalar} shows analogous results for the scalar deformability, while Fig.~\ref{fig:LambdaScalarTensor3} illustrates those for the mixed scalar-tensor quantity. The corresponding results for the dimensionless Love numbers $k_3$ defined in Eq.~\eqref{eq:LoveNumvsTidalDef} are shown in Fig.~\ref{fig:Love3TPlot} for the tensor, Fig.~\ref{fig:Love3SPlot} for the scalar, and Fig.~\ref{fig:Love3STPlot} for the mixed quantities.
\end{justify}
\begin{center}
{\includegraphics[width=0.49\textwidth,clip]{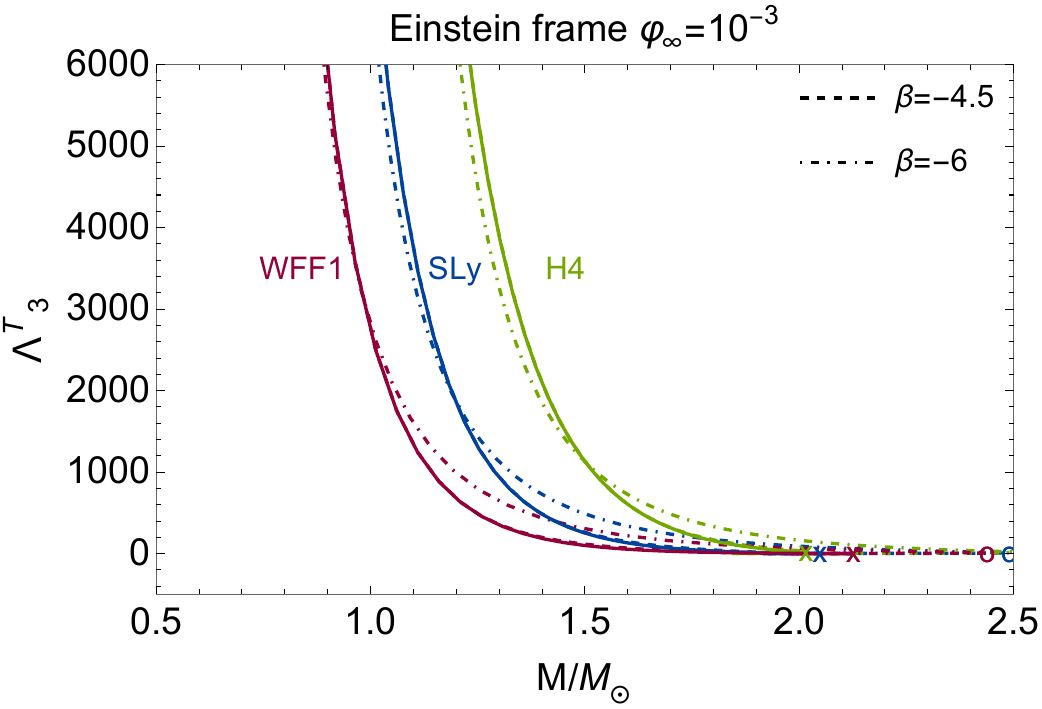}}
{\includegraphics[width=0.49\textwidth,clip]{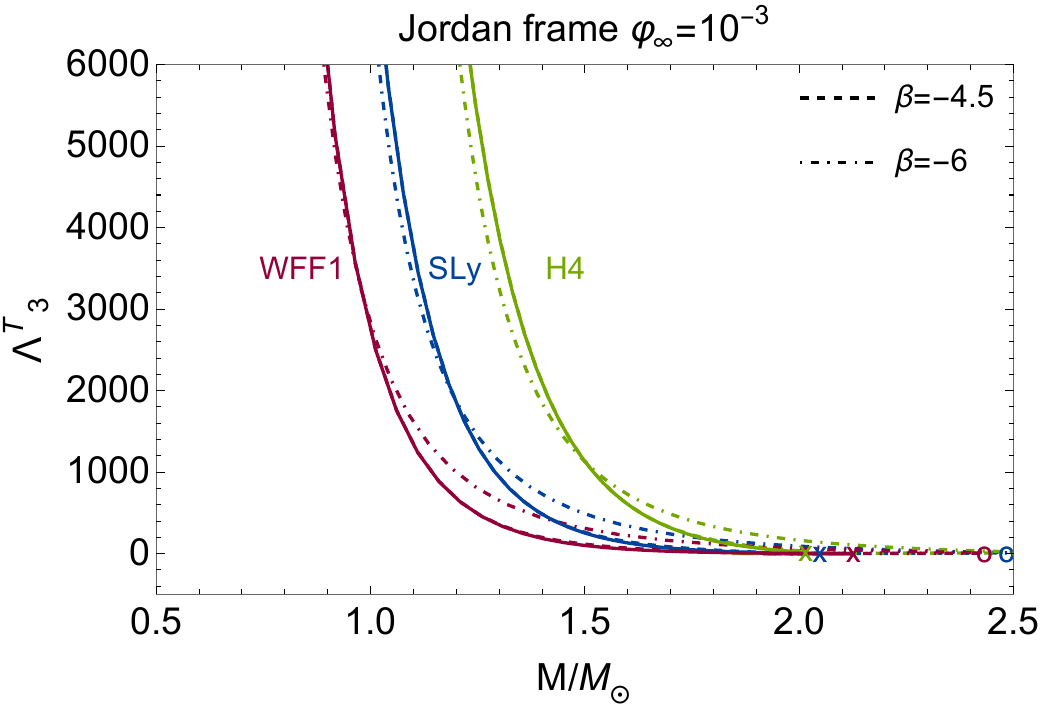}}
\caption{\textbf{\emph{Octupolar tensor adimensional tidal deformabilities}} $\Lambda_3^T$ in the Einstein and Jordan frames for three equations of state (WFF1, SLy, and H4). The Jordan frame plots use the Jordan frame mass. The solid lines represent the GR configurations $\beta=0$ and the dashed and dot-dashed lines are the scalarized configurations with $\beta=-4.5$ and $\beta=-6$, respectively. The cross represents the maximum mass configuration for $\beta=0,-4.5$, and the circle for $\beta=-6$. Both plots correspond to a scalar field at infinity $\varphi_{0\infty}=10^{-3}$.}\label{fig:Lambda3Tensor}
\end{center}
\end{figure*}

\begin{figure*}
\begin{center}
{\includegraphics[width=0.49\textwidth,clip]{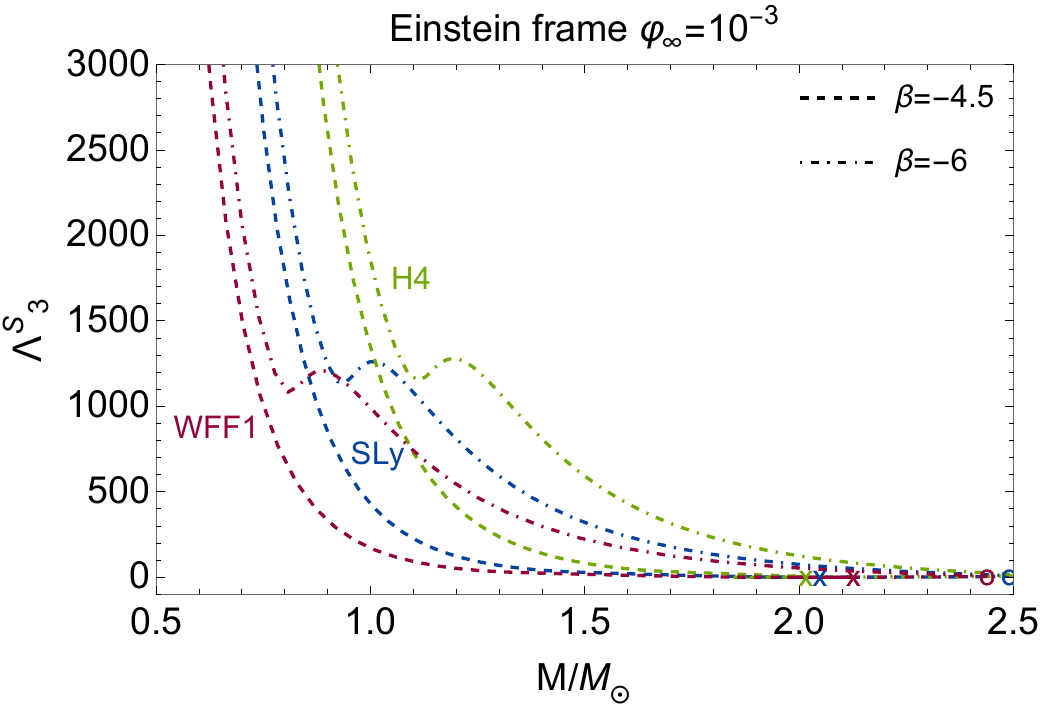}}
{\includegraphics[width=0.49\textwidth,clip]{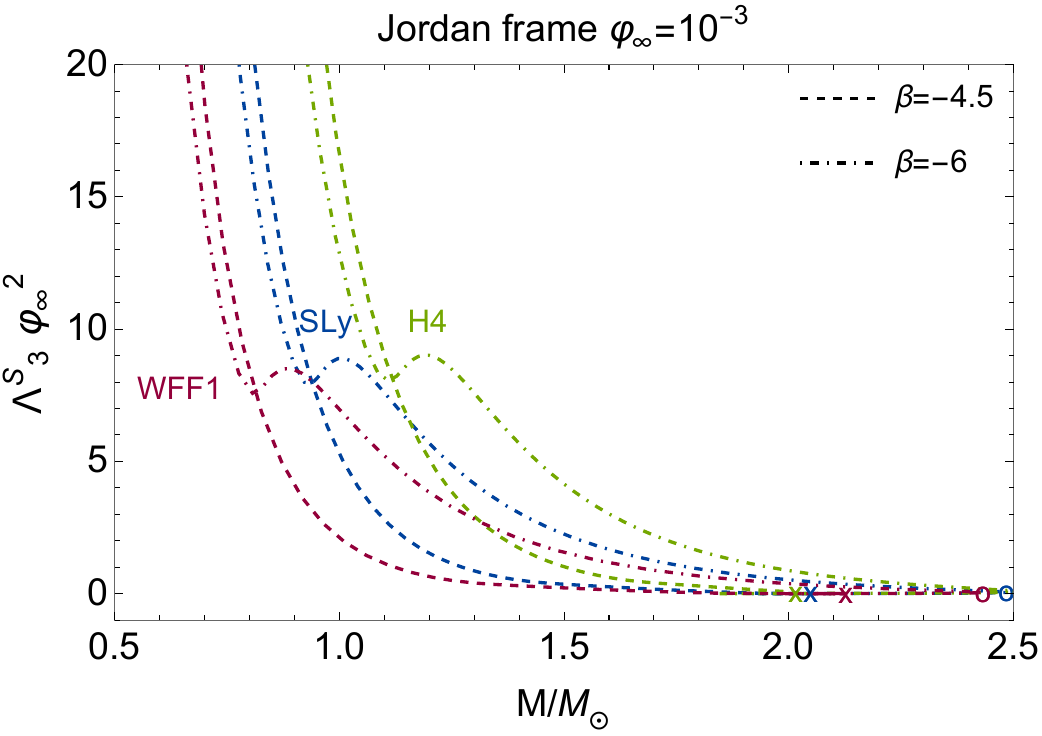}}
\caption{\textbf{\emph{Octupolar scalar adimensional tidal deformabilities}} $\Lambda_3^S$ in the Einstein and Jordan frames for three equations of state (WFF1, SLy, and H4). The Jordan frame plots use the Jordan frame mass. The dashed and dot-dashed lines are the scalarized configurations with $\beta=-4.5$ and $\beta=-6$, respectively. The cross represents the maximum mass configuration for $\beta=0,-4.5$, and the circle for $\beta=-6$. Both plots correspond to a scalar field at infinity $\varphi_{0\infty}=10^{-3}$.}\label{fig:Lambda3Scalar}
\end{center}
\end{figure*}
\begin{figure*}[p]
\begin{center}
{\includegraphics[width=0.49\textwidth,clip]{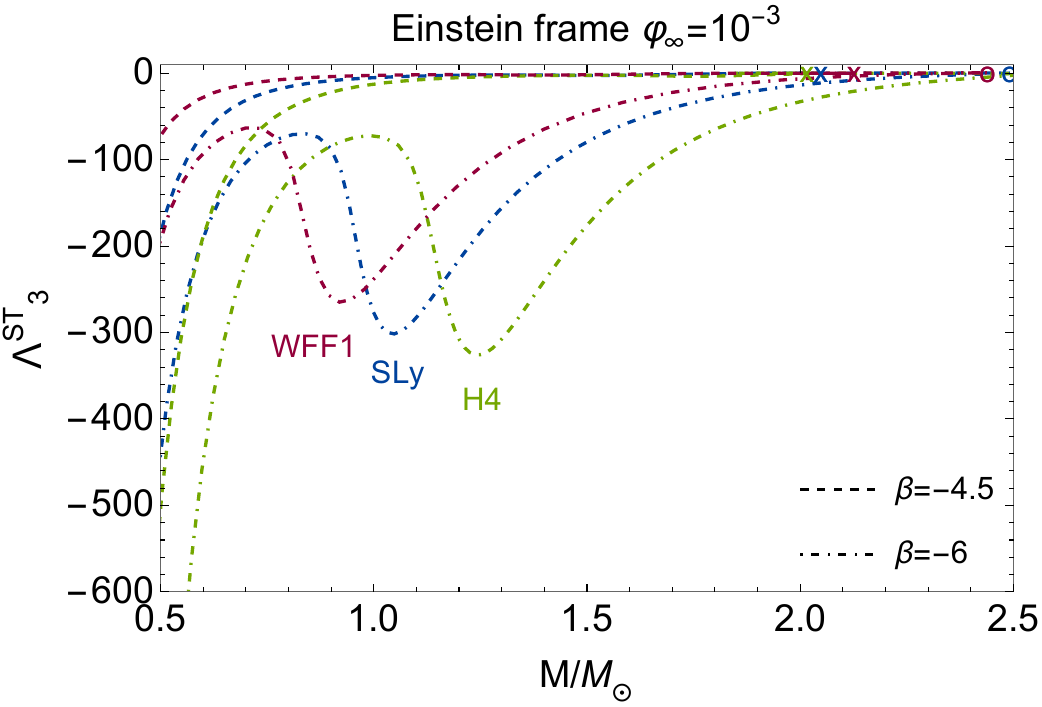}}
{\includegraphics[width=0.49\textwidth,clip]{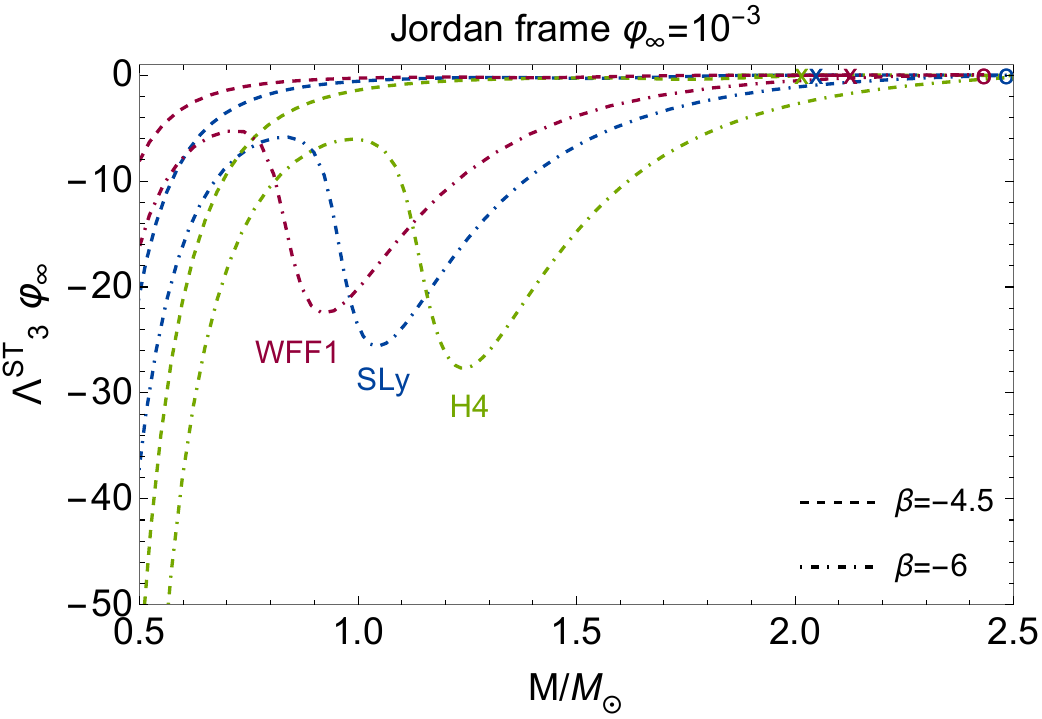}}
{\includegraphics[width=0.49\textwidth,clip]{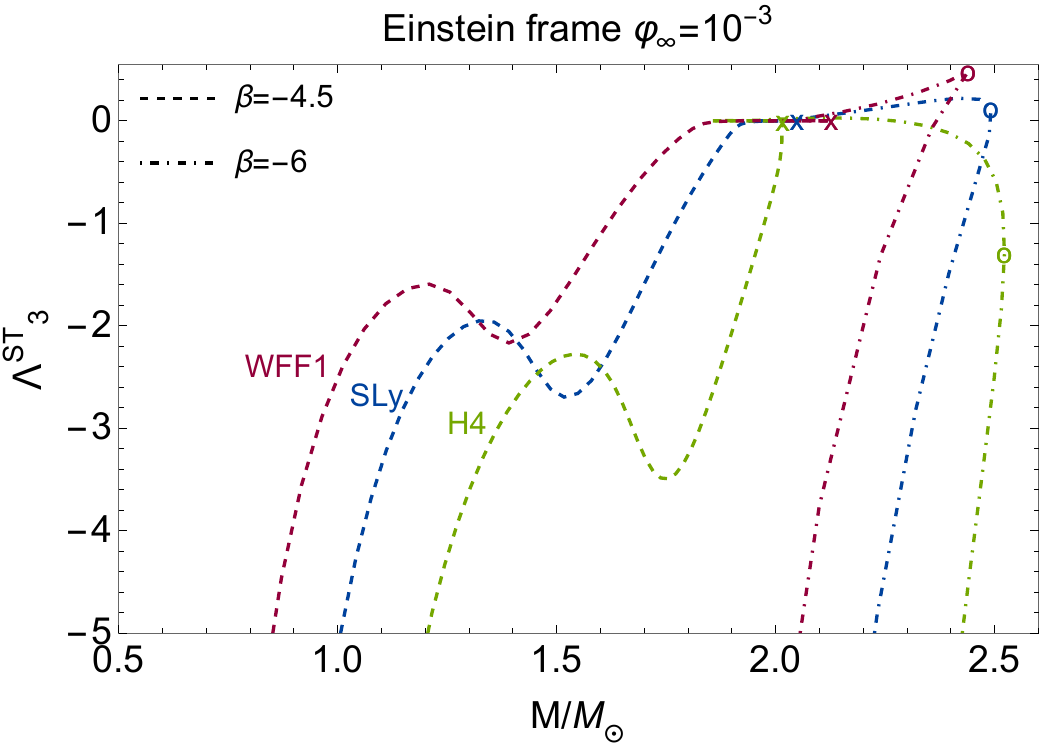}}
{\includegraphics[width=0.49\textwidth,clip]{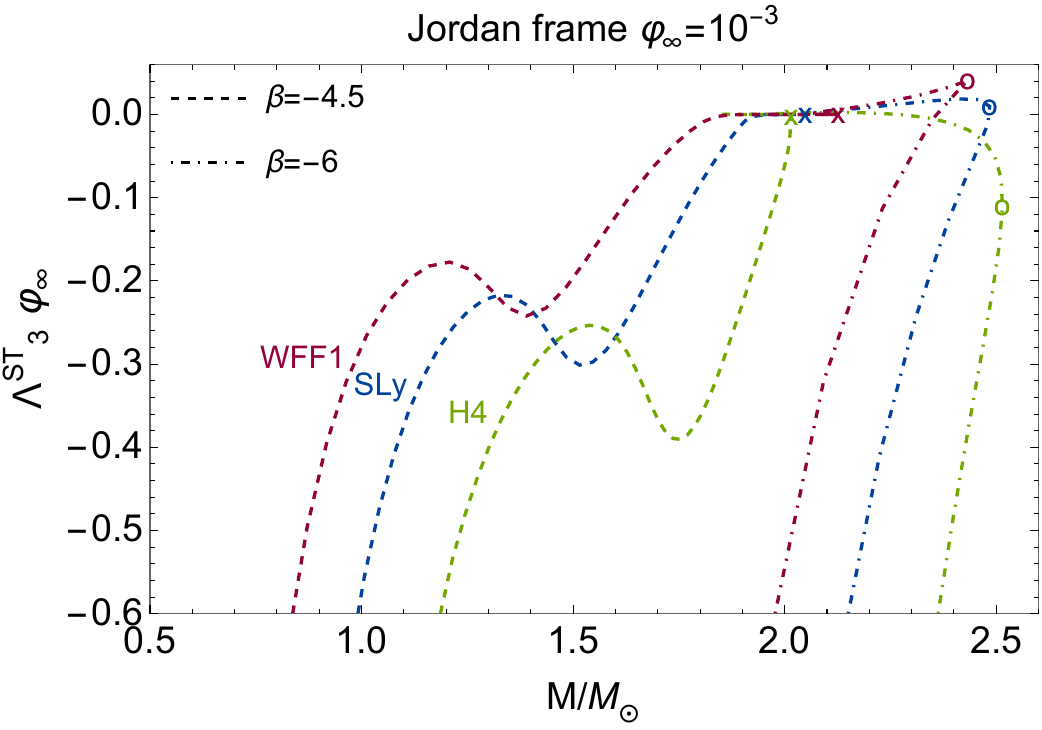}}
\caption{\textbf{\emph{Octupolar scalar-tensor adimensional tidal deformabilities}} $\Lambda_3^{ST}$ in the Einstein and Jordan frames for three equations of state (WFF1, SLy, and H4). The Jordan frame plots use the Jordan frame mass. The solid lines represent the GR configurations $\beta=0$ and the dashed and dot-dashed lines are the scalarized configurations with $\beta=-4.5$ and $\beta=-6$, respectively, and the plots in the bottom row
are enlarged with respect to their counterparts in the top row. The cross represents the maximum mass configuration for $\beta=0,-4.5$, and the circle for $\beta=-6$. All plots correspond to a scalar field at infinity $\varphi_{0\infty}=10^{-3}$.}\label{fig:LambdaScalarTensor3}
\end{center}
\begin{center}
{\includegraphics[width=0.49\textwidth,clip]{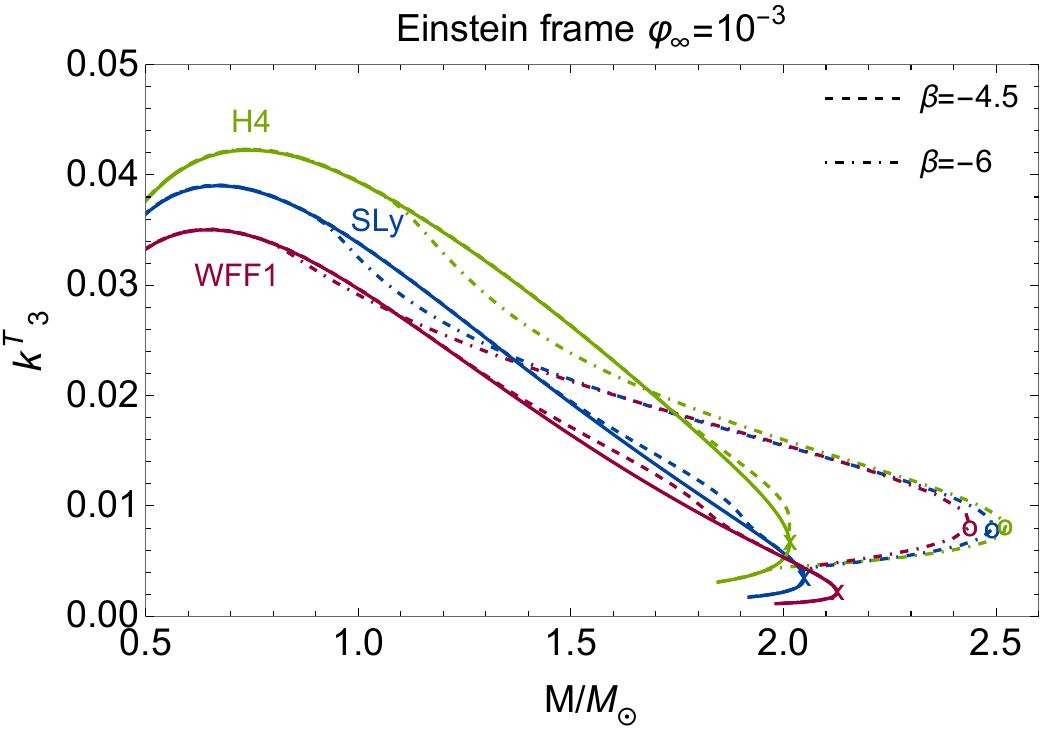}}
\includegraphics[width=0.49\textwidth,clip]{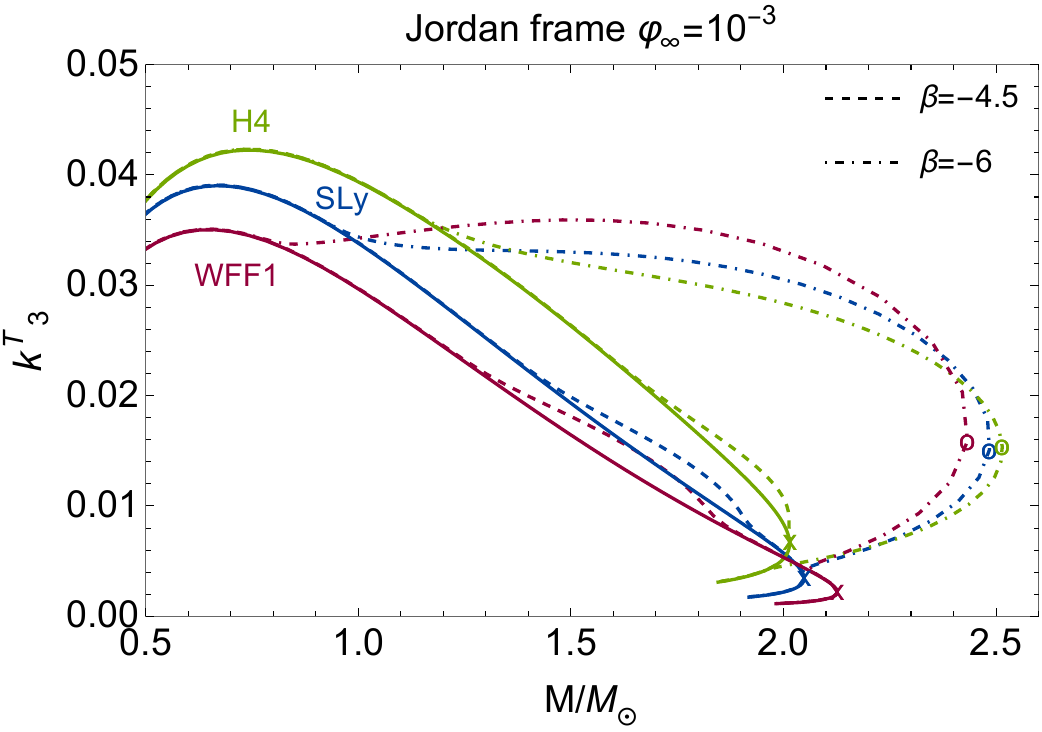}
\caption{\textbf{\emph{Octupolar tensor Love numbers}} $k_3^{T}$ in the Einstein and Jordan frames for three equations of state (WFF1, SLy, and H4). The solid lines represent the GR configurations $\beta=0$ and the dashed and dot-dashed lines are the scalarized configurations with $\beta=-4.5$ and $\beta=-6$, respectively. The cross represents the maximum mass configuration for $\beta=0,-4.5$, and the circle for $\beta=-6$. Both plots correspond to a scalar field at infinity $\varphi_{0\infty}=10^{-3}$.}\label{fig:Love3TPlot}
\end{center}
\end{figure*}

\begin{figure*}[p]
\begin{center}
{\includegraphics[width=0.49\textwidth,clip]{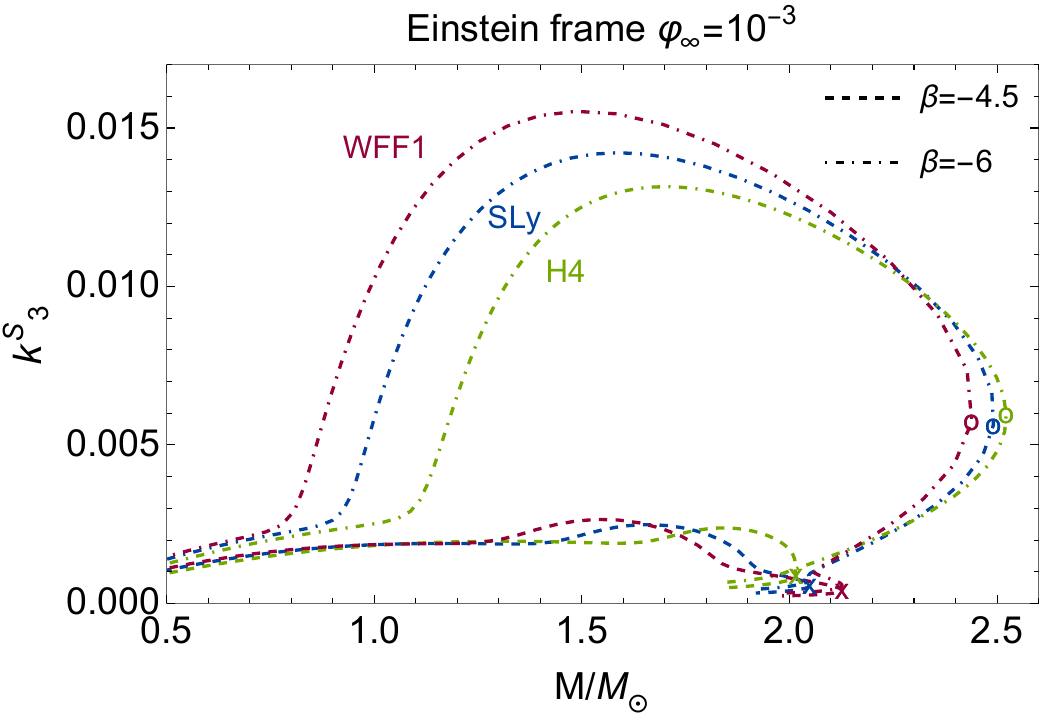}}
{\includegraphics[width=0.49\textwidth,clip]{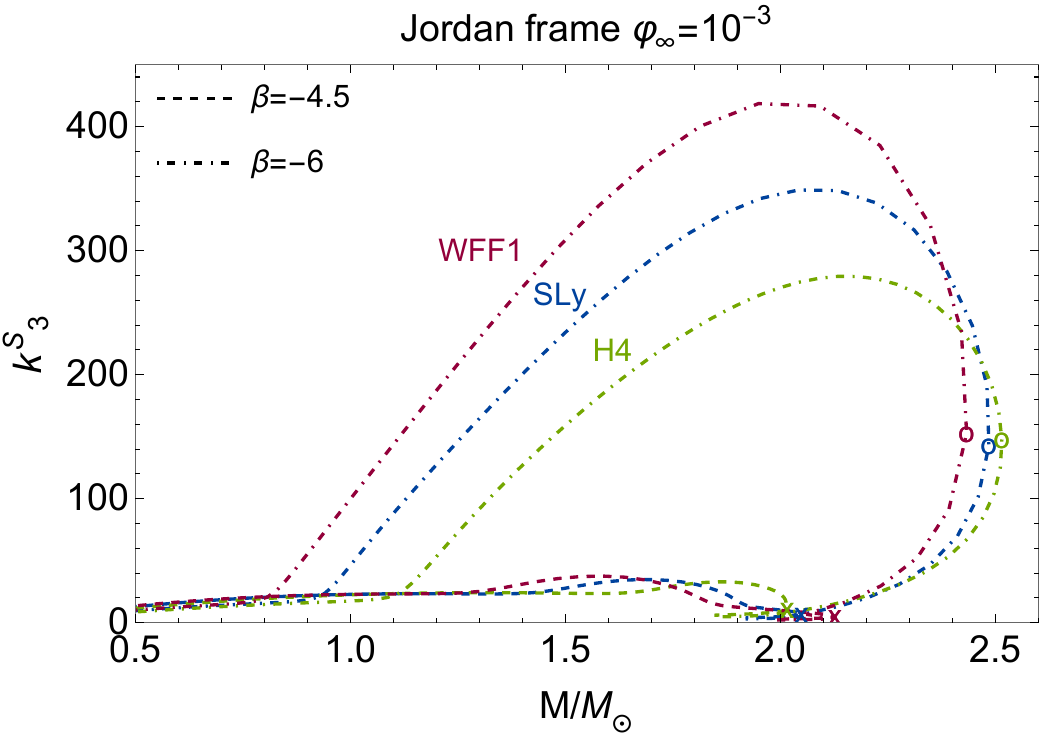}}
{\includegraphics[width=0.49\textwidth,clip]{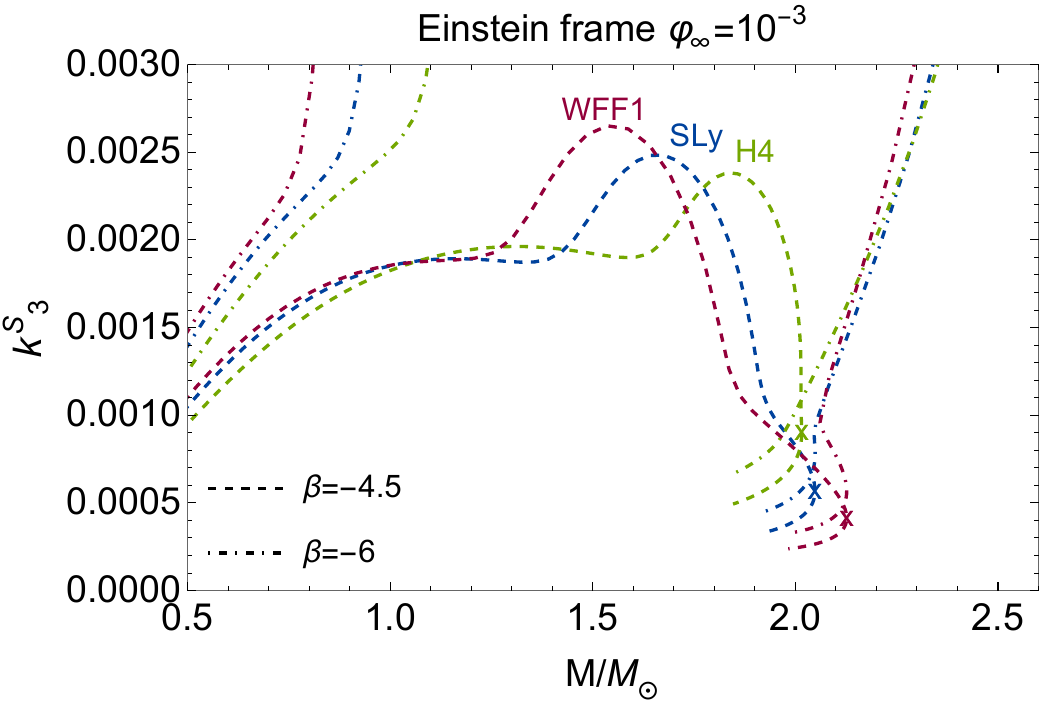}}
{\includegraphics[width=0.49\textwidth,clip]{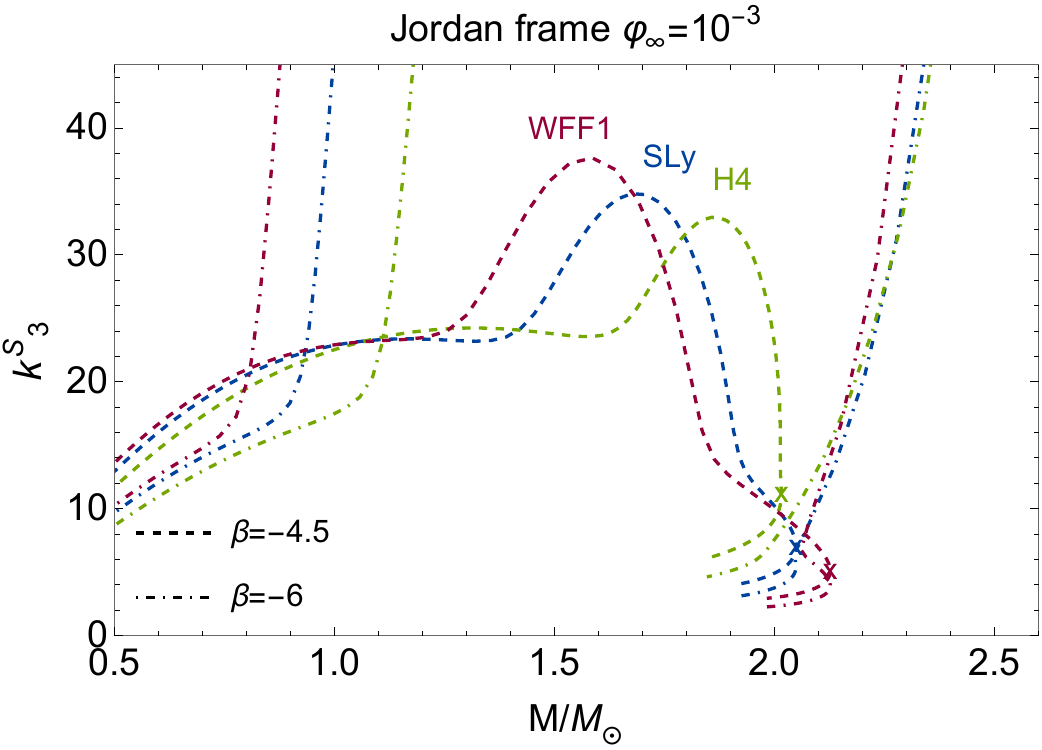}}
\caption{\textbf{\emph{Octupolar scalar Love numbers}} $k_3^{S}$ in the Einstein and Jordan frames for three equations of state (WFF1, SLy, and H4). The Jordan frame plots use the Jordan frame mass. The solid lines represent the GR configurations $\beta=0$ and the dashed and dot-dashed lines are the scalarized configurations with $\beta=-4.5$ and $\beta=-6$, respectively, and the plots in the bottom row are enlarged with respect to their counterparts in the top row. The cross represents the maximum mass configuration for $\beta=0,-4.5$, and the circle for $\beta=-6$. All plots correspond to a scalar field at infinity $\varphi_{0\infty}=10^{-3}$.}\label{fig:Love3SPlot}
\end{center}
\begin{center}
{\includegraphics[width=0.49\textwidth,clip]{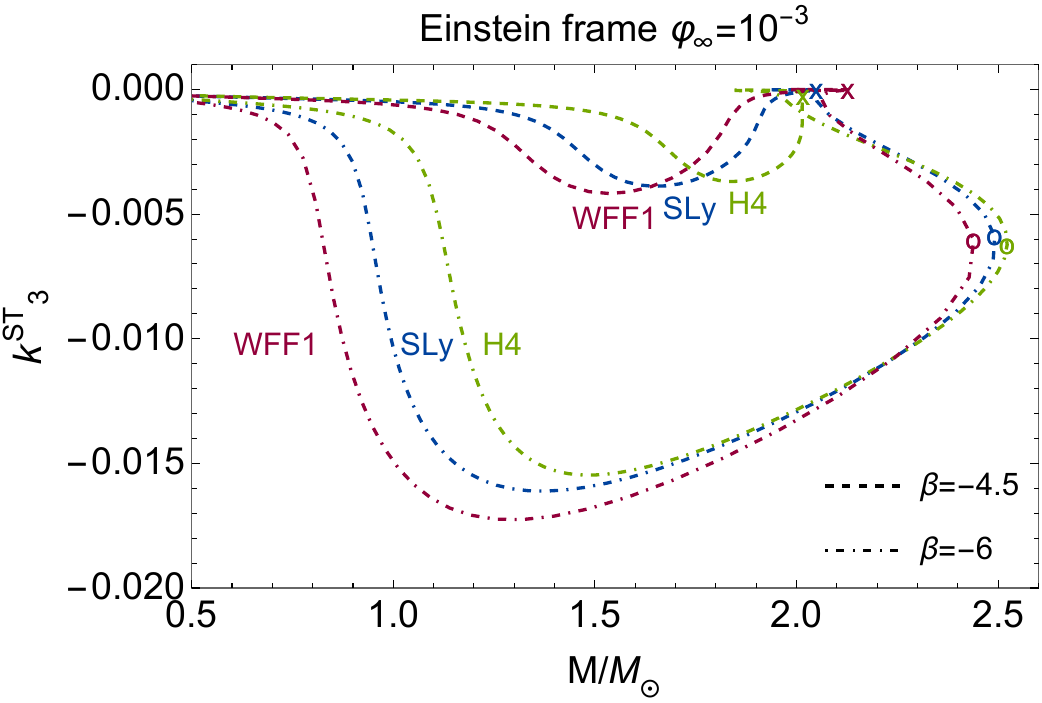}}
{\includegraphics[width=0.49\textwidth,clip]{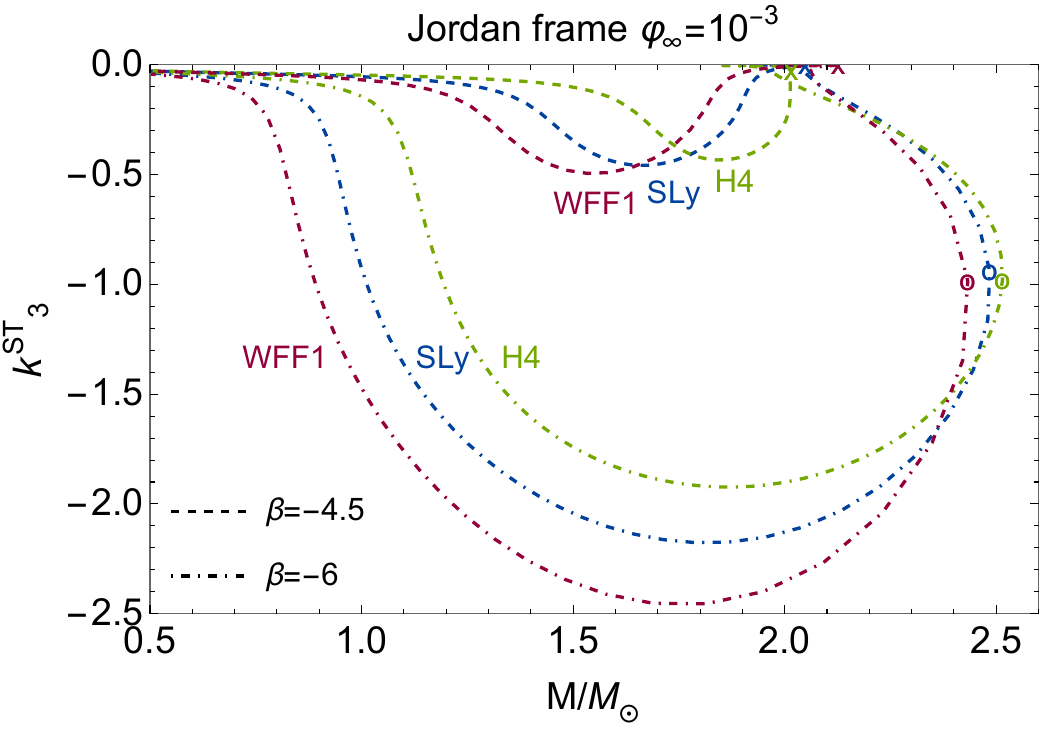}}
\caption{\textbf{\emph{Octupolar scalar-tensor Love numbers}} $k_3^{ST}$ in the Einstein and Jordan frames for three equations of state (WFF1, SLy, and H4). The Jordan frame plots use the Jordan frame mass. The solid lines represent the GR configurations $\beta=0$ and the dashed and dot-dashed lines are the scalarized configurations with $\beta=-4.5$ and $\beta=-6$, respectively. The cross represents the maximum mass configuration for $\beta=0,-4.5$, and the circle for $\beta=-6$. Both plots correspond to a scalar field at infinity $\varphi_{0\infty}=10^{-3}$.}\label{fig:Love3STPlot}
\end{center}
\end{figure*}
\clearpage
\begin{widetext}
\section{Comparison with Brown 2022}\label{app:ComparisonBrown}
In Figure~\ref{fig:AbsRelDiff} we show the absolute relative difference (ARD),
\begin{align}
    {\rm ARD}^{(S/T)}=\left|\frac{{\lambda_2^\ast}^{(S/T)}-{\lambda_2^{\rm Brown}}^{(S/T)}}{{\lambda_2^\ast}^{(S/T)}}\right|~,
\end{align}
with ${\lambda_2^{\rm Brown}}^{(S/T)}$ the quadrupolar tidal deformability from \cite{Brown:2022kbw} computed setting the source terms in \eqref{eq:H0PertEq} and \eqref{eq:PhiPertEq} to zero, $f_s=0=g_s$. In particular, the ARD for the tensor tidal deformability is at most $4\%$ for $\beta=-4.5$ and $26\%$ for $\beta=-6$. For the scalar tidal deformability the ARD is at most $18\%$ for $\beta=-4.5$ and $35\%$ for $\beta=-6$. This implies that neglecting the source terms can introduce noticeable inaccuracies into the scalar and tensor tidal deformabilities in some regions of the parameter space.
\begin{figure*}[htpb!]
\begin{center}
{\includegraphics[width=0.49\textwidth,clip]{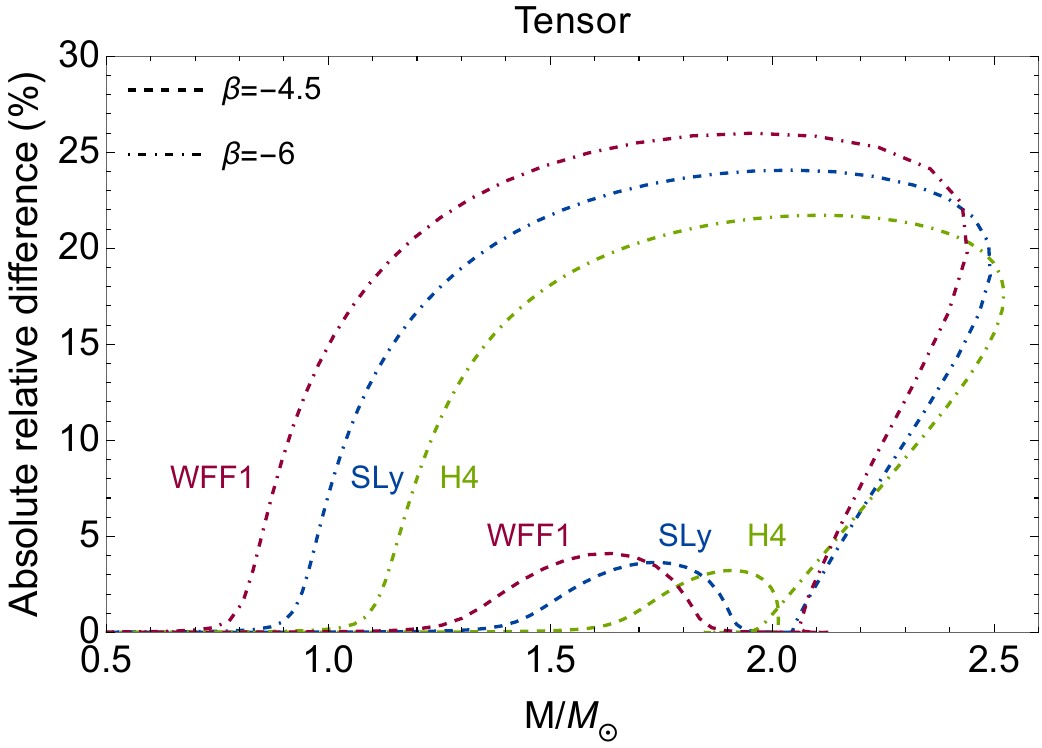}}
{\includegraphics[width=0.49\textwidth,clip]{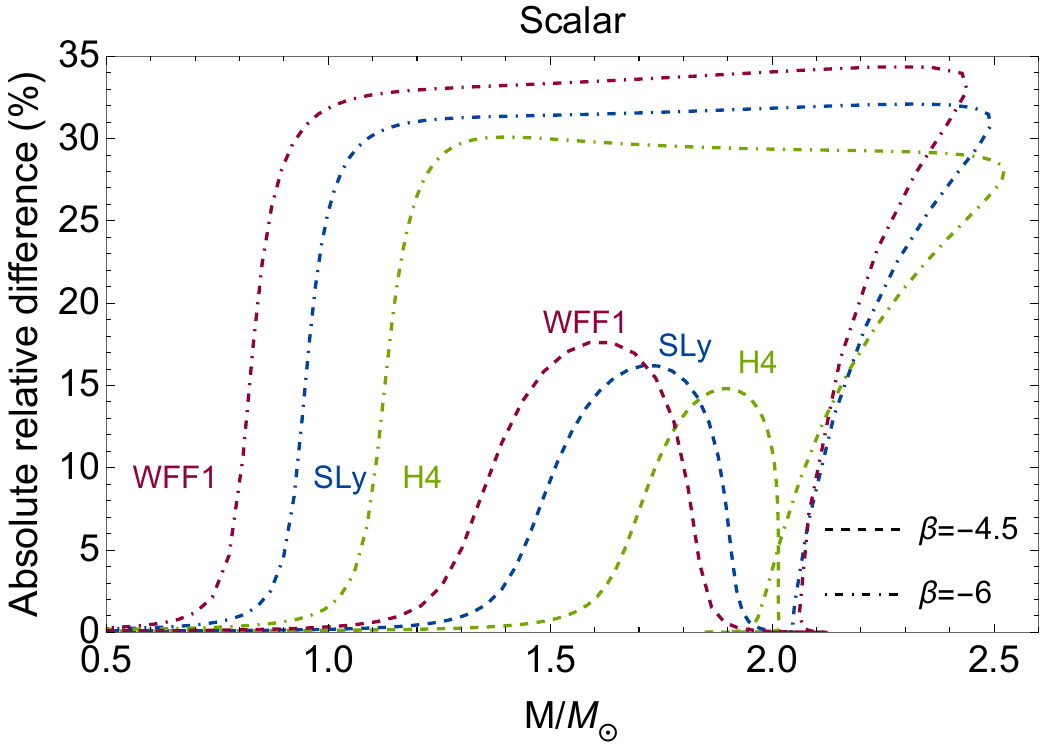}}
\caption{\textbf{\emph{Absolute relative difference}} between the tensor (left) and scalar (right) quadrupolar tidal deformabilities computed with and without setting the source term in the perturbation equations of motion to zero. The mass is the ADM mass in the Einstein frame. We consider three equations of state (WFF1, SLy and H4) and the dashed and dot-dashed lines are the scalarized configurations with $\beta=-4.5$ and $\beta=-6$, respectively. We set a scalar field at infinity $\varphi_{0\infty}=10^{-3}$.}\label{fig:AbsRelDiff}
\end{center}
\end{figure*}
\end{widetext}
\clearpage
\bibliography{Notesbibliography}
\end{document}